\documentclass[journal,twoside,web]{ieeecolor}
\usepackage{tmi}
\usepackage{cite}
\usepackage{amsmath,amssymb,amsfonts}
\usepackage{algorithmic}
\usepackage{graphicx}
\usepackage{textcomp}

\usepackage{lineno,hyperref}
\usepackage{booktabs,makecell,multirow}
\usepackage{amssymb}
\usepackage{graphicx,subfigure}
\usepackage{color}
\usepackage[section]{placeins}
\usepackage{ulem}
\usepackage{grffile}
\usepackage{tabularx}
\usepackage{stfloats}

\def\DELM#1{\iffalse{#1}\fi}
\def\ADDM#1{#1}

\def\DEL#1{\iffalse{#1}\fi}
\def\ADD#1{#1}
\def\MOVE#1{#1}

 \def\BibTeX{{\rm B\kern-.05em{\sc i\kern-.025em b}\kern-.08em
     T\kern-.1667em\lower.7ex\hbox{E}\kern-.125emX}}
\begin{document}
\title{Co-Correcting: Noise-tolerant Medical Image Classification via mutual Label Correction}
\author{Jiarun Liu \IEEEmembership{Member, IEEE}, Ruirui Li \IEEEmembership{Member, IEEE}, Chuan Sun 
\thanks{This work was suppoted by joint project of BRC-BC(Biomedical Translational) Engineering research center of BUCT-CJFH XK2020-07. (Corresponding author: Ruirui Li.)}
\thanks{Jiarun Liu is with College of Information Science and Technology, Beijing University of Chemical Technology (e-mail: jiarunliu@foxmail.com).}
\thanks{Ruirui Li is with College of Information Science and Technology, Beijing University of Chemical Technology (e-mail: ilydouble@gmail.com).}
\thanks{Chuan Sun is with Department  of ophthalmology, China-Japan Friendship Hospital (e-mail: drsunchuan@foxmail.com).}
}

\maketitle

\begin{abstract}
With the development of deep learning, medical image classification has been significantly improved. However, deep learning requires massive data with labels. While labeling the samples by human experts is expensive and time-consuming, collecting labels from crowd-sourcing suffers from the noises which may degenerate the accuracy of classifiers. Therefore, approaches that can effectively handle label noises are highly desired. Unfortunately, recent progress on handling label noise in deep learning has gone largely unnoticed by the medical image. To fill the gap, this paper proposes a noise-tolerant medical image classification framework named Co-Correcting, which significantly improves classification accuracy and obtains more accurate labels through dual-network mutual learning, label probability estimation, and curriculum label correcting. On two representative medical image datasets \ADD{and the  MNIST dataset}, we test \DEL{five}\ADD{six} latest Learning-with-Noisy-Labels methods and conduct comparative studies. The experiments show that Co-Correcting achieves the best accuracy and generalization under different noise ratios in various tasks. Our project can be found at: \href{https://github.com/JiarunLiu/Co-Correcting}{https://github.com/JiarunLiu/Co-Correcting}.
\end{abstract}


\section{Introduction}

    Medical image classification plays an essential role in clinical treatment and teaching tasks. Automated medical image classification is a research hotspot. It is challenging owing to the fine-grained variability in the appearance of lesions. Recently, methods that use deep learning have already achieved impressive, and often unprecedented performance, ranging from early screening to determining subcategories.
    
    The success of deep learning has attributed to the increasing availability of very large and growing datasets as well as more powerful computational hardware. Large data with reliable labels are favorable to the deep neural network to train parameters. However, medical images are hard to collect, as the collecting and labeling of medical data confronted with both data privacy concerns and the requirement for time-consuming expert explanations. In the two general resolving directions, one is to collect more data, such as crowdsourcing \cite{cheplygina_crowd_2018} or digging into the existing clinical reports \cite{wang_tienet_2018}. Another way is studying how to increase the performance with a few labels, which usually applies semi-supervised learning \cite{berthelot_mixmatch_2019} and pseudo labels \cite{arazo_pseudo-labeling_2020}. Data collected through crowdsourcing or pseudo labels typically suffer from very high label noise \cite{kuznetsova_open_2020} and thus they have limited applicability in medical imaging. It is clear that relative small datasets with noisy labels are, and will continue to be, a common scenario in training deep learning models in medical image analysis applications \cite{dgani_training_2018}. Hence, approaches that can effectively handle the label noise are highly desired. 
    
    Essentially, noisy labels systematically corrupt ground-truth labels, which inevitably degenerates the accuracy of classifiers. Such degeneration becomes even more prominent for deep learning models since these complex models can fully memorize noisy labels \cite{arazo_pseudo-labeling_2020}. A promising way to handle noisy labels is training on selected instances, for example, those with small losses. It is observed that deep neural networks memorize clean instances first and gradually adapt to hard instances as training epochs become large. Based on this observation, Co-teaching \cite{han_co-teaching_2018} or Co-teaching+ \cite{yu_how_2019} train two networks simultaneously. In each mini-batch of data, each network views its small-loss instances as useful knowledge and teaches such useful instances to its peer network to update the parameters. Co-teaching or Co-teaching+ could effectively discriminate probably clean labels and keep classifiers away from fitting noisy ones. However, they ignore the samples that are considered unreliable and simply delete many difficult samples which are important to overall accuracy. 
    
    The task of learning with noisy labels (LNL) is particularly significant in medical image classification. Nevertheless, recent progress on handling label noise in deep learning has gone largely unnoticed by the medical image analysis community \cite{karimi_deep_2020}. Medical image classification differs from general image classification in several aspects. Firstly, medical image datasets are relatively small and are expensive to acquire. Hence, it is a big waste to ignore those samples with unreliable labels. Even noisy ones contain a certain level of information for learning. Secondly, it is difficult to classify medical images even if the labels are perfect since there is not always a clear indication of the pathology from the image. Currently, the behavior of deep learning on noisy medical datasets have not been well studied yet. The performance of the state-of-the-art methods, such as Co-teaching or Co-teaching+, are still unknown on noisy medical datasets. There \DEL{is a lack of }\ADD{are insufficient} noise-tolerant approaches specifically designed for medical applications, where labeling requires domain expertise and suffers from high inter- and intra-observer variability, and erroneous predictions may influence decisions that directly impact human health.
    
    To fill the gap, this paper proposes an end-to-end noise-tolerant deep learning framework for medical image classification. Inspired by Co-teaching+, the framework adopts the idea of training two networks simultaneously. It consists of three core components: a dual-network architecture, which can choose reliable instances with small losses by cross-updating the parameters mutually, a new probabilistic model used to uniformly handle noise-contaminated and noise-free labels, and a curriculum which allows the framework to update the labels \DEL{in steps}\ADD{stably}.
    
    The proposed framework called Co-Correcting could take advantage of the noisy labels and handle severe noise situation. We conduct experiments on two public medical image datasets: ISIC-Archive \cite{noauthor_isic_nodate} and PatchCamelyon \cite{ehteshami_bejnordi_diagnostic_2017,veeling_rotation_2018}. Compared with the state-of-the-art methods, Co-Correcting gets the highest accuracy models under noisy ratio from 5\% to 40\%. Our contributions are as follows:
    \begin{enumerate}
        \item This paper proposed a noise-tolerant deep learning framework for medical image classification based on mutual learning and annotation correction. The framework is independent of the backbone network structure and does not need any auxiliary clean dataset, thus it is easy to apply. 
        \item A new label probabilistic model was introduced based on the proposed dual-network architecture. It utilizes the backpropagation of a summarized gradient to probabilistically updates the labels. 
        \item A label correction curriculum learned by unsupervised clustering on deep features is proposed, according to which Co-Correcting can gradually update the labels, from easy to difficult, to improve the stability of the learning. 
        \item On two classical medical image datasets ISIC-Archive and PatchCamelyon, we implemented the latest LNL methods, including Co-Correcting, Co-teaching+, DivideMix etc.. Compared with state-of-the-art methods, Co-Correcting achieves the highest accuracy on noisy medical datasets.
    \end{enumerate}

\section{Related Works}
    \subsection{CNN-based Medical image classification}
        In recent years, the rapid development of deep learning theory has provided a technical approach for solving medical image classification tasks. In 2017, \cite{esteva_dermatologist-level_2017} applied deep neural networks to the classification of skin cancers. In their work, they first built a large dataset containing 129,450 clinical images, which is two orders of magnitude larger than previous datasets. The results showed that the classification level of skin cancer by CNNs reached those of dermatologists. CheXNet \cite{pranav_chexnet_2017}, a CNN with 121 layers trained on a dataset with more than 100,000 frontal-view chest X-rays, achieved a better performance than the average performance of four radiologists. Even though in some applications it has become possible to curate large datasets with reliable labels, in most applications it is very difficult to collect and accurately label datasets large enough to effortlessly train deep learning models. CNN-based methods have various strategies to increase the performance of image classification on small datasets. \cite{wang_effectiveness_2017} researched the effectiveness of data augmentation in image classification. \cite{kermany_identifying_2018} achieved 92\% accuracy on a small pneumonia X-rays image dataset by transfer learning. A promising research direction is to label massive datasets by employing non-expert humans or automated systems with little or no human supervision. However, instances collected using such methods typically suffer from high label noise \cite{kuznetsova_open_2020}, thus they have limited applicability in medical imaging. Therefore, further efforts are needed to study methods that can handle noisy medical datasets.
    \subsection{Deep learning with noisy labels}
        The ability of deep CNNs to overfit\DEL{ to memorize} the corrupted labels can lead to very poor generalization performance\DEL{[17]}. To handle corrupted labels, some deep learning-based methods added regularization in the loss function \cite{miyato_virtual_2019}. They leveraged the regularization bias to overcome the label noise issue. However, they introduce the bias permanently and the learned classifier barely reaches the optimal performance. Other methods estimated transition matrix \cite{patrini_making_2017} which did not introduce regularization bias and thus could improve the accuracy of classifiers. Unfortunately, the transition matrix is hard to estimate especially when the dataset is imbalanced and the number of classes is large. To be free of estimating the noise transition matrix, a promising direction focus on training on selected samples. These works try to select clean instances out of the noisy ones, and then use them to update the network. Intuitively, as the training data becomes less noisy, better performance can be obtained. Table \ref{table:lnl_methods} compares the latest LNL methods in two dimensions: whether accumulate errors and whether correct labels. MentorNet \cite{jiang_mentornet_2018} provides a curriculum for StudentNet to focus on the sample of the label \DEL{of which }is probably correct. It uses a self-paced way to learn the curriculum and has the inferiority of accumulated error caused by sample-selected bias. To solve the problem, Co-teaching trains two deep neural networks simultaneously, and let them teach each other given every mini-batch. In this exchange procedure, the error flows can be reduced by peer network mutually. Co-teaching+ improved Co-teaching by using {``}Update by disagreement{''} strategy to keep two networks diverged within the training epochs. However, they barely make use of the information in corrupted labels. Several works try to correct the errors in the labels. \cite{tanaka_joint_2018} maintain and update the label distributions. PENCIL \cite{yi_probabilistic_2019} does the same job in a principled end-to-end manner. It avoids fitting noisy labels by training with a large learning rate. However, these manners easily introduce sample-selection bias, and thus accumulate errors. Recently, DivideMix \cite{li_dividemix_2020} leverages semi-supervised learning techniques for learning with noisy labels. It randomly mixing two instances proportionally, which may reduce the specificity of medical image classification. 
	\begin{table}[htb!]
        \centering
        \caption{Comparison of LNL Methods.}
        \label{table:lnl_methods}
        \begin{tabular}{ccc}
            \toprule
            Methods             &       Without accumulate errors       &       Correct labels\\
            \midrule
            MentorNet           &       $\times$                    &   $\times$ \\
            Co-teaching     &       \checkmark                      &   $\times$ \\
            PENCIL              &       $\times$                    &   \checkmark \\
            DivideMix           &       \checkmark                      &   $\times$ \\
            Co-Correcting   &       \checkmark                      &   \checkmark \\
            \bottomrule
        \end{tabular}
    \end{table}

    \subsection{Medical image classification with noisy labels}
    Although some approaches have been considered to address the noisy label issue, it is still an ongoing challenge in deep learning in medical field. In fact, not many studies have addressed the medical image noisy label issue. One pioneer work is \cite{dgani_training_2018},  \DEL{they utilized the method of }\ADD{it adapts a general LNL method} \cite{bekker_training_2016} on mammography classification task and outperforms standard training methods, but it is heavily dependent on the noise label distribution assumption. \cite{xue_robust_2019} used a data re-weighting method that amounted to removing data samples with high loss values in each training batch.  \DEL{\cite{le_pancreatic_2019} used a data re-weighting method similar to that proposed by\cite{ren_learning_2019} to deal with noisy annotations in pancreatic cancer detection from whole-slide digital pathology images. }\ADD{\cite{le_pancreatic_2019} deals with noisy annotations in pancreatic cancer detection from whole-slide digital pathology images. It uses a data re-weighting method \cite{ren_learning_2019} that is originally designed for general image classification.} \DEL{All the above methods belong to adaptation and application of general LNL techniques and there is no method specifically designed for medical image classification tasks.}
    
\section{Co-Correccting: Medical image classification framework}\label{chapter:proposed_method}
\begin{figure}[htbp]
        \centering
        \includegraphics[width=8cm]{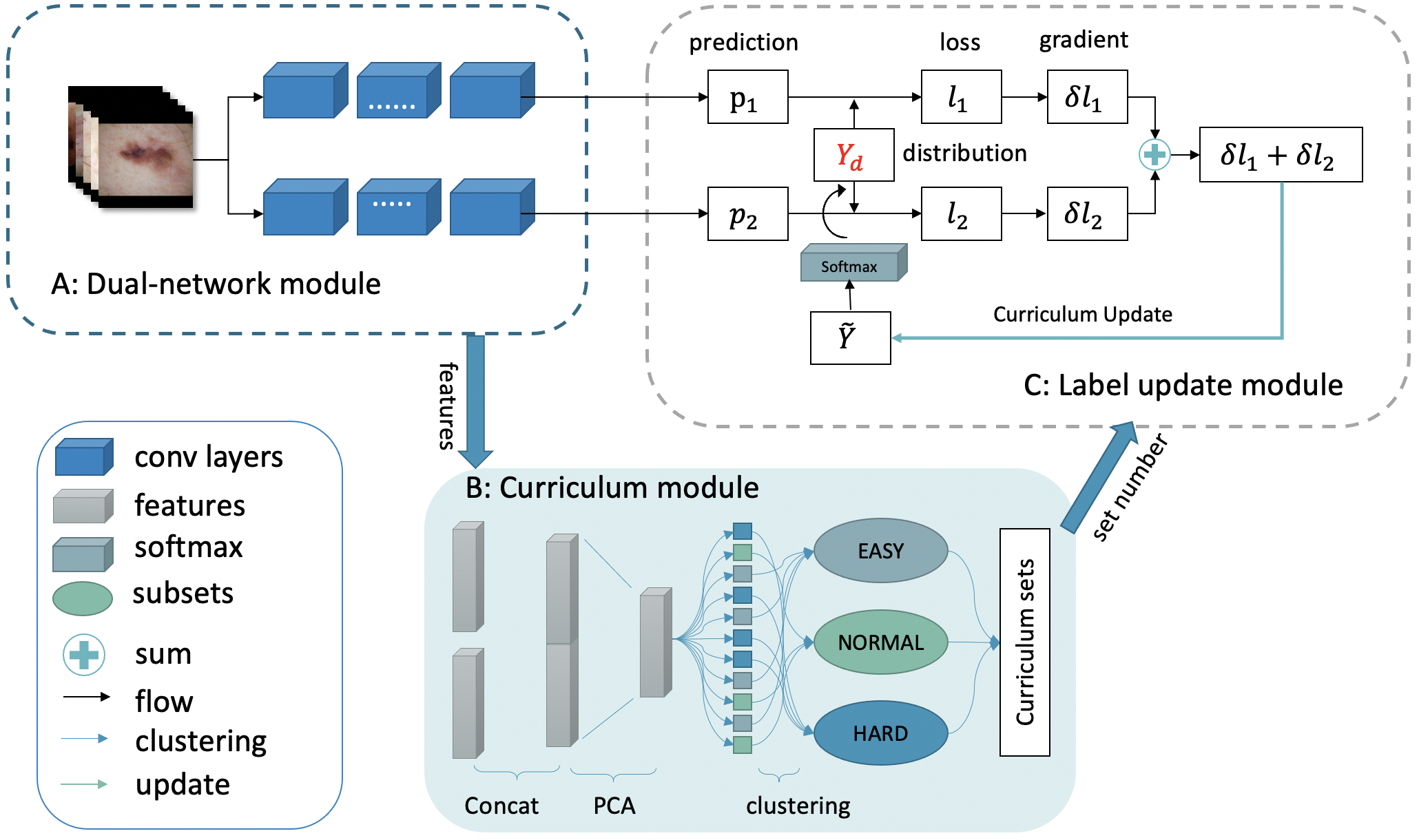}
        \caption{The framework of Co-Correcting, it consists of: A) the dual-network module, B) curriculum learning module, C) label updating module.}
        \label{fig:main}
    \end{figure}

    The proposed method named Co-Correcting is originally designed for medical image classification tasks. These tasks suffer from highly noisy labels and face high inter and intra-observer variability. To deal with the problems, Co-Correcting invented a novel noise-tolerant learning framework which is shown in Figure \ref{fig:main}. It mainly consists of three modules: the dual-network architecture, the curriculum learning module, and the label correction module. 
    
    In Co-Correcting, we use label distribution $y^{d}$ to handle label uncertainty. Based on the observation that mutual training on two networks maintains a better \DEL{estimate}\ADD{estimation} of the labels, we could update the label distribution $y^{d}$ under the supervision of the \DEL{estimate}\ADD{estimation}, and hence noise can be probabilistically corrected. The networks learn the probabilistic label distribution while updating the network parameters. It is considered that the model and the labels are getting better simultaneously. However, it is not always the case. To ensure that the learning is \ADD{stable and} moving in the right direction, we incorporate the concept of curriculum learning. Samples are projected into the embedding spaces and clustered into groups based on the concatenated features. The groups are sorted by densities in descending order. As epochs increase, we gradually add the groups of labels to be updated.
    
    More specifically, we divide the training process into the following four different stages: 
    
        \textit{Stage \uppercase\expandafter{\romannumeral1}:} a warm-up stage that performs mutual learning on Module A.
        
        \textit{Stage \uppercase\expandafter{\romannumeral2}:} calculate the curriculum on Module B.
        
        \textit{Stage \uppercase\expandafter{\romannumeral3}:} on both Module A and C, learning parameters and update the label distribution.
        
        \textit{Stage \uppercase\expandafter{\romannumeral4}:} fine-tuning the network with the corrected labels on Module A.

     In the warm-up stage, we train the models on two networks with exactly the same structure. The network parameters are randomly initialized and updated under the supervision of the information from its peer partner.  \ADD{Different initialization and sample selection bring different decision boundaries, which lead to the divergence between networks.} The mutual information exchange and the divergence between two networks \DEL{enhance}\ADD{enhances} the ability to resist the influence of noisy labels and thus gets a reliable projection into the embedding spaces.  \ADD{It should be noted that the warm-up stage has to be finished before overfitting.} In the following stage, we focus on calculating the curriculum on the embedding spaces, and \DEL{then} in the third stage, we update both the network parameters and the label distribution with the \DEL{guidance}\ADD{help} of the curriculum. Finally, we fine-tune the model with the corrected labels to further improve the accuracy and the generalization. In the following parts, we will give a detailed explanation of the dual network architecture, label probability estimation, curriculum annotation correction, and the loss functions we used in the four stages. Our motivation will also be explained.

    \begin{figure}[htbp]
        \centering
        \subfigure[Co-teaching]{
        \includegraphics[width=2cm]{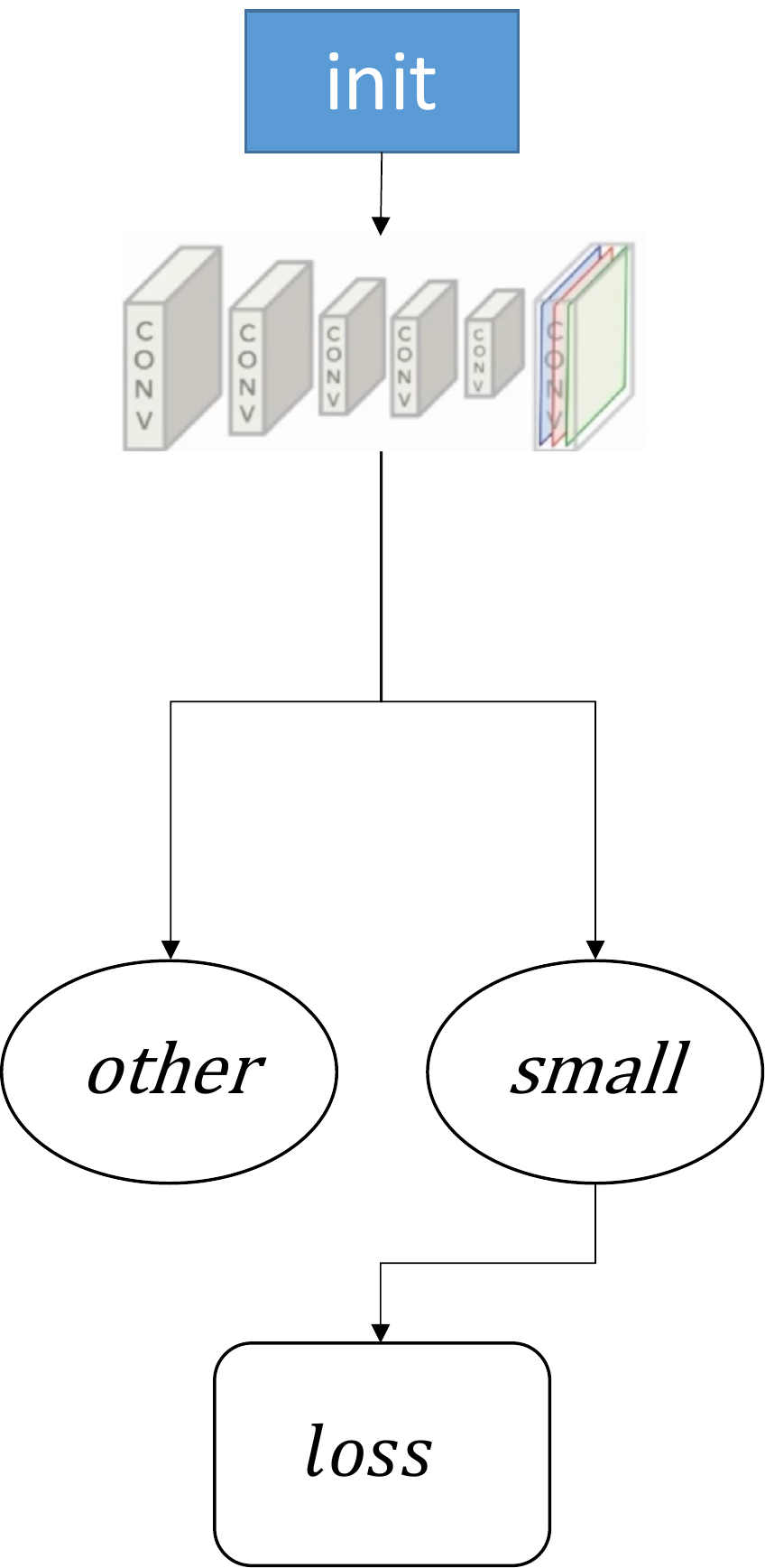} \label{fig:dis_coteaching}
        }  
        \subfigure[Co-teahcing+]{
        \includegraphics[width=2cm]{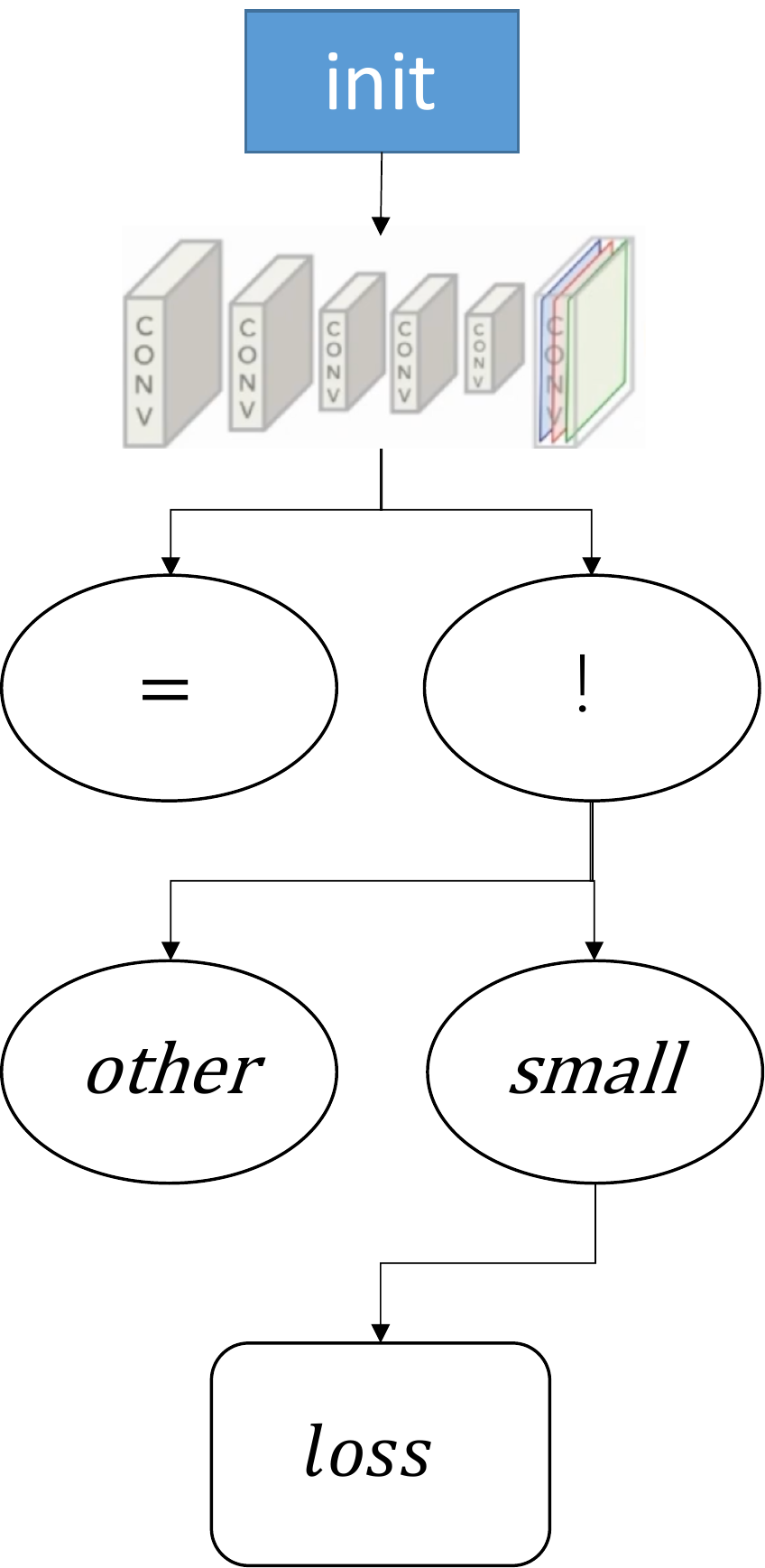} \label{fig:dis_coteaching_plus} 
        }
        \subfigure[Co-Correcting]{
        \includegraphics[width=2cm]{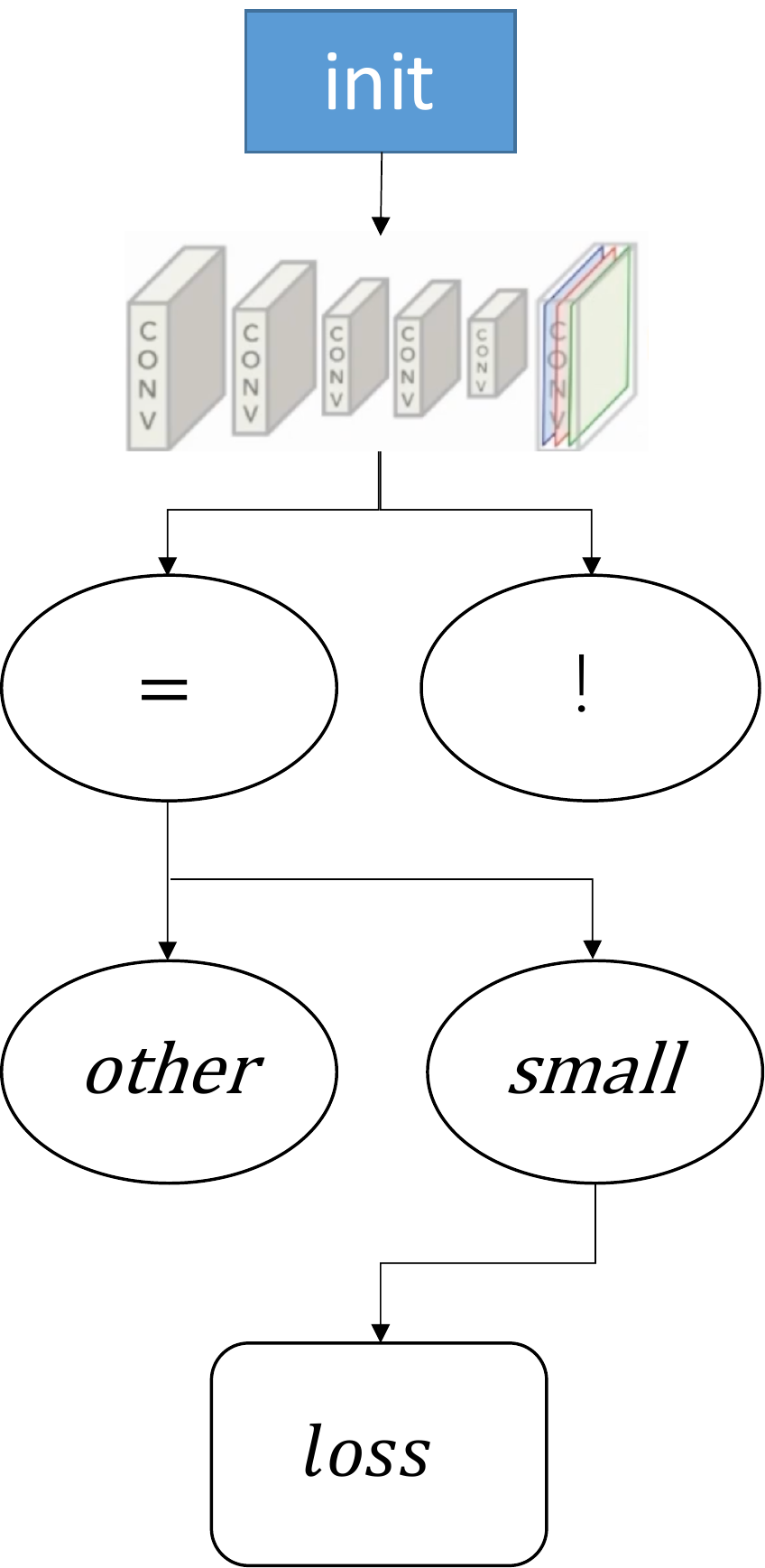} \label{fig:dis_cocorrecting}
        }  
        \caption{\DEL{Different sample select strategy of Co-teaching, Co-teaching+ and Co-Correcting.}\ADD{Sample selection strategies of Co-teaching,  Co-teaching+,  and Co-Correcting: \textit{'init'} represents the initialization of the network; '!' means the disagreement samples,  while \textit{'='} means the agreement samples that are defined by Formula \ref{equ:div}; \textit{'small'} represents the samples which are bottom $R\left(t\right)$ percentage in loss ranking; \textit{'other'} represents the samples other than \textit{'small'}; \textit{'loss'} is the total loss for backward propagation computed from the selected samples. }}
        \label{fig:disagree}
    \end{figure}    
    \subsection{Dual network architecture} \label{section:dual-networks}
        On the research of learning with noisy labels, a challenging issue is to design reliable criteria to select clean samples. Co-teaching and Co-teaching+ train two networks where each network selects small-loss samples in a mini-batch to train the other. Our dual network architecture is inspired by them and use an agreement way to update the parameters illustrated in Figure \ref{fig:disagree}.  
        
        Co-Correcting trains two networks of the same structure. The backbone can be any deep convolutional neural network, for instance, ResNet or DenseNet, depending on the classification tasks. In the forward propagation, the two networks are trained on the same image data individually. In the back-propagation, according to the principle of {``}Update by agreement{''}, the gradient generated by samples with identical predictions will be used to update parameters. It has been shown that DNNs tend to learn simple data first before fitting label noise. Therefore, we treat instances with small loss as clean ones and collect their gradients when agreement occurs. Samples with consistent prediction have greater credibility, which facilitates the screening and correction of noise data.
        
        More specifically, for each mini-batch \ADD{with a batch size of $b$},  assuming $P^{\left( 1 \right)}=\left\{ p_1^{\left( 1 \right)},p_2^{\left( 1 \right)},…,\DEL{p_B^{\left( 1 \right)}}\ADD{p_b^{\left( 1 \right)}}\right\} $ and $P^{\left( 2 \right)}=\left\{ p_1^{\left( 2 \right)},p_2^{\left( 2 \right)},…,\DEL{p_B^{\left( 2 \right)}}\ADD{p_b^{\left( 2 \right)}}\right\}$ are the prediction of the two networks respectively, we select the samples with consistent prediction by the following Formula:
        \begin{equation} \label{equ:div}
        \bar{D}_n=\left[\left(x_i,y_i\right): argmax\left({p_i^{\left(1\right)}}\right) = argmax\left({p_i^{\left(2\right)}}\right)\right]
        \end{equation}
        where $argmax$ is the index of the maximum $ p_i $.

        \ADD{Assuming $e$ is the current epoch number, for each $e$, } we define a memory rate as $R\left(t\right)$ \ADD{to select the samples with small losses}, presented in Formula \ref{equ:rt}.  Among the group of agreement, the samples with the smallest loss in percentage of $R\left(t\right)$ are collected to calculate the gradient \ADD{to update the networks}.
        \begin{equation} \label{equ:rt}
        R\left(t\right) = 1 - \min\left(\frac{e}{E_k}\tau, \tau\right)
        \end{equation}

\begin{figure}[htb!]
            \centering
            \includegraphics[width=5cm]{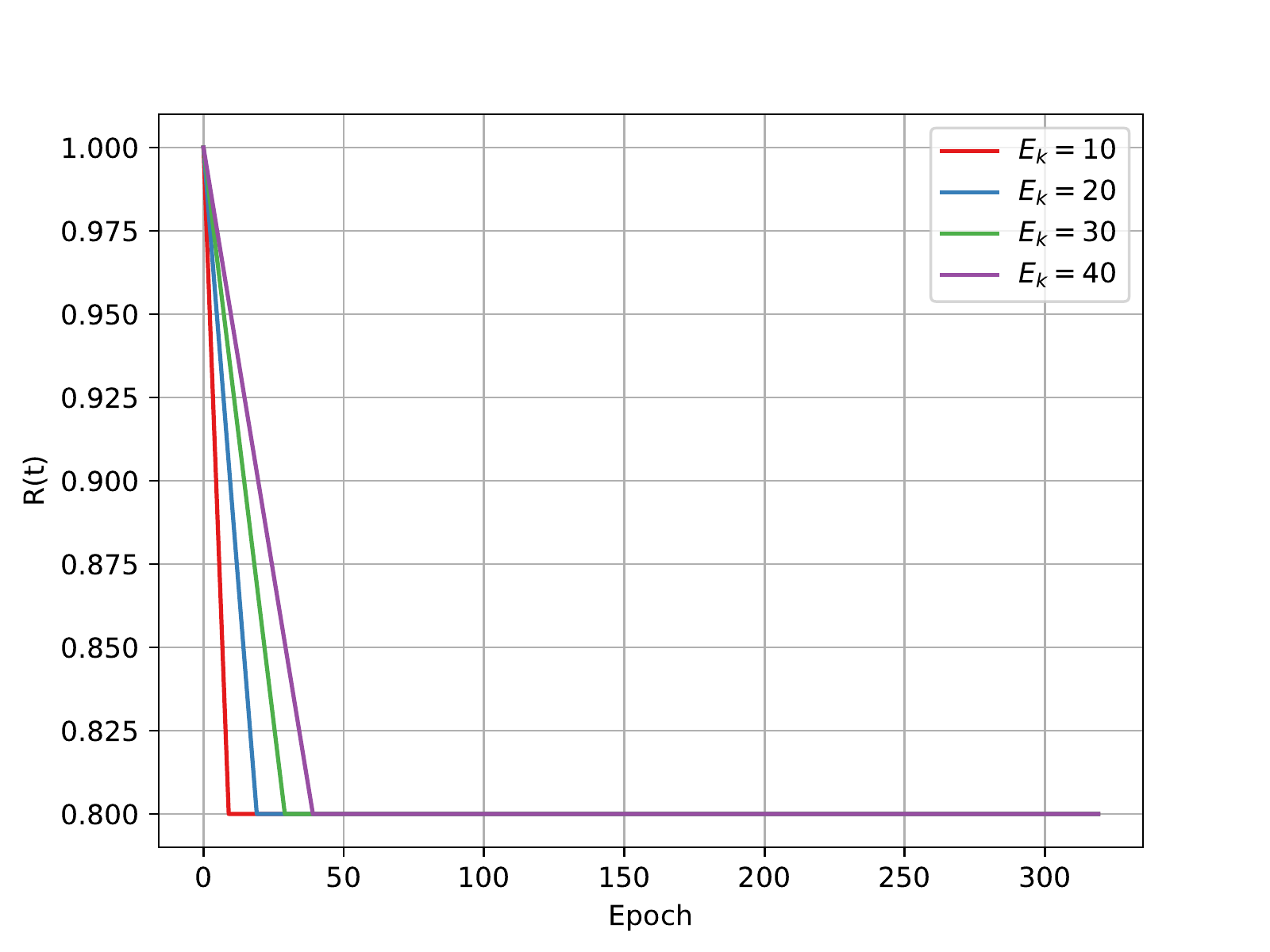}
            \caption{\ADD{The curve of $R(t)$ with $\tau=0.2$ and different $E_k$.}}
            \label{fig:r_t}
        \end{figure}

        In Formula \ref{equ:rt}, $E_k$ and $\tau$ are two hyper-parameters.  \DEL{In our experiments,  }\ADD{As illustrated in Fig. \ref{fig:r_t}, $E_k$ controls the number of epochs to decrease $R(t)$ and $\tau$ is the decrease range for $R(t)$.  Generally, $E_k$ is set to 10 and $\tau$}\DEL{In our experiments,  $E_k$=10 and $\tau$ generally} equal to the noisy rate in the dataset.  \ADD{If we did not know the noisy rate previously,  we}\DEL{We} could also estimate $\tau$ through the methods described in \cite{liu_classification_2016,yu_efficient_2018}. 

        Differed with Co-teaching or Co-teaching+,  we also calculate the losses of potentially noisy samples, though their gradients are set to zero when updating the classification models. These losses are used to update the label distribution, which will be described in the next section. 

        \begin{figure}
            \centering
            \includegraphics[width=8cm]{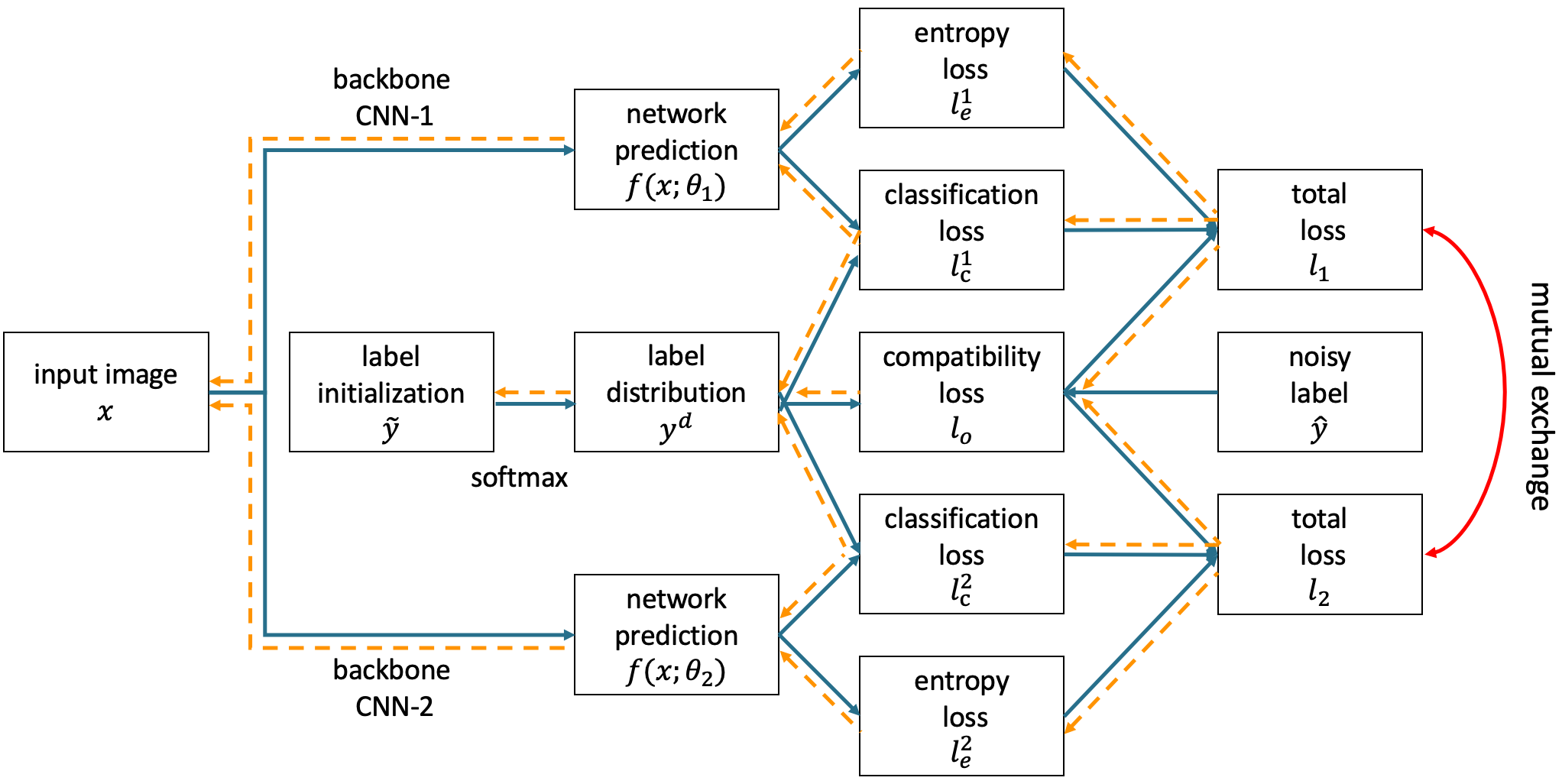}
            \caption{The label distribution was updated by two CNN. The blue line represents the forward propagation path, and the orange dashed line represents the backward propagation path. We use the label distribution $y^d$ to replace the noisy label $\hat{y}$ and calculate the classification loss by network prediction and $y^d$.}
            \label{fig:grad}
        \end{figure}  
          
    \subsection{Label probabilistic modeling and updating} \label{section:label_update}
        Deep label distribution learning \cite{gao_deep_2017} was proposed to handle label uncertainty. We adopt the idea for the purpose such that the label distribution can be updated and hence noise can be probabilistically corrected. Inspired by PENCIL\cite{yi_probabilistic_2019},  \ADD{for a dataset with $n$ image samples.} Co-Correcting maintains a label distribution $ y_i^d \in \left\{y:y \in {\left[ 0, 1 \right]}^c \right\} $ for every image $x_i$\ADD{,  where $i$ is the sample id and $i \in \left[0,n\right)$}. It is a probabilistic \DEL{estimate}\ADD{estimation} of whether the label is noise-free. $y_i^d$ is treated as a pseudo-label that is used in calculation of the losses. 
        
        In Figure \ref{fig:grad}, Co-Correcting estimates and updates $y^d$ by simultaneously training two networks. As previously described, the mutual learning on two networks will not accumulate errors and thus avoiding the interference of the noisy labels to the greatest extend. We define an auxiliary variable $\tilde y$ to assist the label updating which is initialized by \ADD{multiplying} one-hot encoding of the noisy labels \ADD{$\hat{y}$ and a constant $K$}. 
		\begin{equation} \label{equ:init_tilde_y}
        	\ADD{\tilde{y} = K \hat{y}}
        \end{equation}
        
        Assuming the loss on the first network is defined as $l_1$ and that on the second network is defined as $l_2$\DEL{, according toCo-Correcting, the label distribution is updated by the following formulas:}\ADD{. In Co-Correcting,  we update $\tilde{y}$ by gradient descent thus we can update $y^d$ implied:}
        \begin{equation} \label{equ:udpate_y}
            \tilde{y} \gets \tilde{y} - \lambda \left(
            \frac{\delta{l_1}}{\delta{\tilde{y}}} +
            \frac{\delta{l_2}}{\delta{\tilde{y}}}
            \right)
        \end{equation}
        \begin{equation} \label{equ:init_yd}
            y^d = softmax \left( \tilde{y} \right)
        \end{equation}
        where $\lambda$ is the label correction rate\DEL{ and $\delta$ is the gradient operation}\ADD{,  $\delta$ is the gradient operation and $\frac{\delta{l}}{\delta{\tilde{y}}}$ is the gradient of the loss function over the current label.}  \ADD{Usually, $\lambda$ should be chosen according to the noise density and classification number.  A larger $\lambda$ could speed up the gradient decent process, but it may lead to incomplete convergence. The gradient to update labels was calculated by the losses between the predictions and $\tilde{y}$. }Here, the addition operation is chosen by Co-Correcting to fuse the gradients. In the training scenario, if the two networks get the same prediction, the addition operation will speed up the convergence, otherwise, the gradients will offset with each other, which leads to conservatively label updating. 

        Combining the label correction with the dual network architecture is beneficial for learning noise-free hard samples in medical images which have big losses on both of the two networks. Without label correction, the dual-network architecture will not learn too much on hard samples before overfitting to noisy ones, thereby reducing the generalization of the model.
        
    \subsection{\MOVE{Label correction curriculum}} \label{section:curriculum}
        The entanglement of hard samples and the noisy samples has always been a difficult problem in learning with noisy labels. Premature label modification makes mistakes easily on hard samples. It is possible that noise-free samples are permanently modified incorrectly, which negatively affects the accuracy of the model. Curriculum learning \cite{bengio_curriculum_2009} is a learning strategy in machine learning, which starts from easy instances and then gradually handles harder ones. Curriculum learning has been used to handle a massive amount of noisy labels. CurriculumNet \cite{guo_curriculumnet_2018} designs the learning curriculum by measuring the complexity of the data. Inspirited by the idea, Co-Correcting proposed a novel label correction curriculum to avoid premature label modification. It designs a label correction strategy ordered by increasing difficulty, and correction proceeds sequentially from easy tasks to harder ones.
        \begin{figure}
            \centering
            \includegraphics[width=8cm]{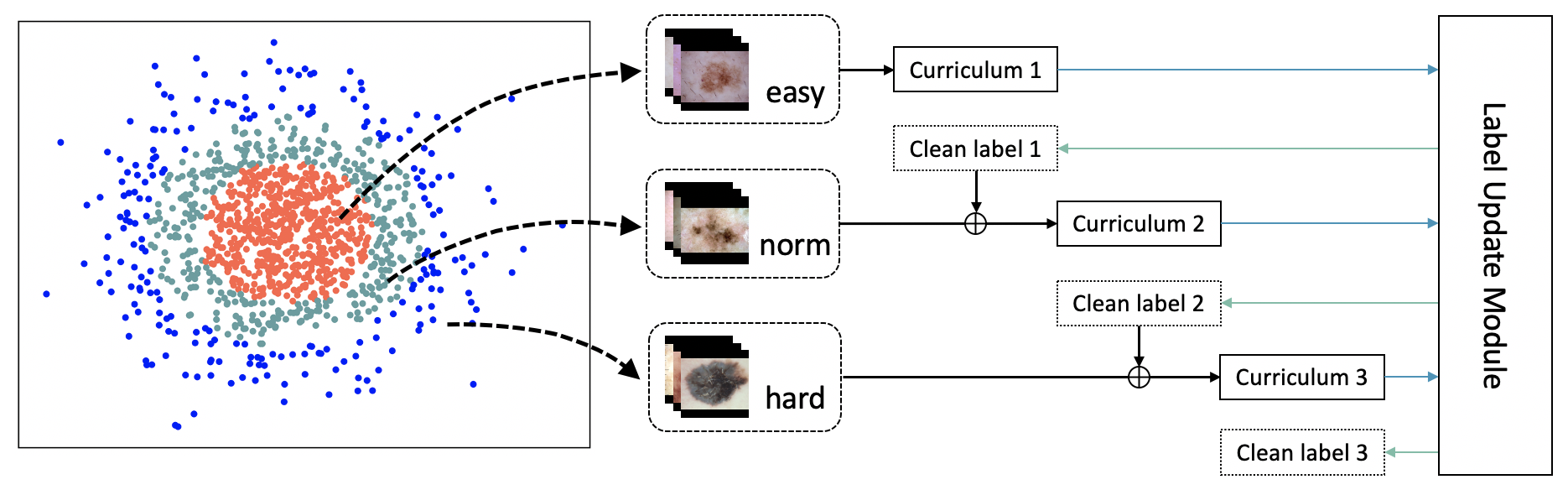}
            \caption{Curriculum based label correction.}
            \label{fig:curriculum}
        \end{figure} 
        
        Therefore, we aim to split the whole training set into a number of subsets, which are ranked from an easy one to a more complex one. The proposed method measures the complexity of data using its distribution densities in the feature space and ranks the complexity in an unsupervised manner. In detail, for each image $x_i$, Co-Correcting projects the image into the deep feature space by the operation $\Gamma$. \DEL{Assuming $f_c \left( x_i,\theta_j \right)$ illustrates the fc-layer features gotten from the warp-up model with the parameters ${\theta}_j$ of the $j_{th}$ network, } \ADD{Assuming $f_c \left( x_i,\theta_m \right)$ illustrates the fc-layer features gotten from the warp-up model with the parameters ${\theta}_m$ of the $m_{th}$ network,  which has already trained in Stage \uppercase\expandafter{\romannumeral1},  }through concatenation and PCA dimension reduction, we get the vector in the feature space by the following function:
        \begin{equation} \label{equ:cat}
            \Gamma \left( x_i \right) \to PCA \left( concat \left(  f_c \left( x_i, {\theta}_1 \right) ,  f_c \left( x_i, {\theta}_2 \right)  \right) \right)
        \end{equation}

        We also define a distance matrix $D \subseteq \mathbb{R}^{n \times n}$, of which each element is the Euclidean distance between $\Gamma \left( x_i \right)$ and $\Gamma \left( x_j \right)$:
        \begin{equation} \label{equ:d}
            D_{ij} = {\left \| \Gamma \left( x_i \right) - \Gamma \left( x_j \right) \right \|}_2
        \end{equation}

        In the feature space, we try to calculate the cluster center for each category. Theoretically, a closer data point to the cluster center has a higher confidence to have an accurate prediction and the error labels could be relatively easily recognized. The approach finds the cluster center by calculating data distribution densities. Specifically, Co-Correcting first \DEL{sort}\ADD{sorts} the elements in the distance matrix $D$ and \DEL{pick}\ADD{picks} up the number which is ranked at $k\%$. \DEL{($k$ between 10-15 in our experiments.)}We called this number as $d_c$.  \ADD{We empirically set $k=60$ in all of our experiments and the final result is insensitive to the value of $k$ just as the paper \cite{guo_curriculumnet_2018} suggested. } Co-Correcting computes the densities by stating the number of the samples whose distances are smaller than $d_c$. Thus, the density \DEL{$p$}\ADD{$s$} is defined by:
        \begin{equation} \label{equ:s}
            \DEL{p_i}\ADD{s_i} = \sum_j X \left( D_{ij} \right),
            X \left( d \right) = \left\{
            \begin{array} {rcl}
                1, & d < d_c; \\ 
                0, & other.
            \end{array} \right.
        \end{equation}

        Based on the density value, a new distance $\varepsilon$ is further calculated by Formula \ref{equ:distance} and the sample whose distance is the maximum is the cluster center.
        \begin{equation} \label{equ:distance}
            \varepsilon = \left\{
            \begin{array} {ll}
                min_{j:\DEL{p_j>p_i}\ADD{s_j>s_i}} \left( D_{ij} \right), & if \  \exists \  j \ s.t. \ \DEL{p_j > p_i}\ADD{s_j > s_i}; \\
                max \left( D_{ij} \right), & otherwise.
            \end{array} \right.
        \end{equation}

        We learn the curriculum in Stage \uppercase\expandafter{\romannumeral2}, which is performed only once during the whole training process. It ranks the samples by $\varepsilon$ to split the dataset into three curriculums. As illustrated in Figure \ref{fig:curriculum}, following the curriculum, Co-Correcting tries to correct simple errors first, then with partially updated samples, it trains the parameters and obtains a more accurate model. After that, Co-Correcting extends the range of the updated data to harder samples. This procedure is iterating in Stage \uppercase\expandafter{\romannumeral3} until all curriculums are dealt with.
        
    \subsection{\MOVE{Loss Functions}} \label{section:loss}

        Co-Correcting uses a compound loss function with three different terms. They are the classification loss $l_c$, the compatibility loss $l_o$ and the entropy loss $l_e$.

        As introduced in Chapter \ref{chapter:proposed_method}, the training process of Co-Correcting is divided into four stages. The first and fourth stages are pure classification training, using only the classification loss. The difference is that the first stage uses hard labels, and the fourth stage uses label distributions that are soft labels. The third stage is the correcting and training stage, using all three kinds of losses together. 

    \subsubsection{Classification Loss}

        For hard labels \ADD{in Stage \uppercase\expandafter{\romannumeral1}},  we adopt CrossEntropy loss as the classification loss, defined by:
        \begin{equation} \label{equ:l_stage1}
            l_{\DEL{stage1}\ADD{Stage \uppercase\expandafter{\romannumeral1}}} = l_c^{hard} = CrossEntropy\left(f\left(x;\theta\right), \hat{y} \right)
        \end{equation}
        where $\hat{y} $ is the original noisy labels and $f(x;\theta)$ is the network prediction with input $x$ and parameters $\theta$.  For soft labels \ADD{in Stage \uppercase\expandafter{\romannumeral3} and \uppercase\expandafter{\romannumeral4}}, we choose flipped Kullback-Leibler divergence as the classification loss. \ADD{Using KL loss could obtain the gradients for both soft label and network parameters, and flip it could obtain a better gradient for label correction\cite{yi_probabilistic_2019}:}
        \begin{equation} \label{equ:l_stage4}
            l_{\DEL{stage4}\ADD{Stage \uppercase\expandafter{\romannumeral4}}} = l_c^{soft} = \frac{1}{\DEL{n}\ADD{b}} \sum_{i=1}^{\DEL{n}\ADD{b}} \sum_{j=1}^c
            f_j \left( x_i ; \theta \right)
            \log \left( \frac{f_j \left( x_i ; \theta \right)}{y_{ij}^d} \right)
        \end{equation}\

    \subsubsection{Compatibility Loss}
        The original annotations play an important role in the supervision of the learning, though they contain partial noises. To avoid the big gap between the corrected labels and the original labels, we introduce the compatibility loss in Stage \uppercase\expandafter{\romannumeral3}. The compatibility loss is in the pattern of Kullback-Leibler divergence. It is defined as follows:
        \begin{equation} \label{equ:l_o}
            l_o \left( \hat{y}, y^d \right) = - \frac{1}{\DEL{n}\ADD{b}} \sum_{i=1}^{\DEL{n}\ADD{b}} \sum_{j=1}^c \hat{y}_{ij} \log{y_{ij}^d}
        \end{equation}
        where $\hat{y}$ is the original noisy label, $y^d$ is the label distribution and $c$ is the number of classes.

    \subsubsection{Entropy Loss}
        When the network prediction is very close to the soft label $y^d$, the gradients will disappear and the training will stop updating both the labels and the parameters. To prevent the training fall into such local dilemma, we introduce an additional regular term called the entropy loss $l_e$. The entropy loss can force the network to peak at only one category rather than being flat because the one-hot distribution has the smallest possible entropy value. This property is advantageous for classification problems. The entropy loss is defined as:
        \begin{equation} \label{equ:l_e}
            l_e \DEL{\left( f \left( x; \theta \right) \right)} = - \frac{1}{\DEL{n}\ADD{b}} \sum_{i=1}^{\DEL{n}\ADD{b}} \sum_{j=1}^c f_j \left( x_i ; \theta \right) \log f_j \left( x_i ; \theta \right)
        \end{equation}

        Based on the above loss definitions, in Stage \uppercase\expandafter{\romannumeral3}, the overall loss function is showed as follows:
        \begin{equation} \label{equ:l_stage3}
            l_{\ADD{Stage\uppercase\expandafter{\romannumeral3}}} = l_c^{soft} \left( f \left( x; \theta \right), Y^d \right) + \alpha l_o \left( \hat{Y}, Y^d\right) + \beta l_e \left( f \left( x; \theta \right) \right)
        \end{equation}
        where $\alpha$ and $\beta$ are two hyper-parameters taking the response of the weights of $l_o$ and $l_e$ respectively.  \ADD{$\alpha$ should be appropriately reduced under extreme noise and $\beta$ general equals to 0.1.}

\section{Experiments and Analysis}

    \subsection{Experimental setup}
        In order to verify the effectiveness of the proposed method, we conducted comparative experiments on two challenging medical imaging datasets: ISIC-Archive \cite{noauthor_isic_nodate} and PatchCamelyon \cite{ehteshami_bejnordi_diagnostic_2017,veeling_rotation_2018}. These two datasets will be described in detail in Part \ref{section:isic_dataset} and Part \ref{section:pcam_dataset} in this section respectively. They were collected from different medical image analysis tasks and are distinct in data sizes, image types, and classifying difficulties, which can verify the generalization when applying Co-Correcting.
        
        To prove the robustness in learning with noisy labels, we simulated the real situation by corrupting the labels. In binary image classification, we added noises by randomly flipping the labels.  \ADD{All the noise simulation was added on the training set. We use the original label in all the experiment for test set.}
        
    We chose \DEL{five}\ADD{six} latest Learning-with-Noisy-Labels (LNL) classification methods as baselines. They will be explained in Part \ref{section:baselines}. We first applied the baseline methods to the medical image classification tasks on ISIC-Archive and PatchCamelyon. Then we compared Co-Correcting with these methods in the same environment under various noise ratios, from 5\% to 40\%. To further explain the reason why Co-Correcting is effective, ablation studies were conducted to see the influence of factors such as the adaptation of dual network architecture and the usage of curriculum.  \ADD{We also conducted experiments on MNIST,  it}\DEL{It} will help us to understand the internal mechanism of Co-Correcting and its scope of application.        
        
        Our experiments were carried out on the software platform of Ubuntu 18.04 LTS and PyTorch 1.4.0, using the hardware of Intel i7-8700k CPU, NVIDIA RTX2080Ti GPU, and 32GB RAM.

    \subsection{Benchmark comparison methods} \label{section:baselines}
        The methods selected for the comparative studies are listed as follows:
        \begin{enumerate}
            \item \textbf{Standard} This is the common training procedure without using any LNL strategy.
            \item \textbf{Joint Optim} This is the method proposed by \cite{tanaka_joint_2018}. They jointly learned the label distribution and the classification model, and invented a label update method to confront noisy annotations.
            \item \textbf{PENCIL}   This is the method proposed by \cite{yi_probabilistic_2019}. They introduced an end-to-end label updating strategy. 
            \item \textbf{Co-teaching} This is the method proposed by \cite{han_co-teaching_2018} which is based on the dual-network architecture. They selected samples with smaller losses and updated the network mutually.
            \item \textbf{Co-teaching+} This is the method proposed by \cite{yu_how_2019}. It is based on Co-teaching \cite{han_co-teaching_2018} and adopts "Update by Disagreement" strategy which keeps the two networks divergent.
            \item \textbf{DivideMix} This is the latest LNL method proposed by \cite{li_dividemix_2020}. It adopts the technique of semi-supervised learning MixMatch \cite{berthelot_mixmatch_2019} to improve the robustness.
            \ADD{\item \textbf{Reweight} This is the method proposed by \cite{ren_learning_2019}.  It adopts meta-learning and uses a small verification set with the same distribution as the test set to guide the learning of the network.}
        \end{enumerate}
        
        When applying these methods, we directly used the codes provided by the original authors to ensure the fairness of the comparison. It should be noted that due to the limitation of GPU memory, the maximum batch size is set to 64 when testing DivideMix on PatchCamelyon.
        
    \subsection{Evaluation metric}
        We use two metrics to evaluate the models: classification accuracy and labeling accuracy. Classification accuracy evaluates the quality of the entire model, while \DELM{the }labeling accuracy evaluates the effectiveness of the label correction.  \ADDM{Notably, we binarize the soft labels to compute labeling accuracy since the ground truth of soft labels is incapable.} In our experiments, it is not always that better models owe to better annotations. The overall accuracy is relevant to many factors. However, on the other hand, better annotations do help to get better models. 
        The definition of the classification accuracy and the labeling accuracy are shown as follows:
        \begin{equation} \label{equ:acc}
            Acc_{class} = \frac{ number \; of \; correct \; predictions }{ number\; of \; samples \; in \; test \; dataset }
        \end{equation}
        \begin{equation} \label{equ:acc_label}
            Acc_{label} = \frac{ number \; of \; correct \; labels }{ number\; of \; samples \; in \; train \;dataset }
        \end{equation}

    \subsection{Experiments on ISIC-Archive} \label{section:isic_dataset}
        \subsubsection{ISIC-Archive Dataset}
            The ISIC-Archive dataset \cite{noauthor_isic_nodate} comes from the Melanoma project initiated by the International Skin Imaging Collaboration. As of July 3rd, 2020, there are 23,906 skin lesion images in this archive. Most images are obtained through dermoscopy. In the experiments, we selected 3324 samples whose annotations were further confirmed by pathologists. The chosen samples are balanced in positive and negative categories. They are challenging since the confirmed samples are usually difficult to distinguish by people. We \DEL{divide them into the training set, the testing set, and the validation set}\ADD{split them into the training, test, and validation sets} according to the ratio of 6:3:1 \ADD{to ensure the test data is sufficient}.  All images are resized to $224px * 224px$ and are augmented by randomly adopting horizontal flipping, vertical flipping, or image rotation.

        \subsubsection{Implementation details}
            We selected ResNet-50 as the backbone network for all the experiments on ISIC-Archive. We choose SGD as the optimizer with a momentum of 0.9 and a weight decay of 0.001. Due to the GPU memory space limitation, we set the maximum batch size to 32 and the training epoch to 320. 
            
            The hyperparameter $\lambda$ is a performance-sensitive parameter. It controls the rate of label correction. Theoretically, \ADDM{$\lambda$ needs to be a large value to update labels under the low iteration frequency efficiently.  Subsequently, }$\lambda$ should be set larger as the noise ratio increases. The choices of $\lambda$ used in our experiments are listed in Table \ref{table:lambda_setting}.
            \begin{table}[htbp]
                \centering
                \caption{Choice of $\lambda$ in different dataset.}
                \label{table:lambda_setting}
                \begin{tabular}{ccccccccc}
                    \toprule
                    Noise                   &   $clean$ &   0.05            &   0.1         &   0.2         &   0.3         &   0.4\\
                    \midrule        
                    ISIC-Archive        &   50          &   100         &   200         &   300         &   350         &   400\\
                    PatchCamelyon&  50          &   100         &   200         &   300         &   350         &   400\\
                    \ADD{MNIST}           	&   \ADD{500}         &   \ADD{1500}        &   \ADD{2000}        &   \ADD{2500}        &   \ADD{3000}        &   \ADD{4000}\\
                    \bottomrule
                \end{tabular} 
            \end{table}
        \begin{table}[htb!]
            \centering
            \caption{$ACC_{class}$ under different noise levels in ISIC-Archive dataset.}
            \label{table:isic_acc}
            \begin{tabular}{ccccccc}
                \toprule
                Methods             &   $clean$                                         &   0.05                                                &   0.1                                             &   0.2                                             &   0.3                                             &   0.4\\
                    \midrule        
                Standard                &   64.32                                           &   62.97                                           &   63.00                                           &   57.18                                           &   55.40                                           &   52.09\\
                Joint Optim         &   42.80                                           &   54.10                                           &   52.00                                           &   39.69                                           &   57.54                                           &   55.24\\
                PENCIL              &   \textcolor{blue}{78.26}             &   \textcolor{blue}{78.34}             &   75.45                                           &   \textcolor{blue}{75.23}             &   \textcolor{blue}{70.77}                 &   \textcolor{blue}{59.28}\\
                Co-teaching         &   71.39                                           &   72.14                                           &   71.85                                           &   69.88                                           &   65.17                                           &   51.98\\
                Co-teaching+        &   75.73                                           &   77.03                                           &   76.94                                           &   71.30                                           &   64.24                                           &   56.88\\
                DivideMix               &   76.71                                           &   76.35                                           &   \textcolor{blue}{77.04}                 &   74.29                                           &   55.04                                           &   50.00\\
                \ADD{Reweight}				& 73.64	& 74.95 & 70.10 & 66.33 & 61.83 & 57.54\\
                Co-Correcting       &   \textcolor{red}{\textbf{78.83}}     &   \textcolor{red}{\textbf{78.50}}     &   \textcolor{red}{\textbf{77.26}}     &   \textcolor{red}{\textbf{76.35}}     &   \textcolor{red}{\textbf{71.50}}     &   \textcolor{red}{\textbf{62.13}}\\
                \bottomrule
            \end{tabular}
        \end{table}
		\begin{figure}[htb!]
            \centering
            \subfigure[$clean$]{
                \includegraphics[width=2.9cm]{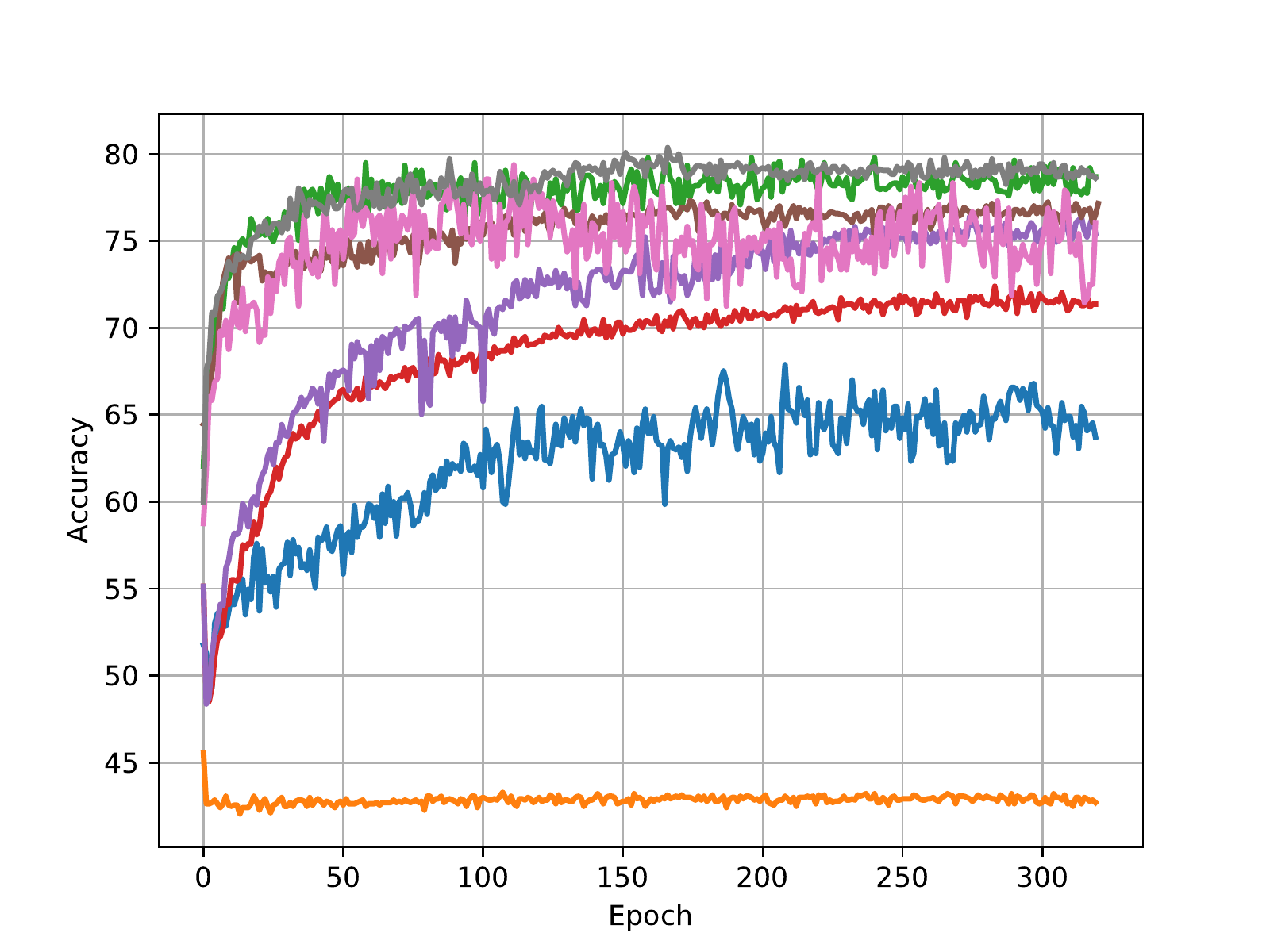} \label{fig:isic_acc_1}
            }  
            \hspace{-6mm}
            \subfigure[$noise=0.05$]{
                \includegraphics[width=2.9cm]{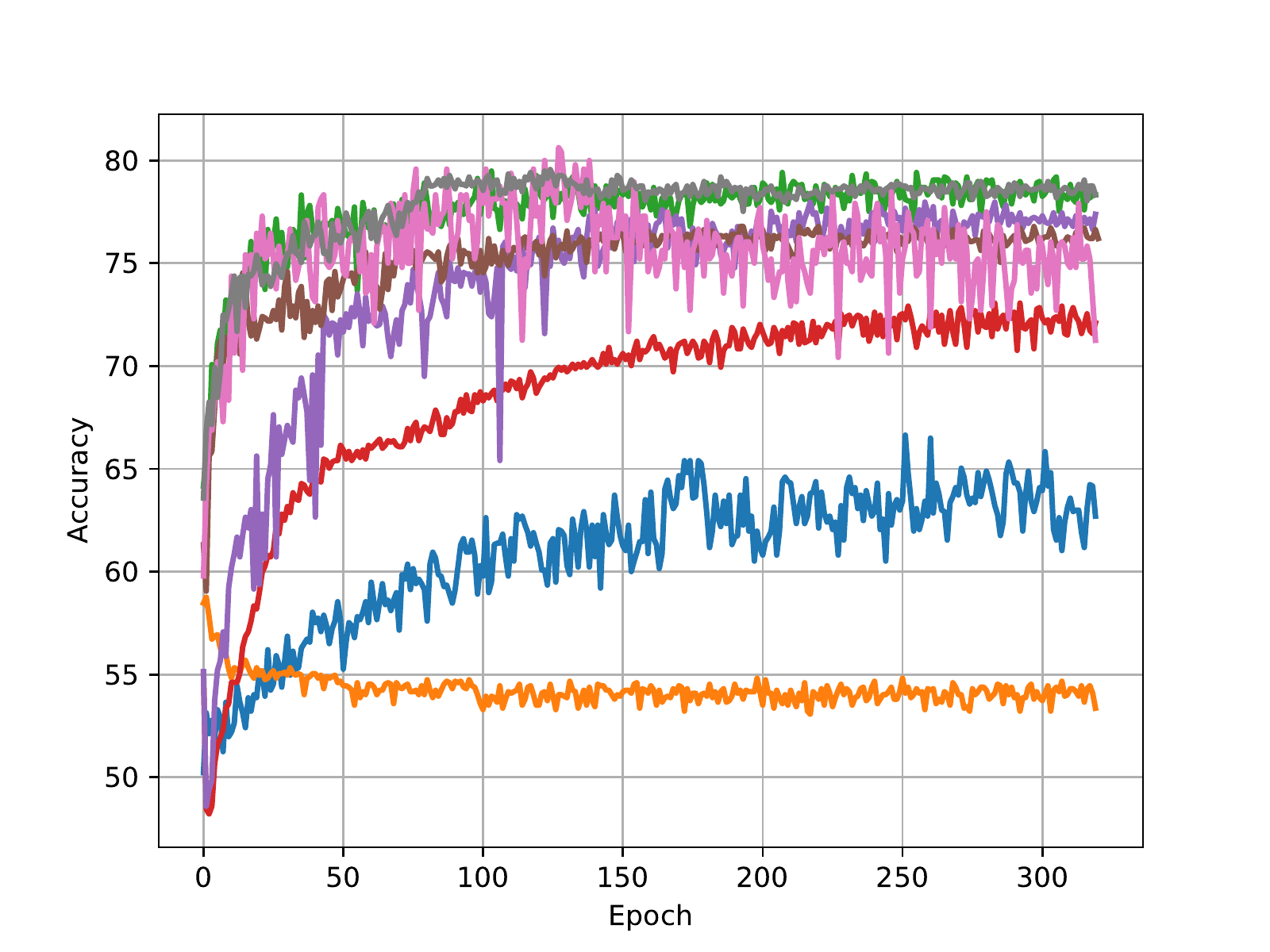} \label{fig:isic_acc_2} 
            }
            \hspace{-6mm}
            \subfigure[$noise=0.1$]{
                \includegraphics[width=2.9cm]{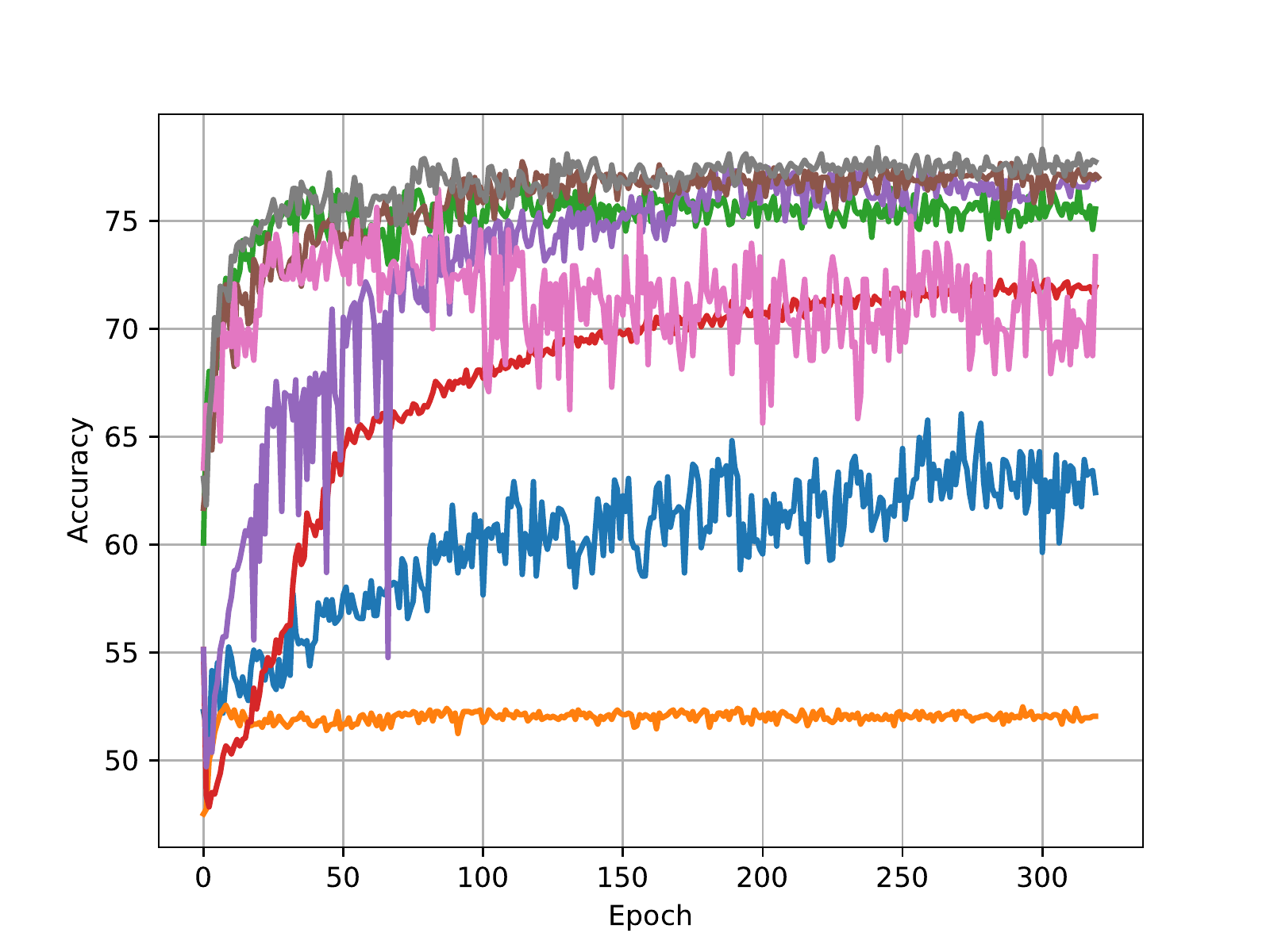}\label{fig:isic_acc_3}
            }
            \hspace{-6mm}
            \subfigure[$noise=0.2$]{
                \includegraphics[width=2.9cm]{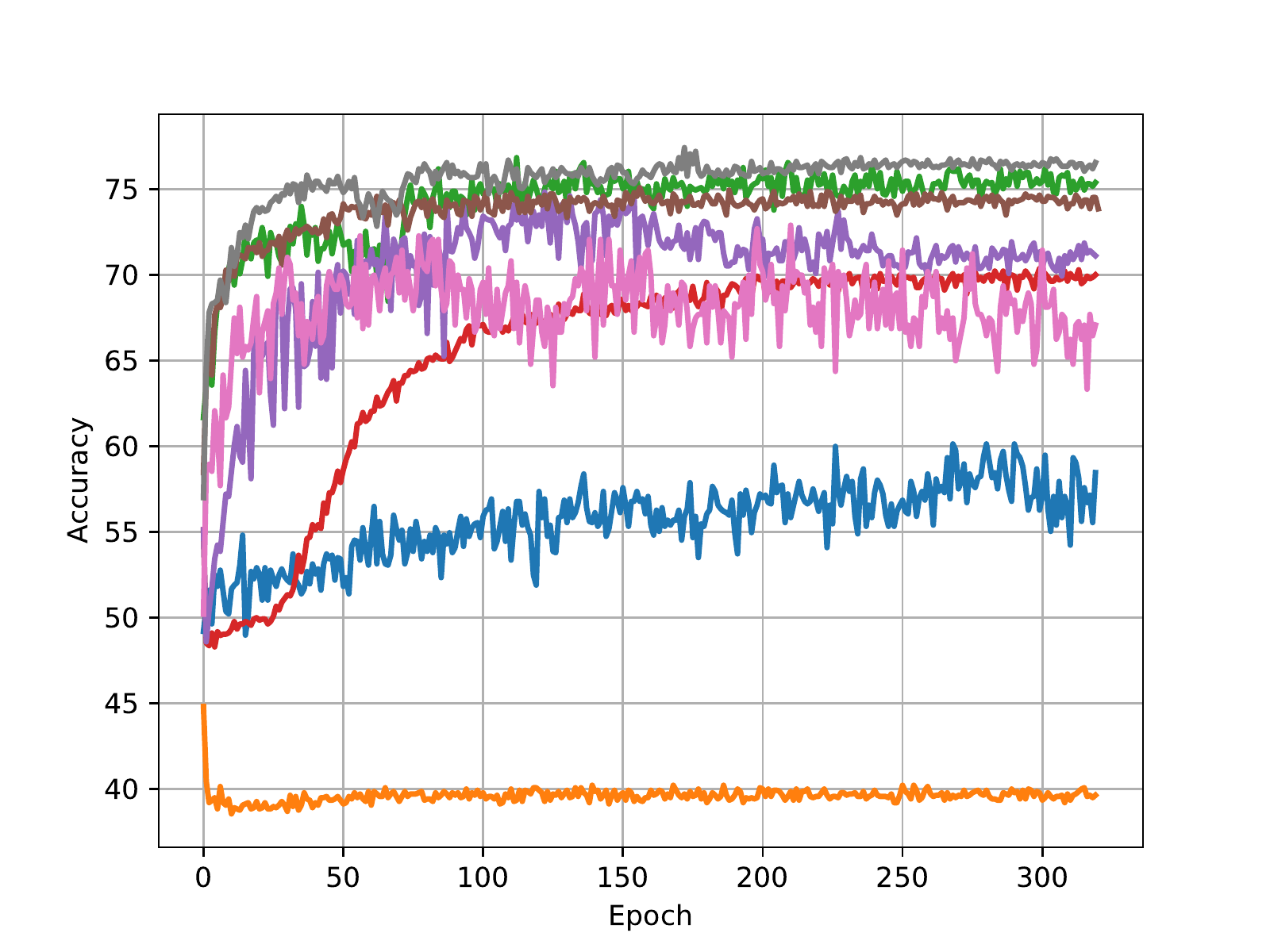}\label{fig:isic_acc_4}
            }
            \hspace{-6mm}
            \subfigure[$noise=0.3$]{
                \includegraphics[width=2.9cm]{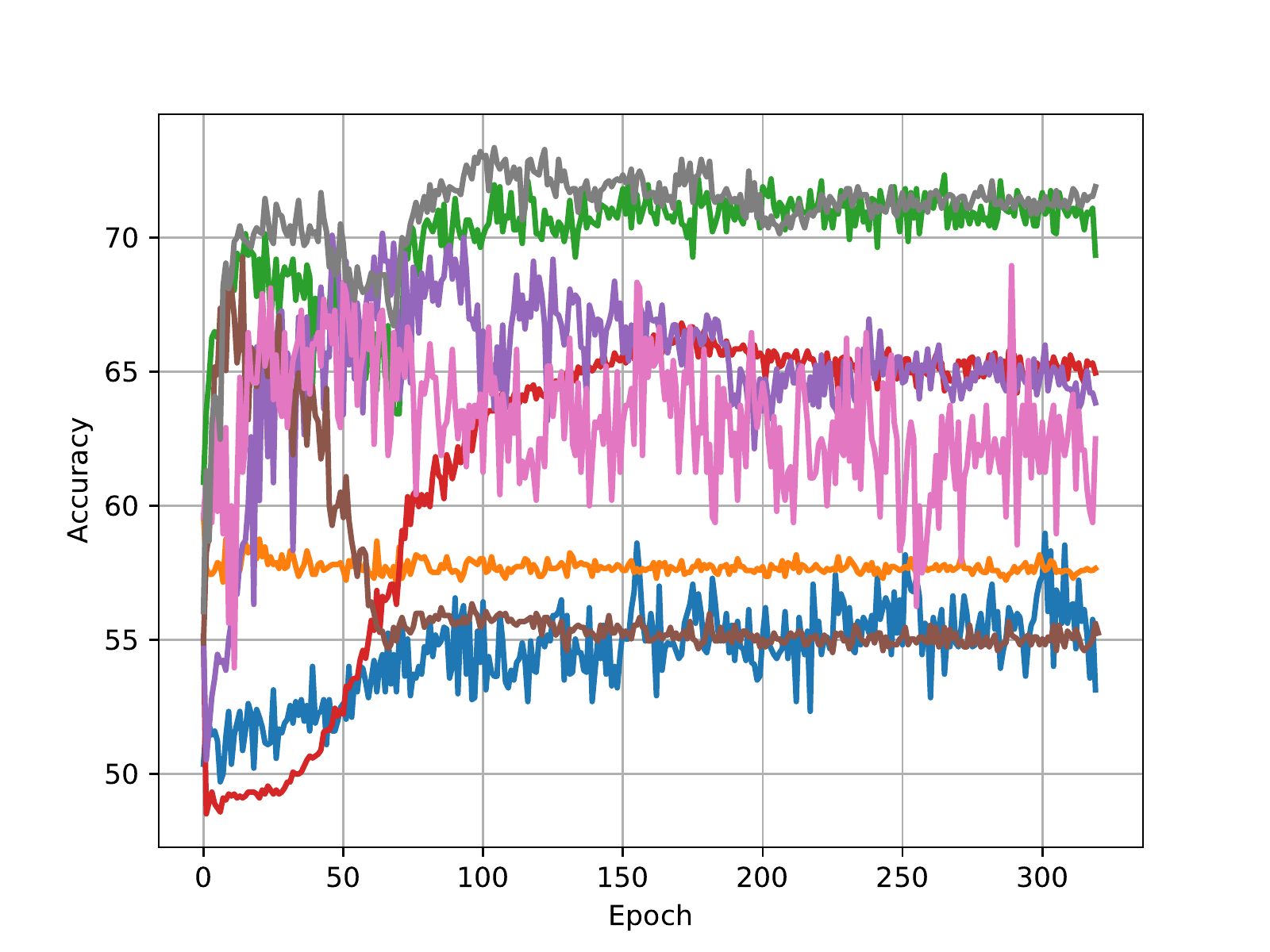}\label{fig:isic_acc_5}
            }
            \hspace{-6mm}
            \subfigure[$noise=0.4$]{
                \includegraphics[width=2.9cm]{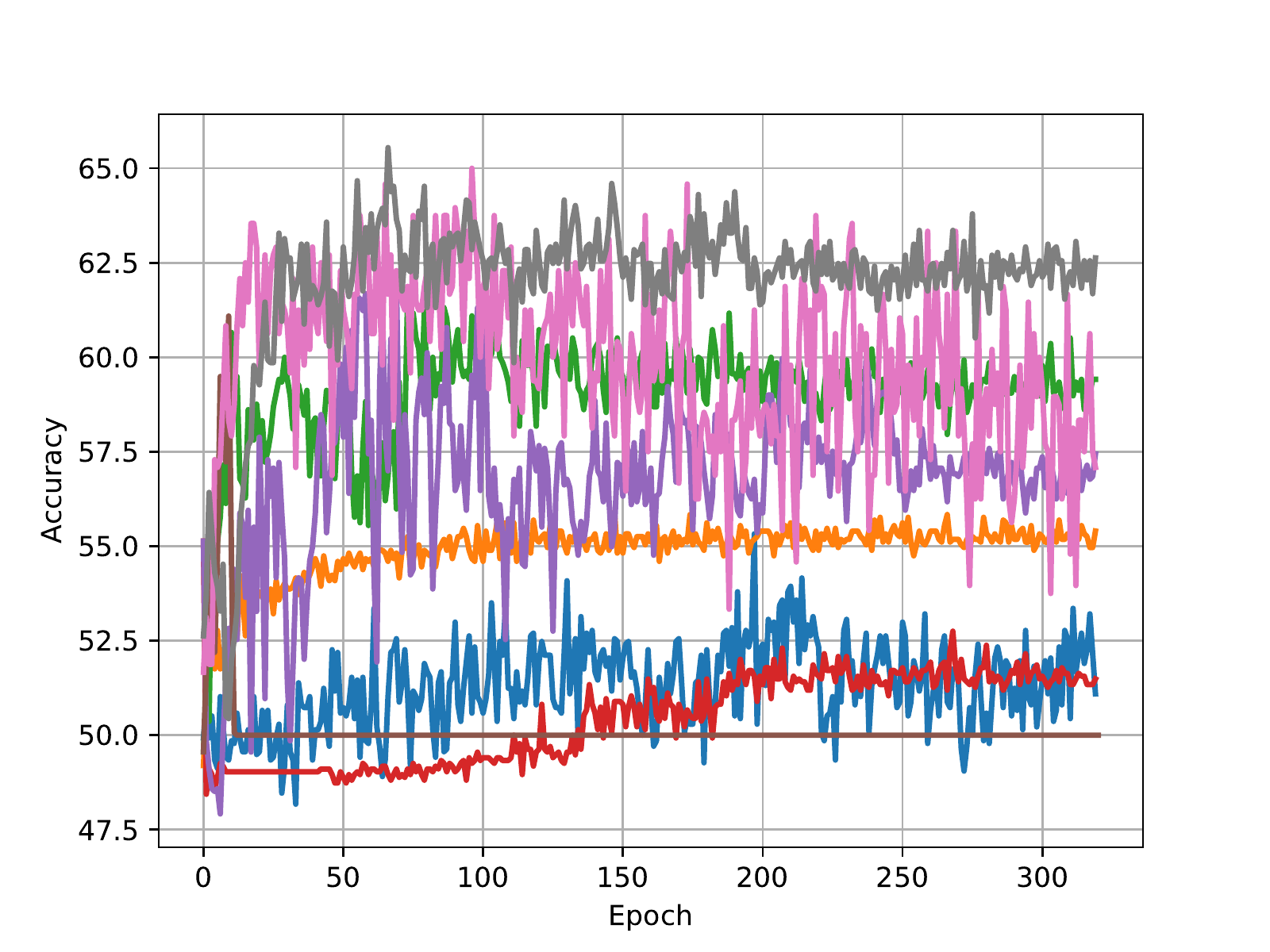}\label{fig:isic_acc_6}
            }               
            \hspace{-6mm}
            \subfigure{
                \includegraphics[width=8cm]{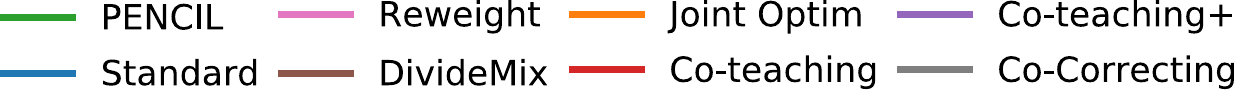}\label{fig:isic_acc_legend}
            }
            \caption{$ACC_{class}$ on ISIC-Archive. Co-Correcting won't remember the noise while Standard and PENCIL will.} 
            \label{fig:isic_acc}
        \end{figure}
            
        \subsubsection{Results and analysis}
            Table \ref{table:isic_acc} shows the classification accuracy of all comparison methods on ISIC-Archive. Co-Correcting won the first place under various noise ratios.
            
\begin{figure}
                \centering
                \subfigure[ISIC-Archive]{
                \includegraphics[width=4.1cm]{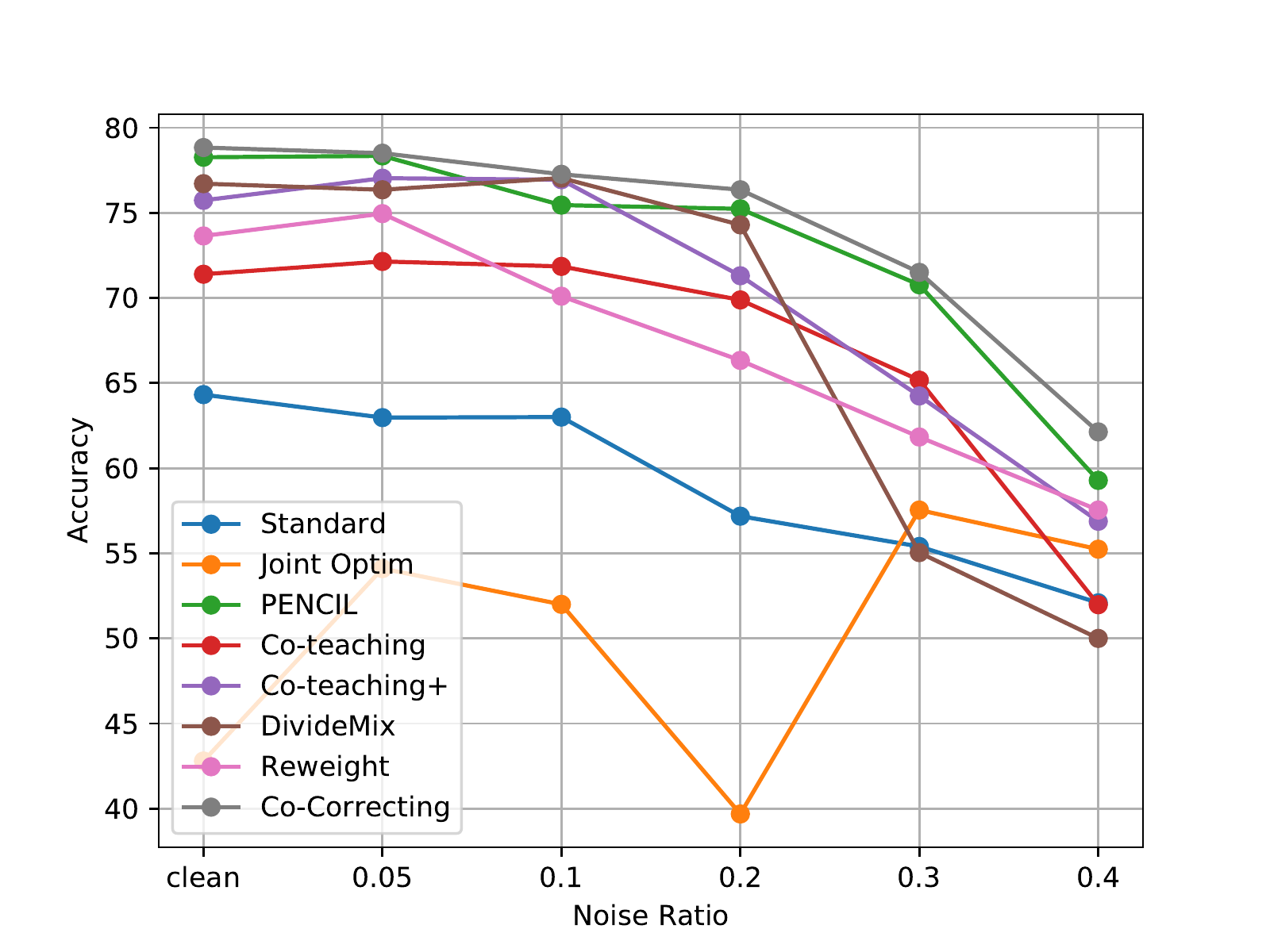} \label{fig:acc-noise_isic}
                } 
                \subfigure[PatchCamelyon]{
                \includegraphics[width=4.1cm]{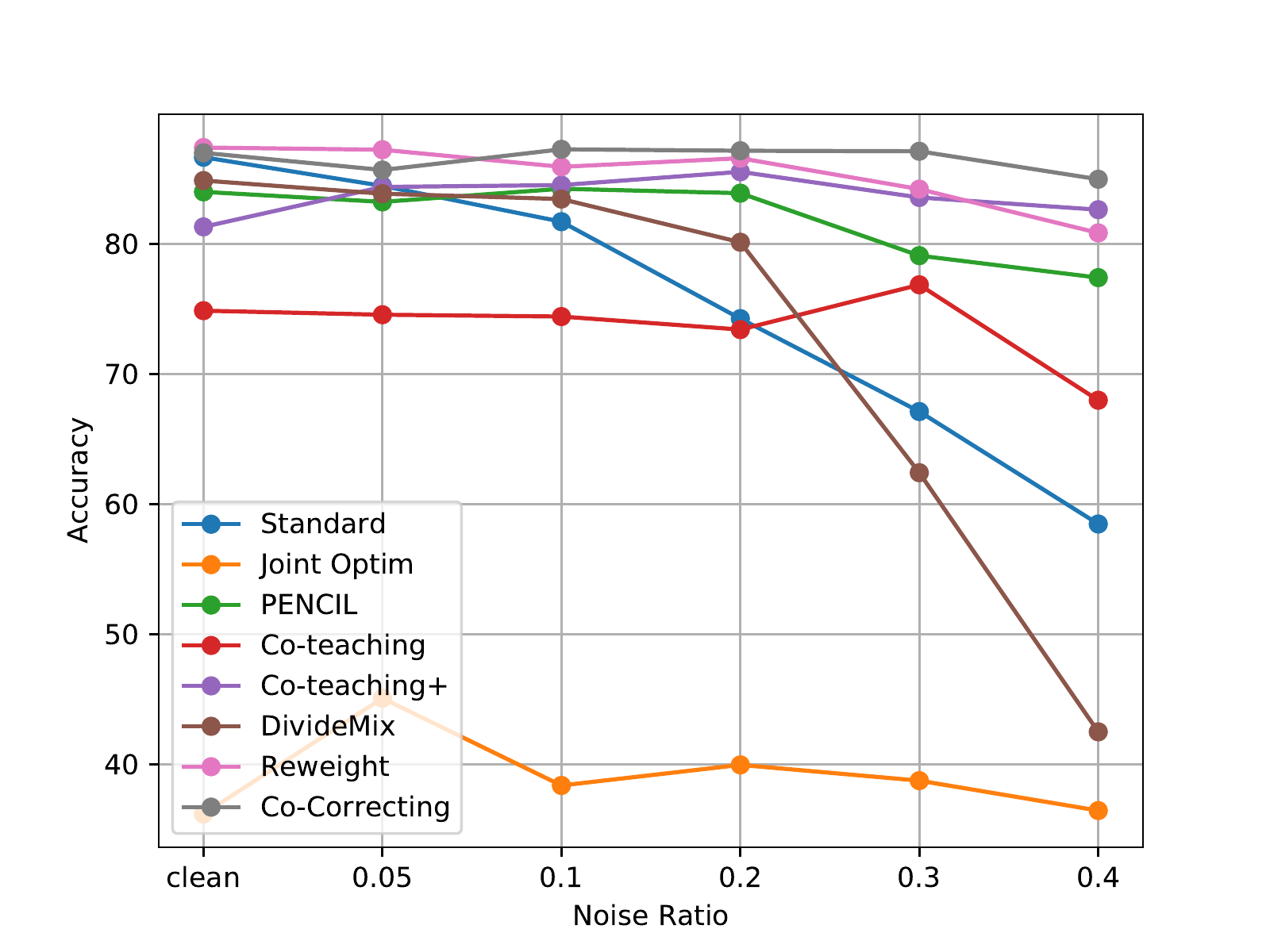} \label{fig:acc-noise_pcam}
                } 
                 \caption{$ACC_{class}$ in various noise ratios on ISIC-Archive and PatchCamelyon. Co-Correcting obtains the highest accuracy under different noise ratios.}
            \end{figure}

            From experimental results, we found that noisy labels do degenerate the accuracy of DNN classifiers. In Figure \ref{fig:acc-noise_isic}, as the noise ratio increases, the accuracy of almost all the comparison methods decrease in varying degrees. The Standard method that uses none LNL strategy gets much worse. Its accuracy is reduced by 19\% when the noise ratio equals 0.4.

            Among LNL methods, Co-Correcting obtains the highest accuracy under different noise ratios. Even under severe noise, its accuracy is still close to Standard{’}s results on clean labels. PENCIL ranks second. In contrast, the methods that do not update the label have a significant decrease in accuracy as the proportion of noises increases. These methods try to avoid learning noisy samples. Therefore as the noise ratio increases, the samples actually used for training will be reduced accordingly.  \ADD{It can be found that the accuracy of Join Optim is much lower than the other methods. This phenomenon may be caused by the following two reasons: 1) Due to the limited data, the network of Joint Optim may not be able to learn a good model in the first stage.  2) Joint Optim accumulates errors,  which will lead to lower accuracy in medical image classification.}
        
            Another interesting phenomenon is that the accuracy of most LNL methods exceeds that of Standard even on clean labels. Co-Correcting can not only deal with the noisy labels but also can improve the accuracy and generalization on noise-free data. In the experiment, Co-Correcting obtains an accuracy of 78.83\% which is 22.6\% higher than that of Standard on clean labels. In our analysis, it may be due to three reasons. First of all, there may exist fuzzy and inaccurate annotations in the ground truth labels. Secondly, the proposed LNL mechanism inhibits the network to early overfit to a few of hard samples which could improve the generalization of the model. Lastly, Co-Correcting adopts soft labels which could more accurately represent the label distribution. 
        
            The learning process with noisy labels has memorization effects which are very similar to the overfitting phenomenon according to which the accuracy decreases after it reaches the peak value. The memorization effect is caused by accumulating errors. For example, as Figure \ref{fig:isic_acc_6} shown, the purple line declines after 150 epochs. From Figure \ref{fig:isic_acc}, we could observe that Co-Correcting (the \DEL{pink}\ADD{gray} line) is scarcely affected by the memorization effect. Therefore, it does not accumulates errors when learning with noisy labels.
        
    \subsection{Experiments on PatchCamelyon}   \label{section:pcam_dataset}
        \subsubsection{PatchCamelyon Dataset}
            PatchCamelyon \cite{ehteshami_bejnordi_diagnostic_2017,veeling_rotation_2018} is a new medical image classification dataset. It consists of 327,680 color images ($96px * 96 px$) extracted from histopathologic scans of lymph node sections. Each image is annotated with a binary label indicating the presence of metastatic tissue. We selected 32,766 samples from PatchCamelyon and \DEL{split it into training, testing, and validating sets}\ADD{split them into the training, test, and validation sets in the ratio of 8:1:1, which is following the data author’s suggestion.} The numbers of samples are 26214, 3276, 3276 respectively. The data is augmented in the same way as ISIC-Archive experiments. DenseNet-169 was selected as the backbone network. Except for setting the batch size to 128, all other parameters are the same as those in the ISIC-Archive experiment.
        \begin{table}[htbp]
            \centering
            \caption{$ACC_{class}$ under different noise levels in PatchCamelyon.}
            \label{table:pcam_acc}
            \begin{tabular}{ccccccc}
                \toprule
                Methods             &   clean                                           &   0.05                                                &   0.1                                             &   0.2                                             &   0.3                                             &   0.4\\
                \midrule        
                Standard                &   \textcolor{blue}{86.64}             &   \textcolor{blue}{84.44}             &   81.69                                           &   74.24                                           &   67.10                                           &   58.46\\
                Joint Optim         &   36.18                                           &   45.07                                           &   38.37                                           &   39.95                                           &   38.74                                           &   36.44\\
                PENCIL              &   83.98                                           &   83.21                                           &   84.22                                           &   83.88                                           &   79.08                                           &   77.39\\
                Co-teaching         &   74.85                                           &   74.54                                           &   74.40                                           &   73.40                                           &   76.84                                           &   67.97\\
                Co-teaching+        &   81.29                                           &   84.36                                           &   84.50             &   \textcolor{blue}{85.52}             &   83.55             &   \textcolor{blue}{82.61}\\
                DivideMix               &   84.85                                           &   83.84                                           &   83.43                                           &   80.11                                           &   62.40                                           &   42.49\\
                \ADD{Reweight}				& 81.64	& 74.06 & \textcolor{blue}{85.90} & 84.56 & \textcolor{blue}{84.19} & 80.83\\
                Co-Correcting       &   \textcolor{red}{\textbf{86.98}} &   \textcolor{red}{\textbf{85.66}} &   \textcolor{red}{\textbf{87.25}}     &   \textcolor{red}{\textbf{87.14}}     &   \textcolor{red}{\textbf{87.10}} &   \textcolor{red}{\textbf{84.95}}\\
                \bottomrule
            \end{tabular}
        \end{table}
        \subsubsection{Results and analysis}    
            The experimental results on PatchCamelyon are listed in Table \ref{table:pcam_acc} and the classification accuracy is shown in Figure \ref{fig:acc-noise_pcam}. Just as the experiments on ISIC-Archive, Co-Correcting achieved the best result under different noise ratios.  Co-Correcting is much better than \DEL{second-place} \ADD{Co-teaching,} Co-teaching+ \ADD{and Reweight}.  At the noise ratio of 0.4, Co-Correcting surpasses Co-teaching+ in the accuracy by 2.34 percentage points. On clean data, while most LNL methods failed to maintain the accuracy level, Co-Correcting still outperforms Standard. It confirms that Co-Correcting can improve the classification accuracy under clean samples which makes it very versatile.
            
            The experimental results on PatchCamelyon differed with those on ISIC-Archive in three aspects. Figure \ref{fig:acc-noise_pcam} shows the curves. Due to the larger amount of data in this experimental group, deep neural networks have a certain degree of robustness. It has been observed that the performance of Standard does not decline significantly if the noise ratio is less than 0.2. Secondly, the performance of LNL methods decreases more slowly as the noise ratio increases, which means that the advantages of Co-Correcting can be better utilized if given more training data. Thirdly, when the ratio of label noises is low, some LNL methods, such as Co-teaching or Co-teaching+, are not as accurate as Standard. This may be because that they select samples with small losses to train the model, which conversely suppress the learning of hard samples, thereby affecting the overall accuracy.

\begin{figure}[htb!]
                \centering
                \subfigure[$clean$]{
                \includegraphics[width=2.9cm]{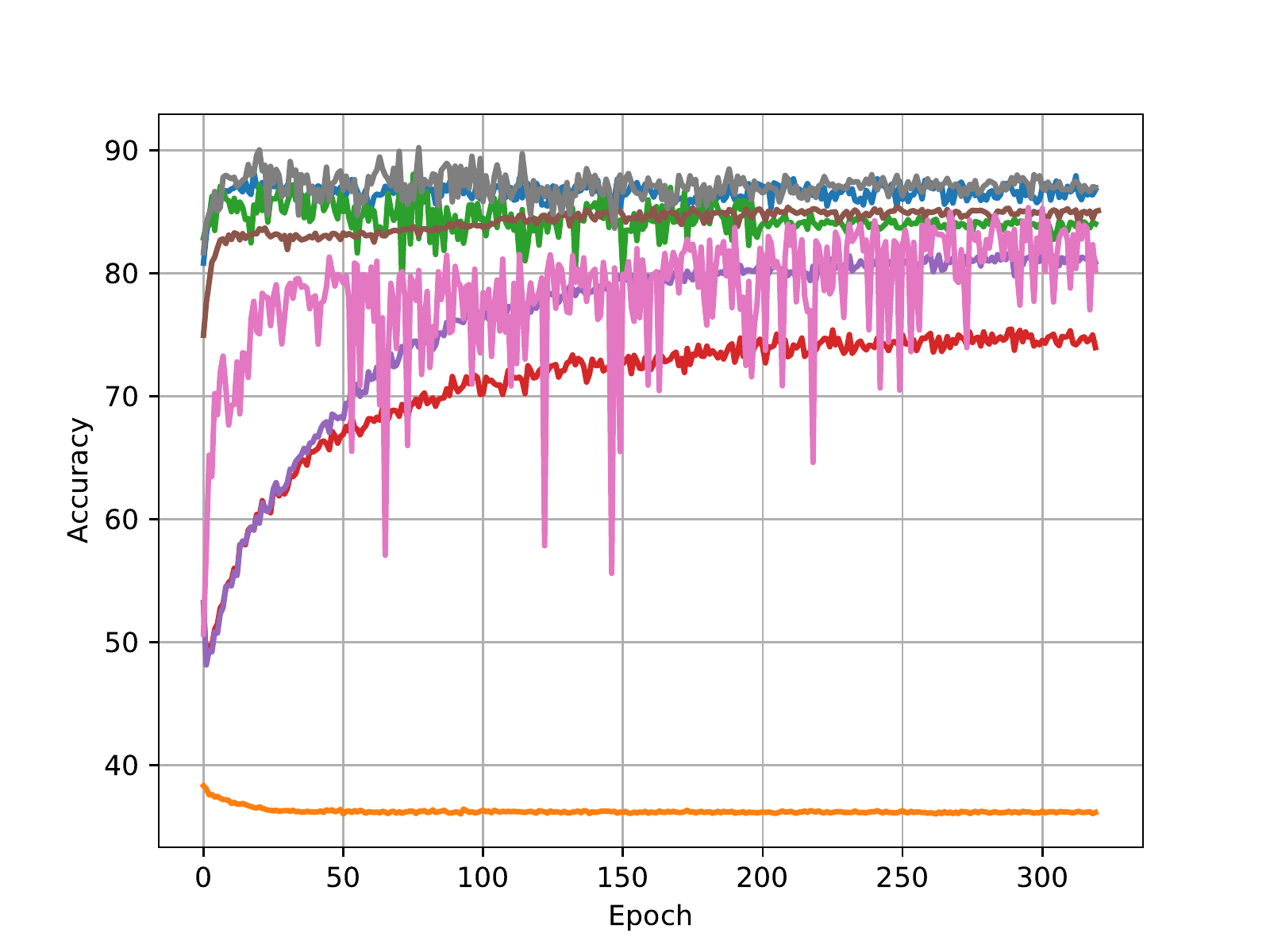} \label{fig:pcam_acc_1}
                } 
                \hspace{-6mm}
                \subfigure[$noise=0.05$]{
                \includegraphics[width=2.9cm]{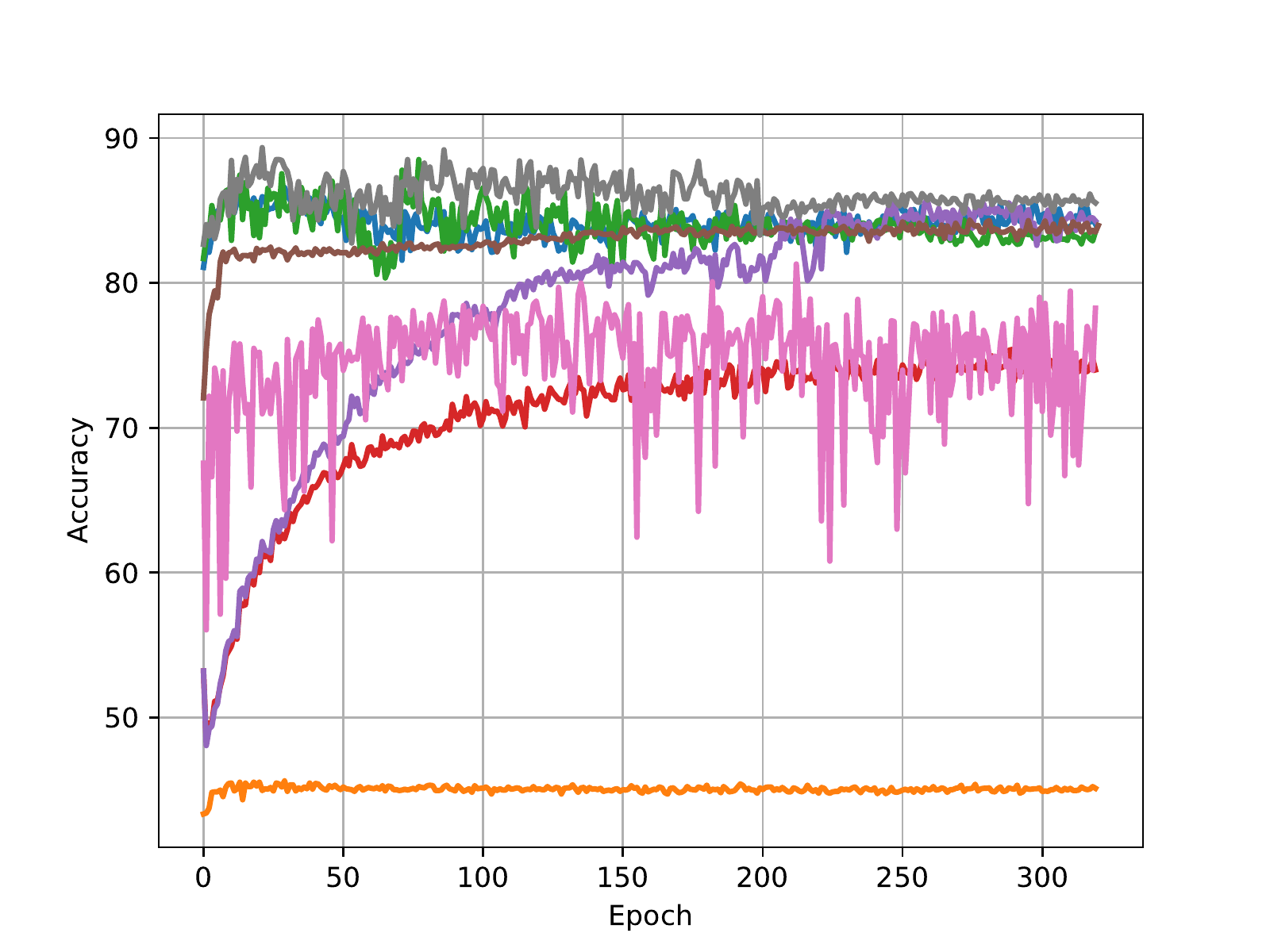} \label{fig:pcam_acc_2} 
                }
                \hspace{-6mm}
                \subfigure[$noise=0.1$]{
                \includegraphics[width=2.9cm]{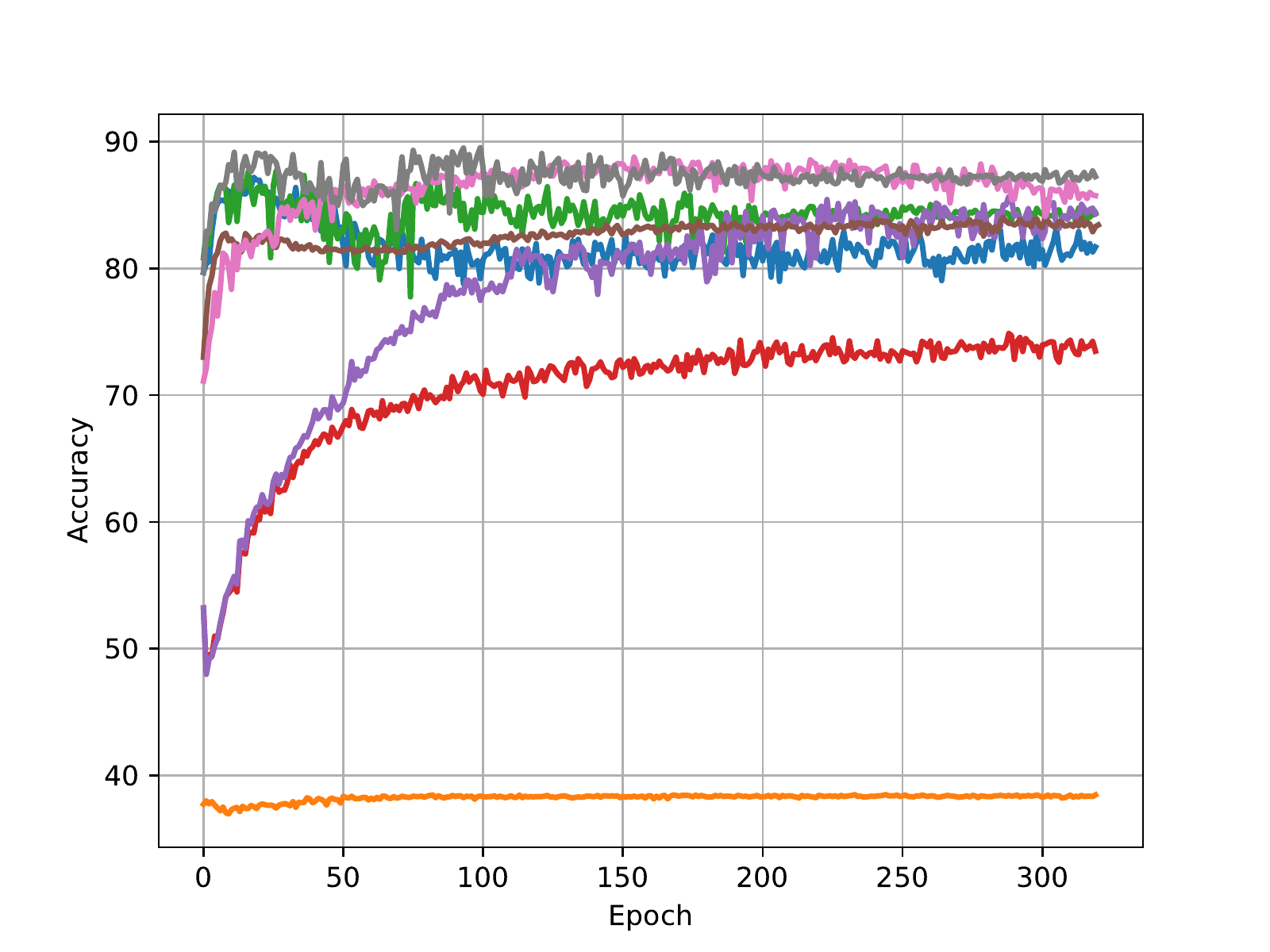}\label{fig:pcam_acc_3}
                }
                \hspace{-6mm}
                \subfigure[$noise=0.2$]{
                \includegraphics[width=2.9cm]{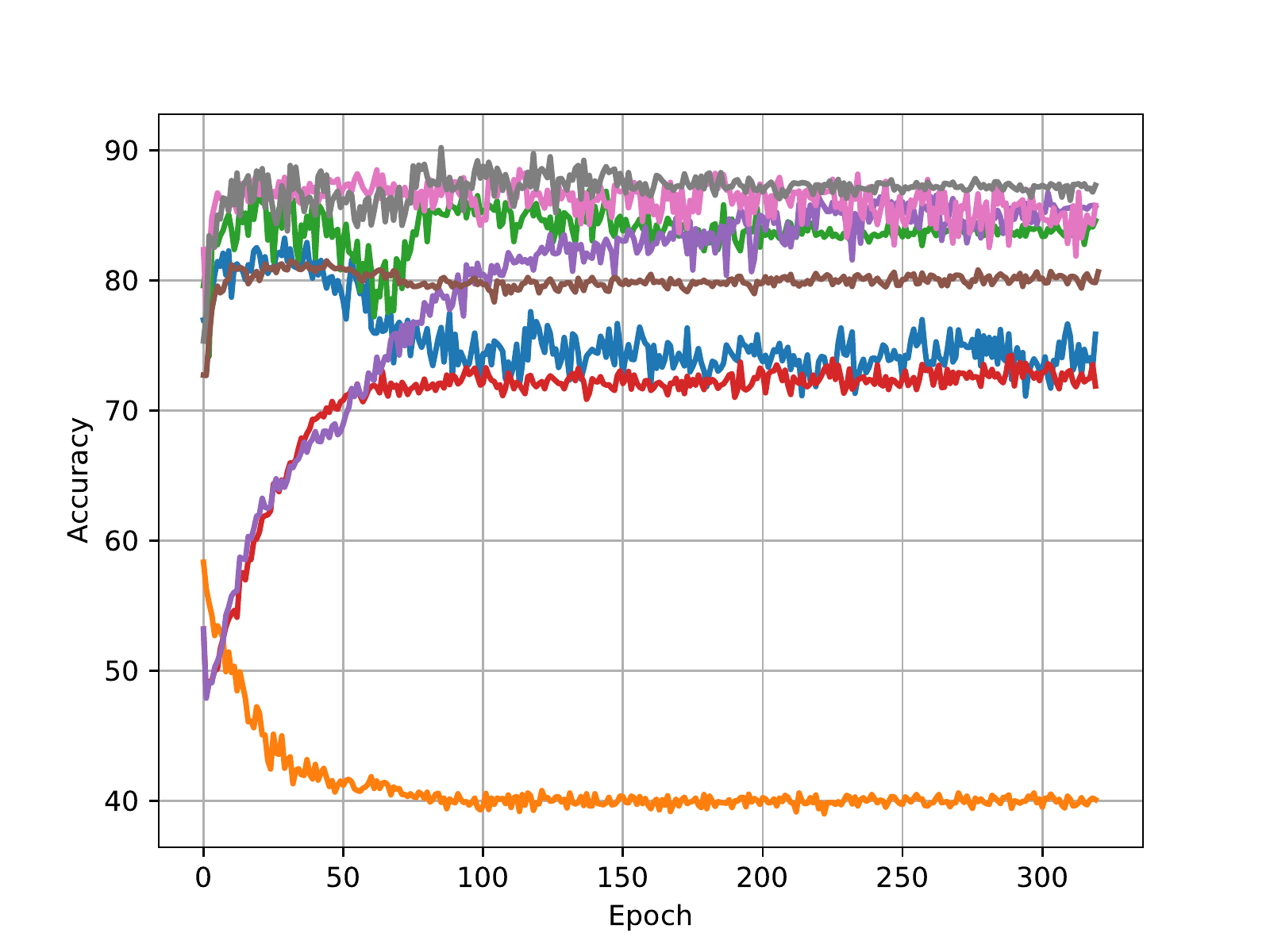}\label{fig:pcam_acc_4}
                }
                \hspace{-6mm}
                \subfigure[$noise=0.3$]{
                \includegraphics[width=2.9cm]{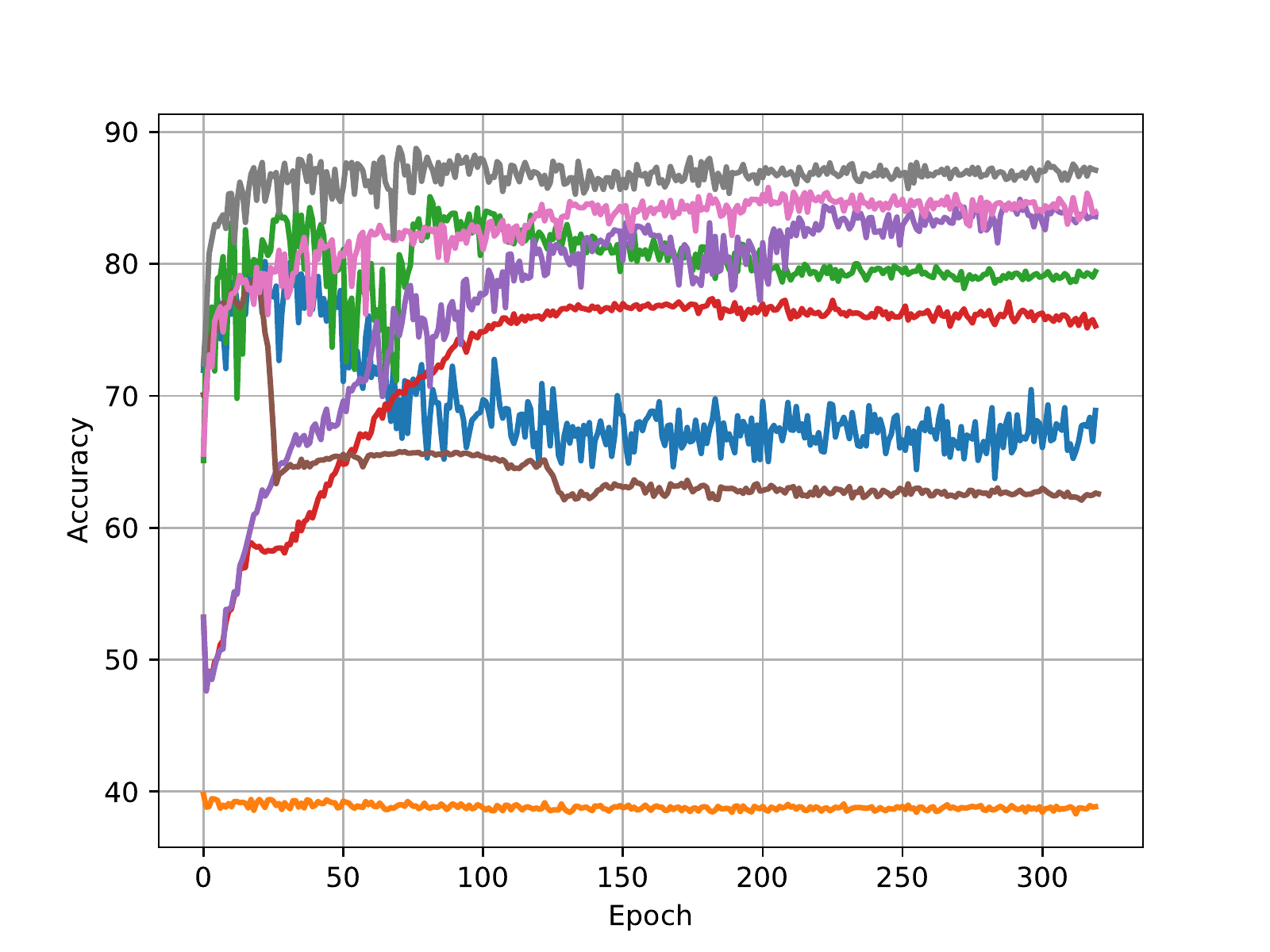}\label{fig:pcam_acc_5}
                }
                \hspace{-6mm}
                \subfigure[$noise=0.4$]{
                \includegraphics[width=2.9cm]{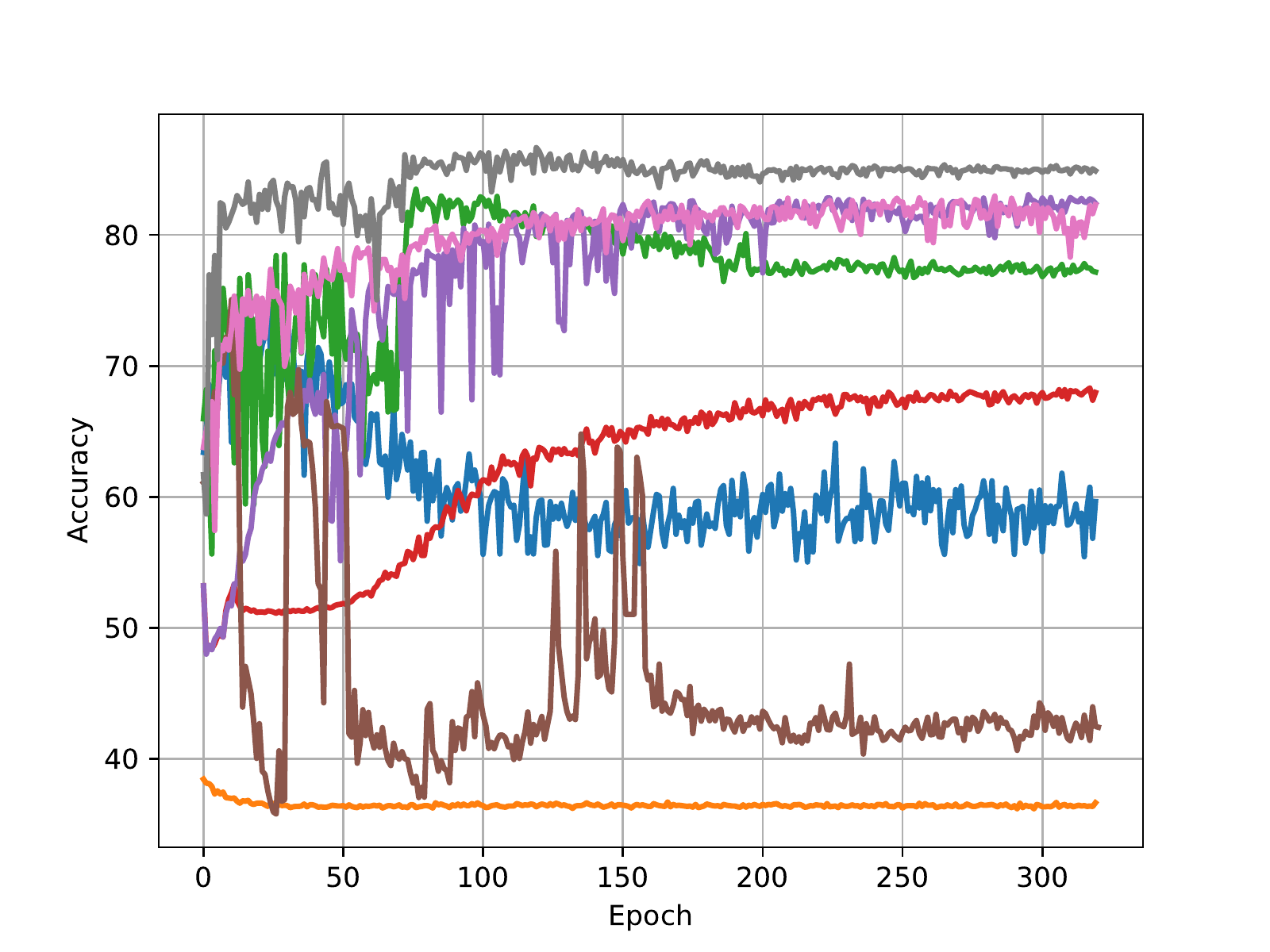}\label{fig:pcam_acc_6}
                }
                \hspace{-6mm}
                \subfigure{
                \includegraphics[width=8cm]{{legend_acc}.pdf}\label{fig:pcam_acc_legend}
                }
                \caption{$ACC_{class}$ on PatchCamelyon.}
                \label{fig:pcam_acc}
            \end{figure}

            Figure \ref{fig:pcam_acc} is the learning curves under different noisy ratios. Figure \ref{fig:pcam_acc_5}$\sim$\ref{fig:pcam_acc_6} represent more obvious memorization effects. In the cases when the noise ratio is greater than 0.3, it can be observed that both the accuracy of Standard and PENCIL increase first and then decrease obviously. Co-teaching and Co-teaching+ learn on small-loss samples conservatively. Their curves continue to rise until they reach their peak. Unfortunately, their peaks are always lower than those of Co-Correcting. Co-Correcting reaches the optimal model quickly and then remains stable.
            
    \subsection{The analysis of label correction}
\begin{figure}[htb!]
            \centering
            \subfigure[ISIC-Archive]{
            \includegraphics[width=3cm]{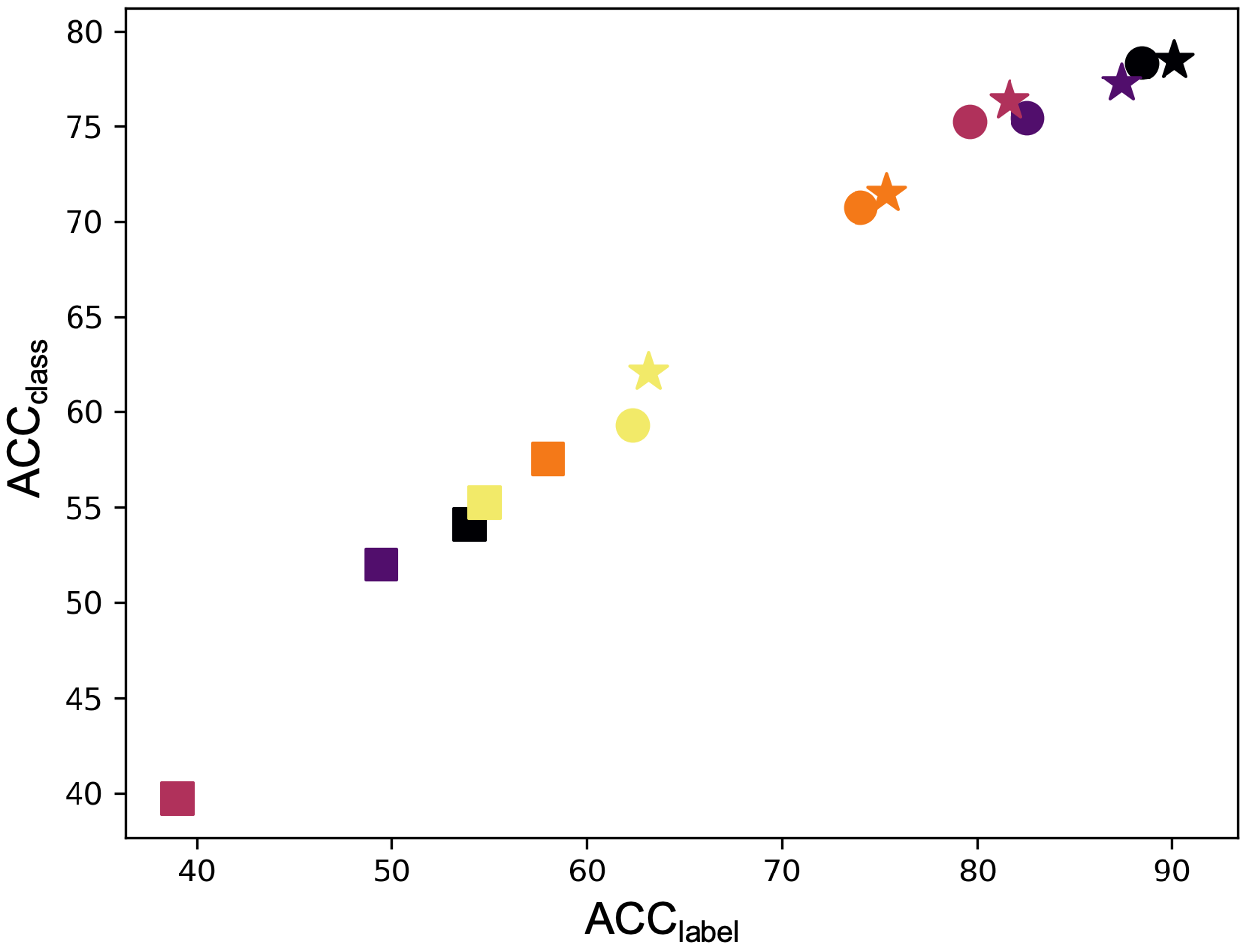} \label{fig:acc_depend_label_isic}
            }  
            \hspace{-3mm}
            \subfigure[PatchCamelyon]{
            \includegraphics[width=3cm]{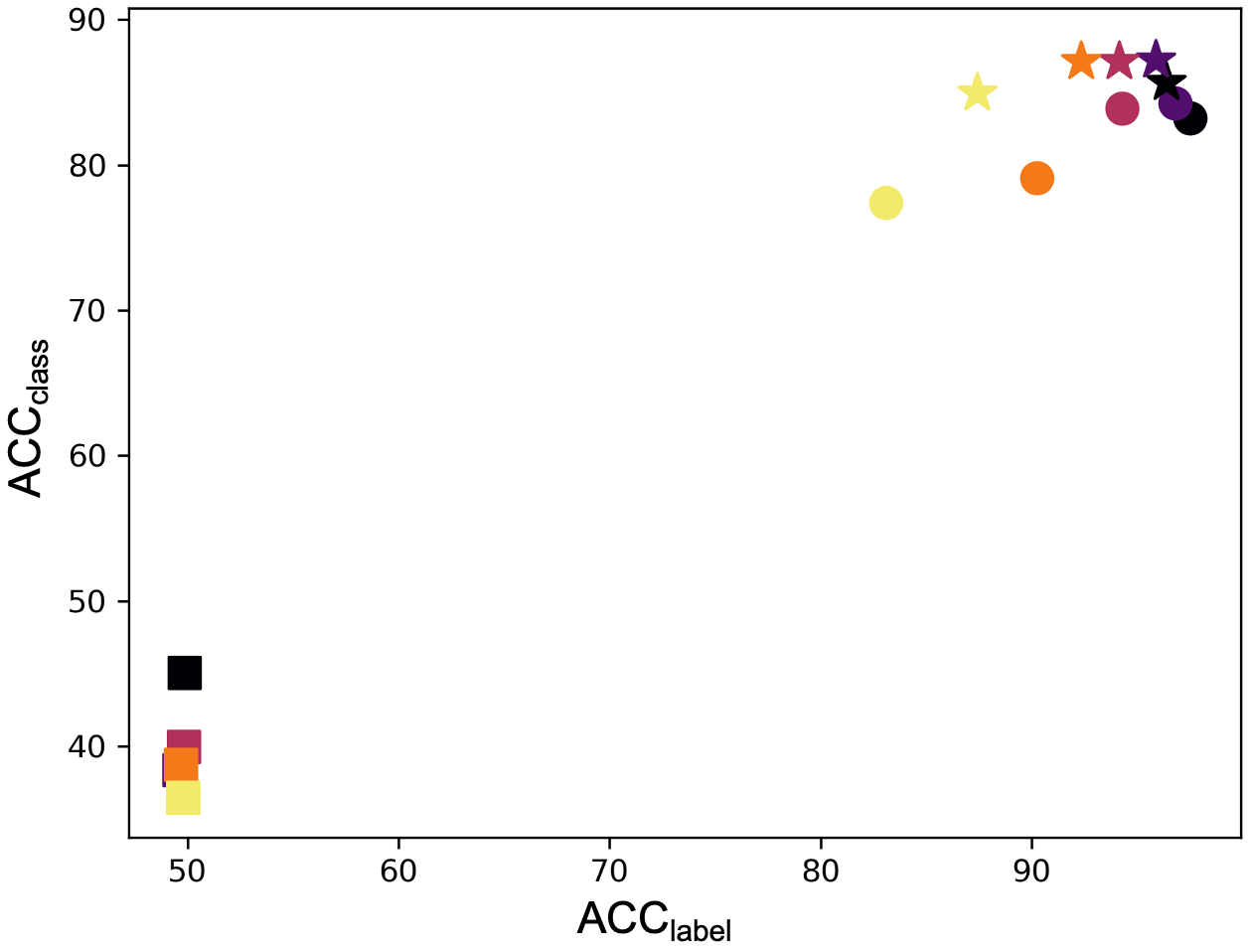} \label{fig:acc_depend_label_pcam} 
            }
            \vspace{-3mm}
            \subfigure{
            \includegraphics[width=6cm]{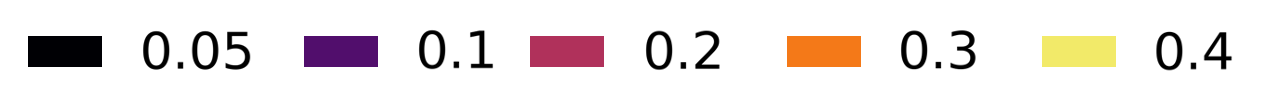}\label{fig:acc_depend_label_lengend_noise}
            }
            \subfigure{
            \includegraphics[width=6cm]{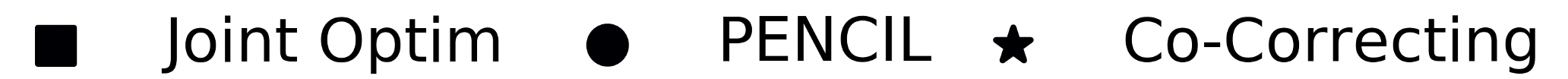}\label{fig:acc_depend_label_lengend_method}
            }
            \caption{Performance depends on the quality of the label correction. In figures, colors represent the noise levels, and the shapes represent different methods. The abscissa is $ACC_{label}$ and the ordinate is $ACC_{class}$. It can be seen that $ACC_{class}$ gets higher with the increase of $ACC_{label}$ in most cases. Co-Correcting can always get the highest $ACC_{class}$ in various noise levels.}
            \label{fig:acc_depend_label}
        \end{figure}

        The accuracy of the models is positively correlated with the quality of the label correction. As shown in Figure \ref{fig:acc_depend_label}, a higher labeling accuracy leads to higher model accuracy in most cases. 

\begin{figure}[htb!]
            \centering
            \subfigure[epoch 0]{
            \includegraphics[width=2.9cm]{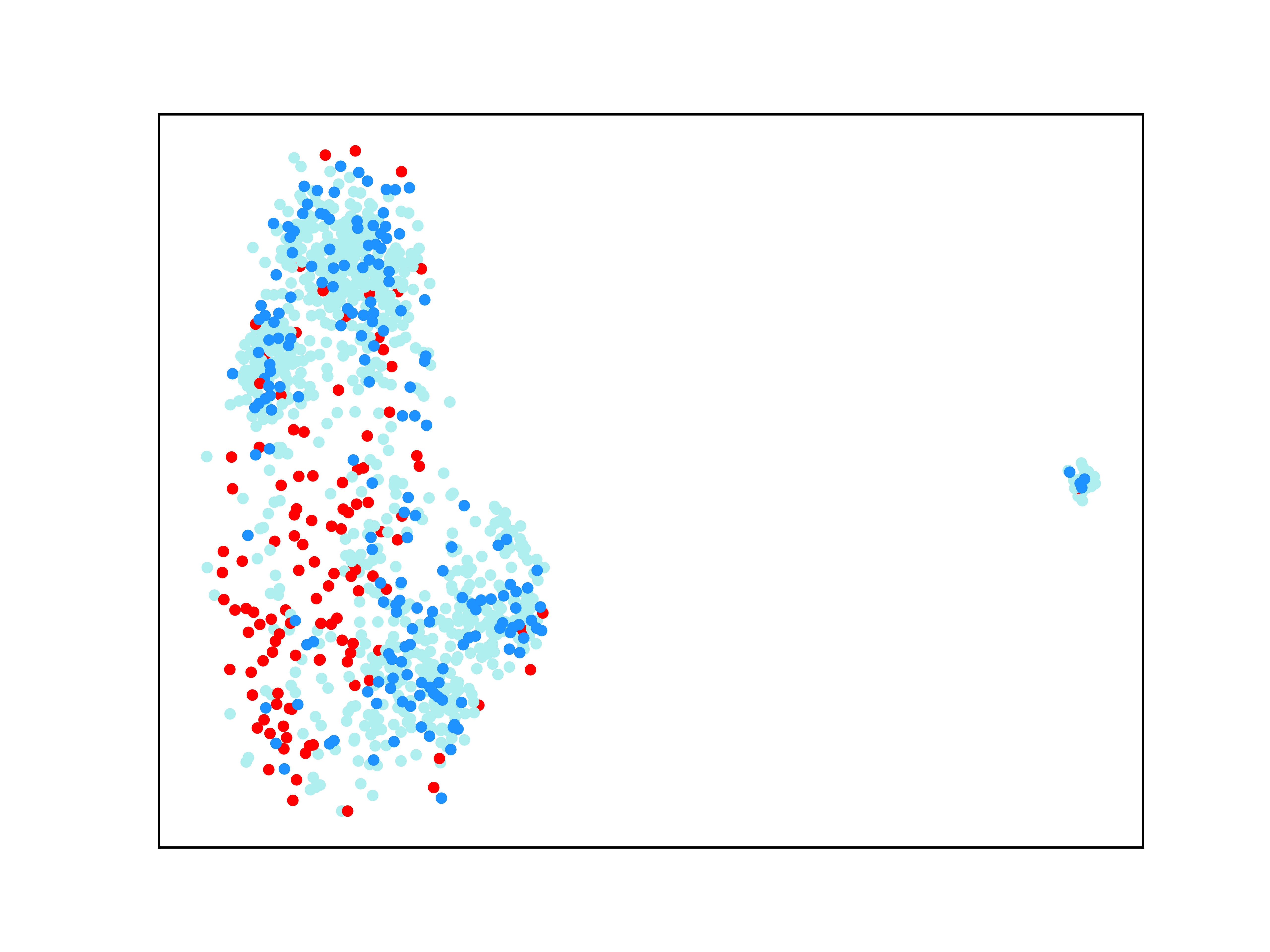} \label{fig:tsne_0}
            }
            \hspace{-6mm}
            \subfigure[epoch 70]{
            \includegraphics[width=2.9cm]{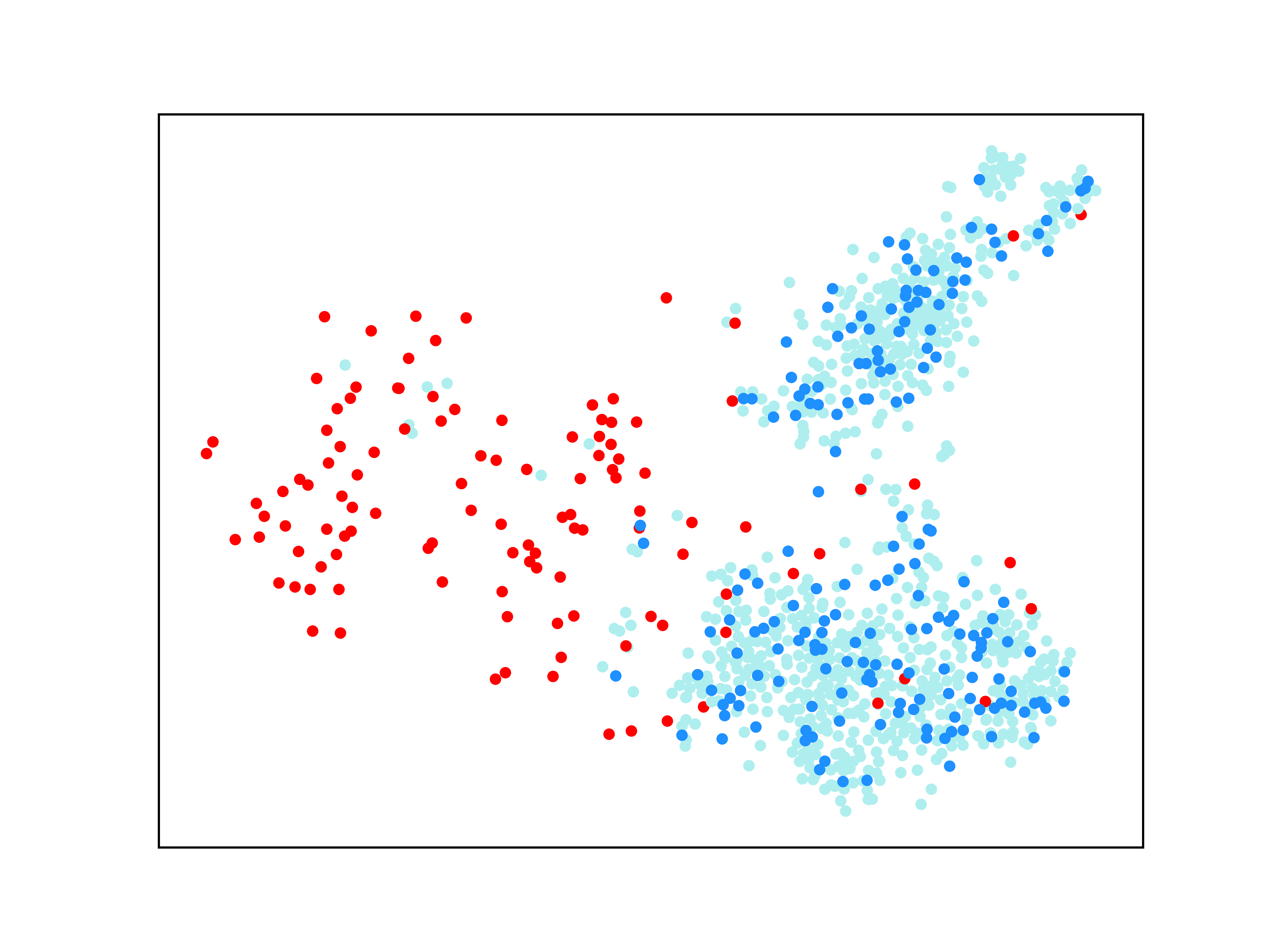} \label{fig:tsne_1}
            }
            \hspace{-6mm}
            \subfigure[epoch 200]{
            \includegraphics[width=2.9cm]{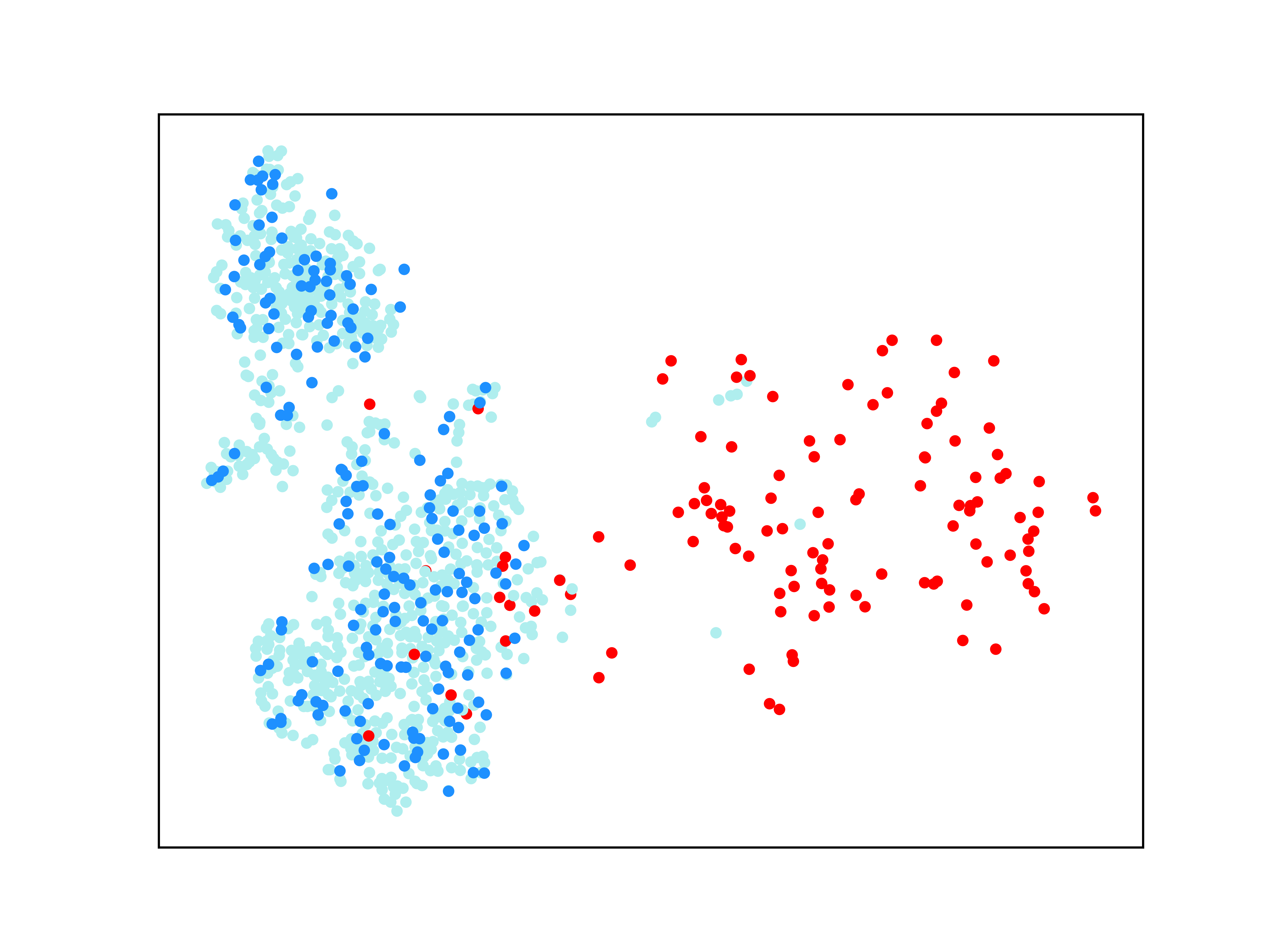} \label{fig:tsne_2}
            }
            \subfigure{
            \includegraphics[width=4.5cm]{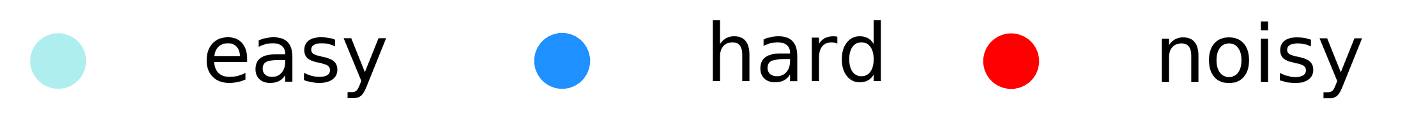} \label{fig:tsne_lengend}
            }
            \caption{t-SNE plot in feature space with 0.3 noise on PatchCamelyon.}
            \label{fig:tsne}
        \end{figure}

        Co-Correcting performs well in annotation correction. As Co-Correcting iteratively updates the labels, it gradually increases the distance between the hard samples and the noisy samples in the embedding space, which will provide more opportunities to the model to learn more hard samples and improve the generalization. This can be confirmed by observing the t-SNE visualization. In Figure \ref{fig:tsne}, red points represent noisy samples and blue points represent hard samples. The two sets of samples tend to be far away from each other.
        
        \begin{table}[htb!]
            \centering
            \caption{Ablation study methods.}
            \label{table:isic_ablation_baselines}
            \begin{tabular}{cc}
                \toprule
                Methods                 &   Description\\
                \midrule        
                Co-Correcting       &   Dual-NN + Curriculum + Label update + Sum gradient\\
                Co-Cor-A                &   Single-NN + Label update + Curriculum\\
                Co-Cor-B                &   Dual-NN + Label update + Sum gradient\\
                \bottomrule
            \end{tabular}
        \end{table}
    
    \FloatBarrier
    \subsection{Ablation studies}
        In order to understand the role of each component of Co-Correcting better, we perform ablation studies on both ISIC and PatchCamelyon. It is worth noting that if none of dual-network architecture and label correction curriculum is adopted, Co-Correcting will degenerate into PENCIL. For the convenience of explanation, we define the abbreviations of different versions in Table \ref{table:isic_ablation_baselines}.
        
        Table \ref{table:ablation_acc} show the classification accuracy results on the ISIC-Archive and PatchCamelyon datasets. Table \ref{table:ablation_label_accu} show the label accuracy after label correction. 
            \begin{table}[htb!]
                \centering
                \caption{$ACC_{class}$ on ISIC-Archive and PatchCamelyon.}
                \label{table:ablation_acc}
                \begin{tabularx}{8.6cm}{p{1.65cm}<{\centering}p{1.65cm}<{\centering}X<{\centering}X<{\centering}X<{\centering}X<{\centering}X<{\centering}}
                    \toprule
                    Datasets & Methods             &   0.05                                                &   0.1                                         &   0.2                                             &   0.3                                             &   0.4\\
                    \midrule      
                    \multirow{3}*{ISIC-Archive} & Co-Correcting       &   \textcolor{red}{\textbf{78.50}}     &   \textcolor{red}{\textbf{77.26}} &   \textcolor{red}{\textbf{76.35}}     &   \textcolor{red}{\textbf{71.50}}     &   \textcolor{red}{\textbf{62.13}}\\
                    ~ & Co-Cor-A                &   78.50                                           &   75.11                                           &   69.36                                           &   62.58                                           &   55.18\\
                    ~ & Co-Cor-B                &   77.28                                           &   75.68                                       &   68.76                                           &   65.25                                           & 59.10\\
                    \midrule
                    \multirow{3}*{PatchCamelyon} & Co-Correcting       &   85.66                                           &   \textcolor{red}{\textbf{87.25}} &   \textcolor{red}{\textbf{87.14}} &   \textcolor{red}{\textbf{87.10}} &   \textcolor{red}{\textbf{84.95}}\\
                    ~ & Co-Cor-A                &   83.26                                           &   83.23                                       &   82.55                                       &   79.35                                       &   78.27\\
                    ~ & Co-Cor-B                &   \textcolor{red}{\textbf{86.56}} &   86.84                                       &   86.60                                       &   87.08                                       &   83.87\\
                    \bottomrule
                \end{tabularx}
            \end{table}
            \begin{table*}[!h]
                \centering
                \caption{$ACC_{label}$ \ADD{and $ACC_{label}$ improvement (in brackets)} on ISIC-Archive and PatchCamelyon. }
                \label{table:ablation_label_accu}
                \begin{tabular}{ccccccc}
                    \toprule
                    Datasets & Methods             &   0.05                                                &   0.1                                             &   0.2                                             &   0.3                                             &   0.4\\
                    \midrule        
                    \multirow{3}*{ISIC-Archive} & Co-Correcting       &   90.11 (-4.89)                                          &   87.38 (-2.62)                                          &   \textcolor{red}{\textbf{81.65 (+1.65)}}     &   \textcolor{red}{\textbf{75.34 (+5.34)}}     &   \textcolor{red}{\textbf{63.12 (+3.12)}}\\
                    ~ & Co-Cor-A                &   \textcolor{red}{\textbf{93.98 (-1.02)}} &   \textcolor{red}{\textbf{88.87 (-1.13)}}     &   79.24 (-0.76)                                           &   68.92 (-1.08)                                          &   59.26 (-0.74)\\
                    ~ & Co-Cor-B                &   87.52 (-7.48)                                          &   82.23 (-7.77)                                           &   72.28 (-7.72)                                           &   66.26 (-3.74)                                           &   61.30 (+1.30)\\
                    \midrule    
                    \multirow{3}*{PatchCamelyon} & Co-Correcting       &   96.37 (+1.37)                                           &   95.87 (+5.87)                                           &   94.15 (+14.15)                                           &   92.33 (+22.33)                                           &   87.39 (+27.39)\\
                    ~ & Co-Cor-A                &   \textcolor{red}{\textbf{97.79 (+2.79)}}     &   \textcolor{red}{\textbf{96.37 (+6.37)}} &   92.13 (+12.13)                                           &   85.94 (+15.94)                                           &   79.74 (+19.74)\\
                    ~ & Co-Cor-B                &   96.47 (+1.47)                                          &   95.90 (+5.90)                                           &   \textcolor{red}{\textbf{94.48 (+14.48)}} &   \textcolor{red}{\textbf{92.60 (+22.60)}} &   \textcolor{red}{\textbf{87.46 (+27.46)}} \\
                    \bottomrule
                \end{tabular}
            \end{table*}
            \begin{figure}[!h]
                \centering
                \subfigure[$clean$]{
                \includegraphics[width=2.9cm]{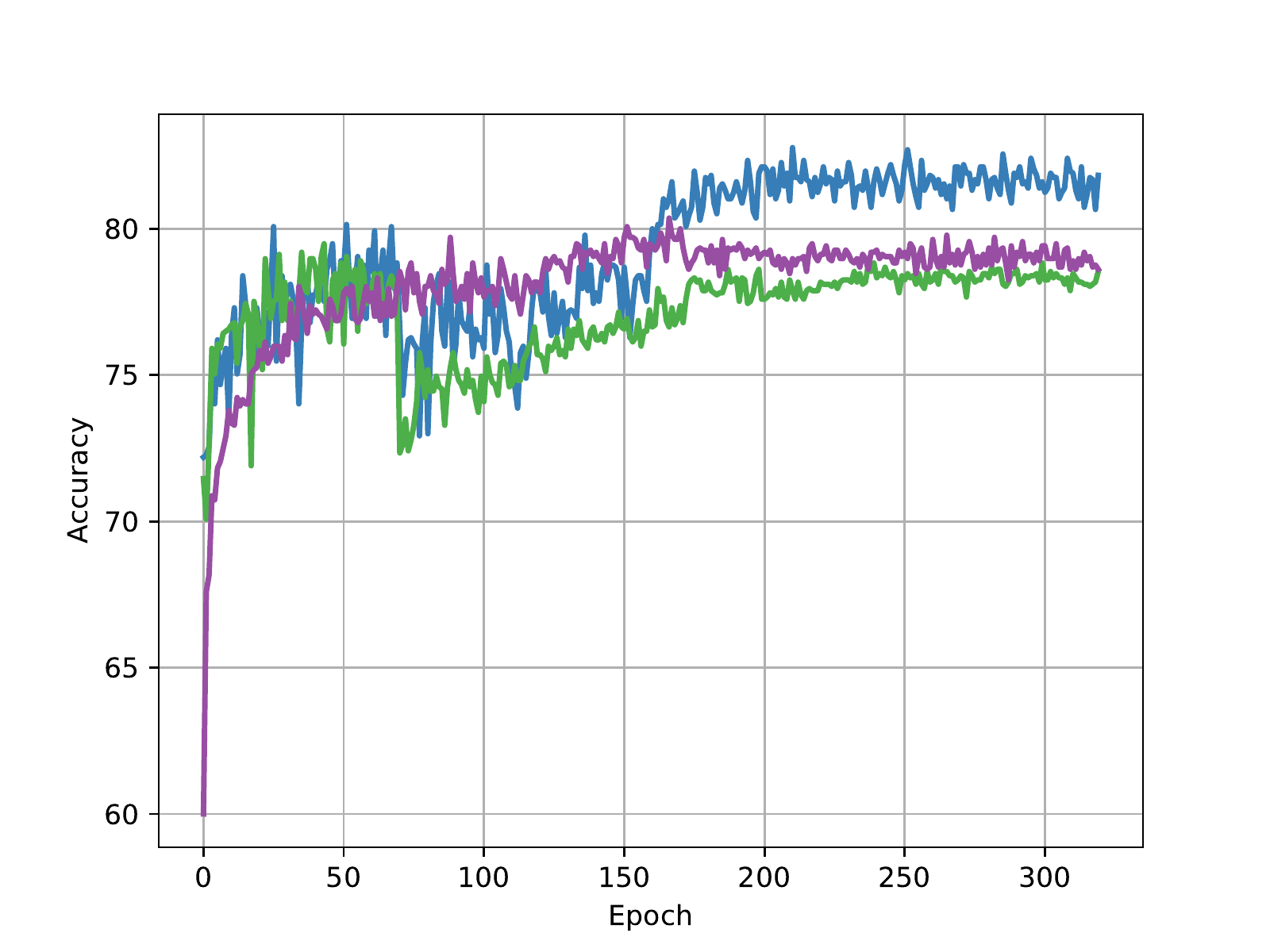} \label{fig:isic_ablation_acc_1}
                }  
                \hspace{-6mm}
                \subfigure[$noise=0.05$]{
                \includegraphics[width=2.9cm]{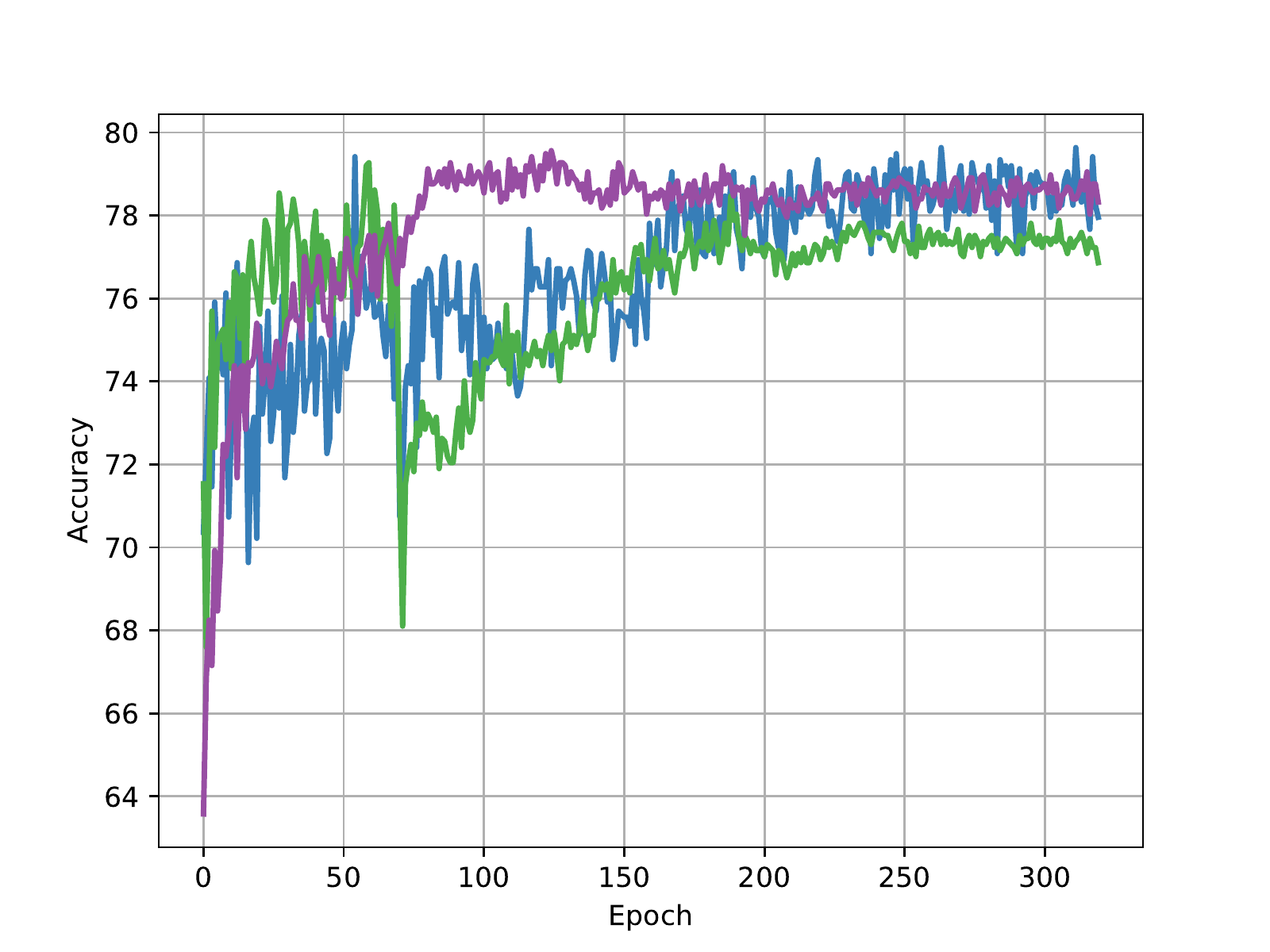} \label{fig:isic_ablation_acc_2} 
                }
                \hspace{-6mm}
                \subfigure[$noise=0.1$]{
                \includegraphics[width=2.9cm]{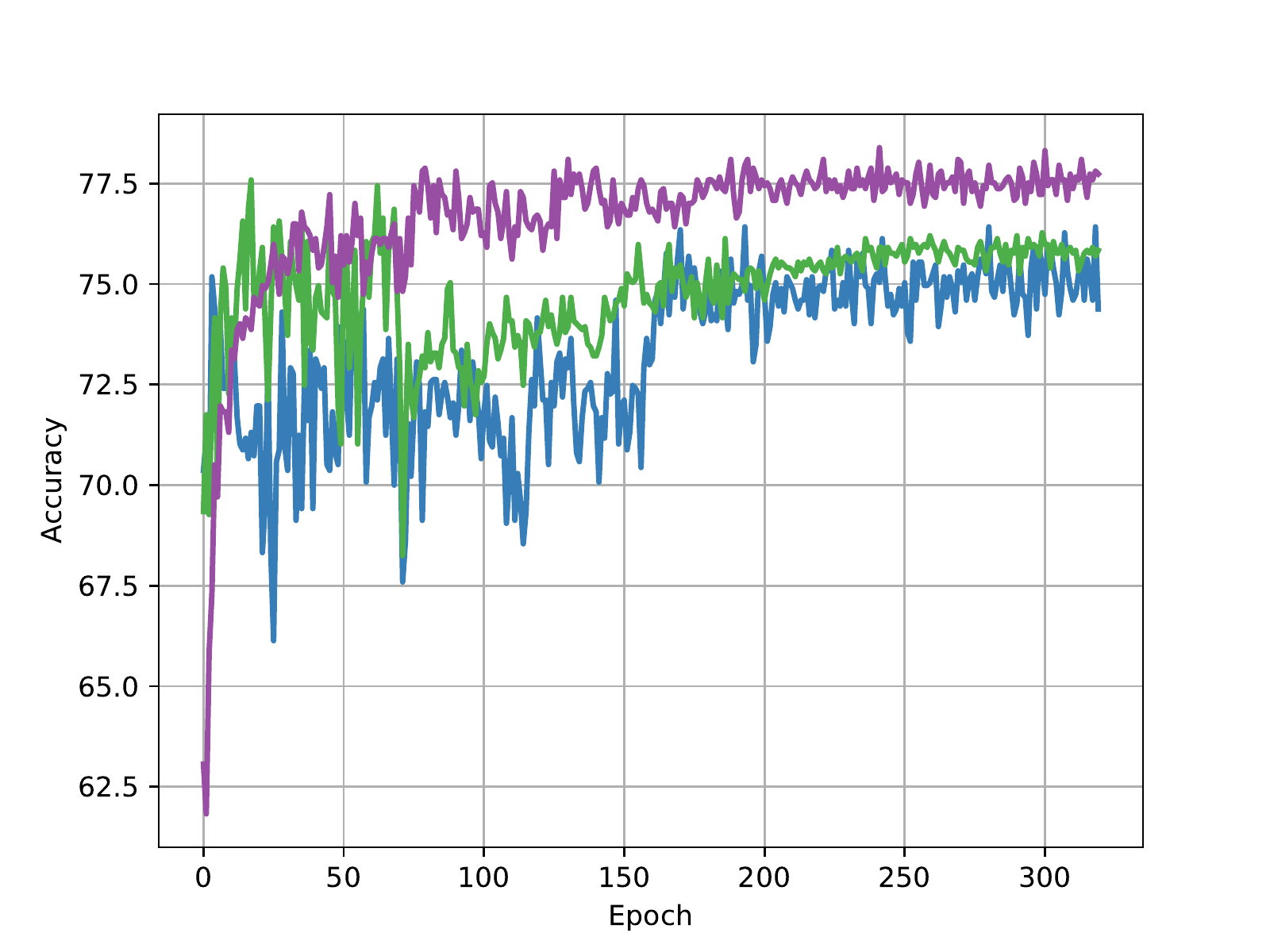}\label{fig:isic_ablation_acc_3}
                }
                \hspace{-6mm}
                \subfigure[$noise=0.2$]{
                \includegraphics[width=2.9cm]{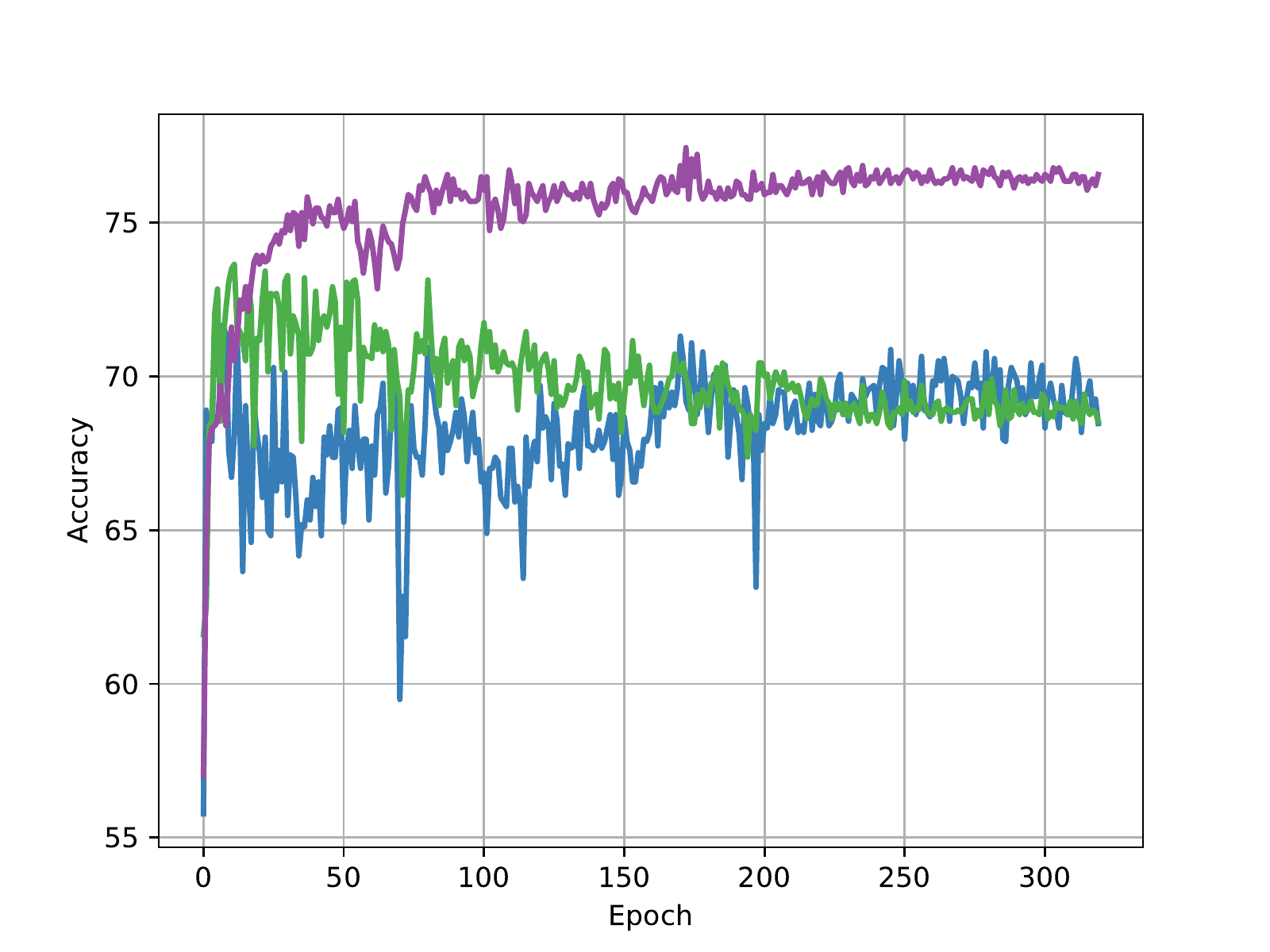}\label{fig:isic_ablation_acc_4}
                }
                \hspace{-6mm}
                \subfigure[$noise=0.3$]{
                \includegraphics[width=2.9cm]{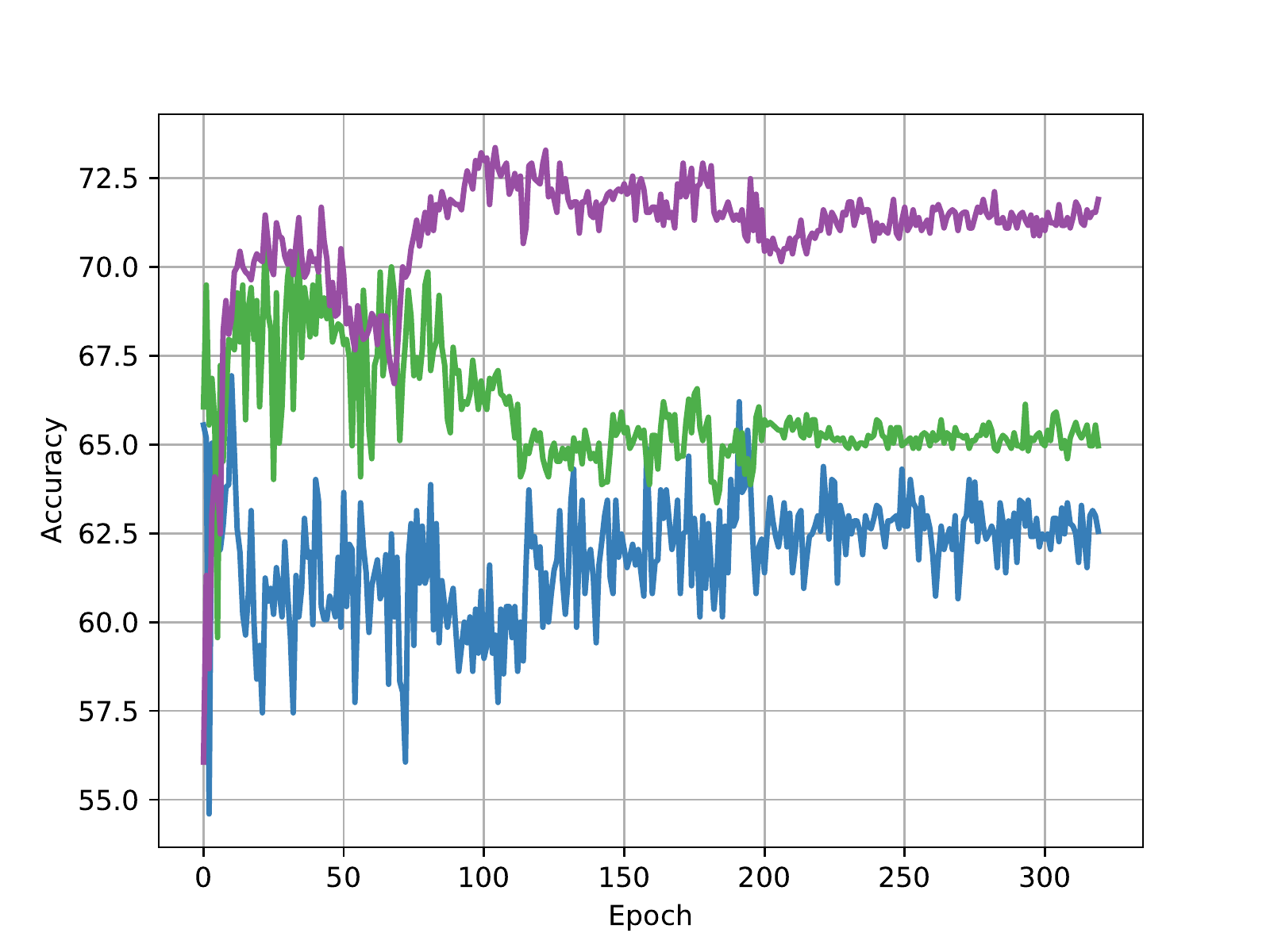}\label{fig:isic_ablation_acc_5}
                }
                \hspace{-6mm}
                \subfigure[$noise=0.4$]{
                \includegraphics[width=2.9cm]{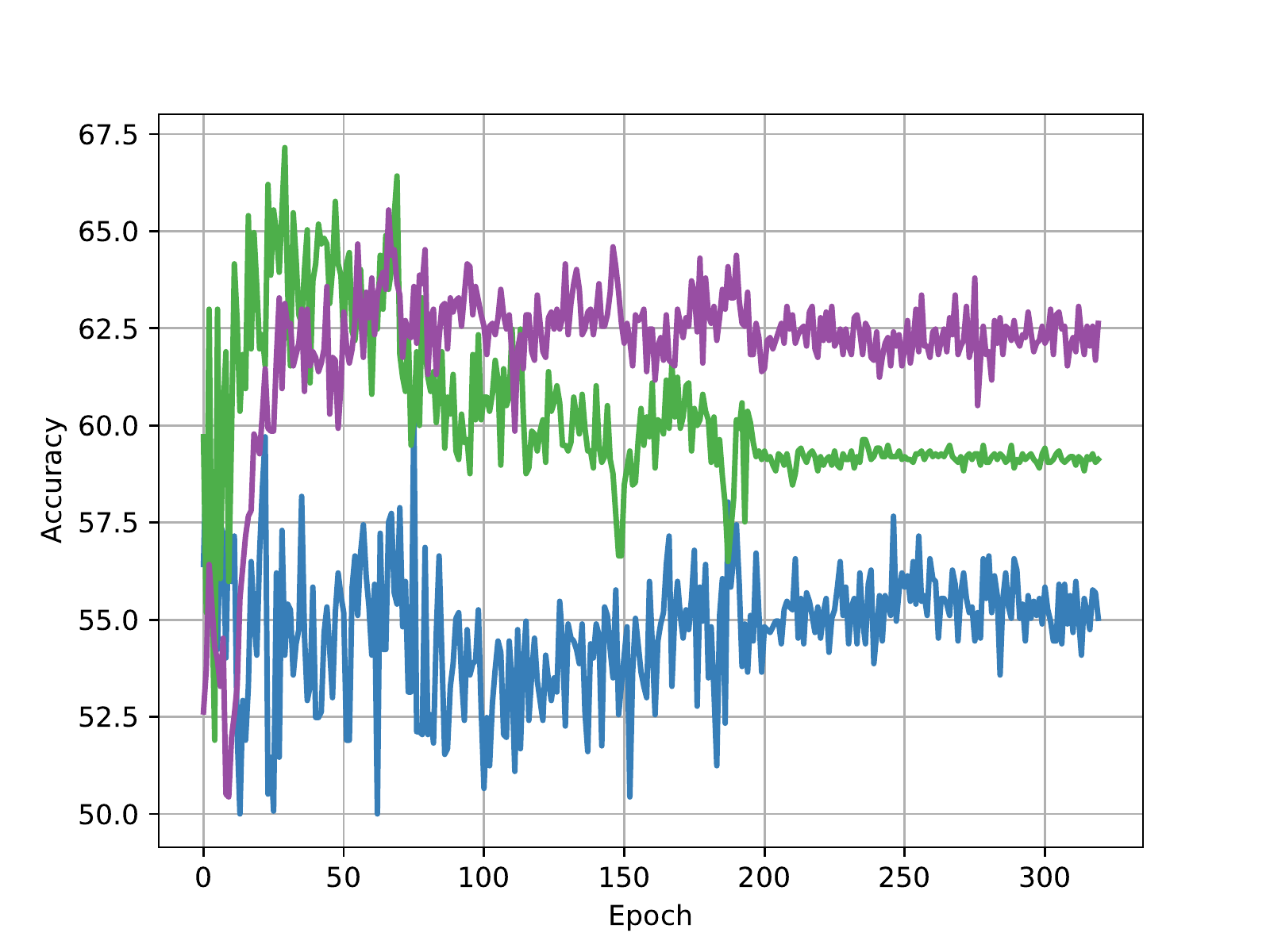}\label{fig:isic_ablation_acc_6}
                }               
                \hspace{-6mm}
                \subfigure{
                \includegraphics[width=6cm]{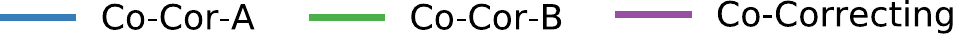}\label{fig:isic_ablation_acc_legend}
                }
                \caption{$ACC_{class}$ on ISIC-Archive.}
                \label{fig:isic_ablation_acc}
            \end{figure}
            
            \begin{figure}[!htp]
                \centering
                \subfigure[clean]{
                \includegraphics[width=2.9cm]{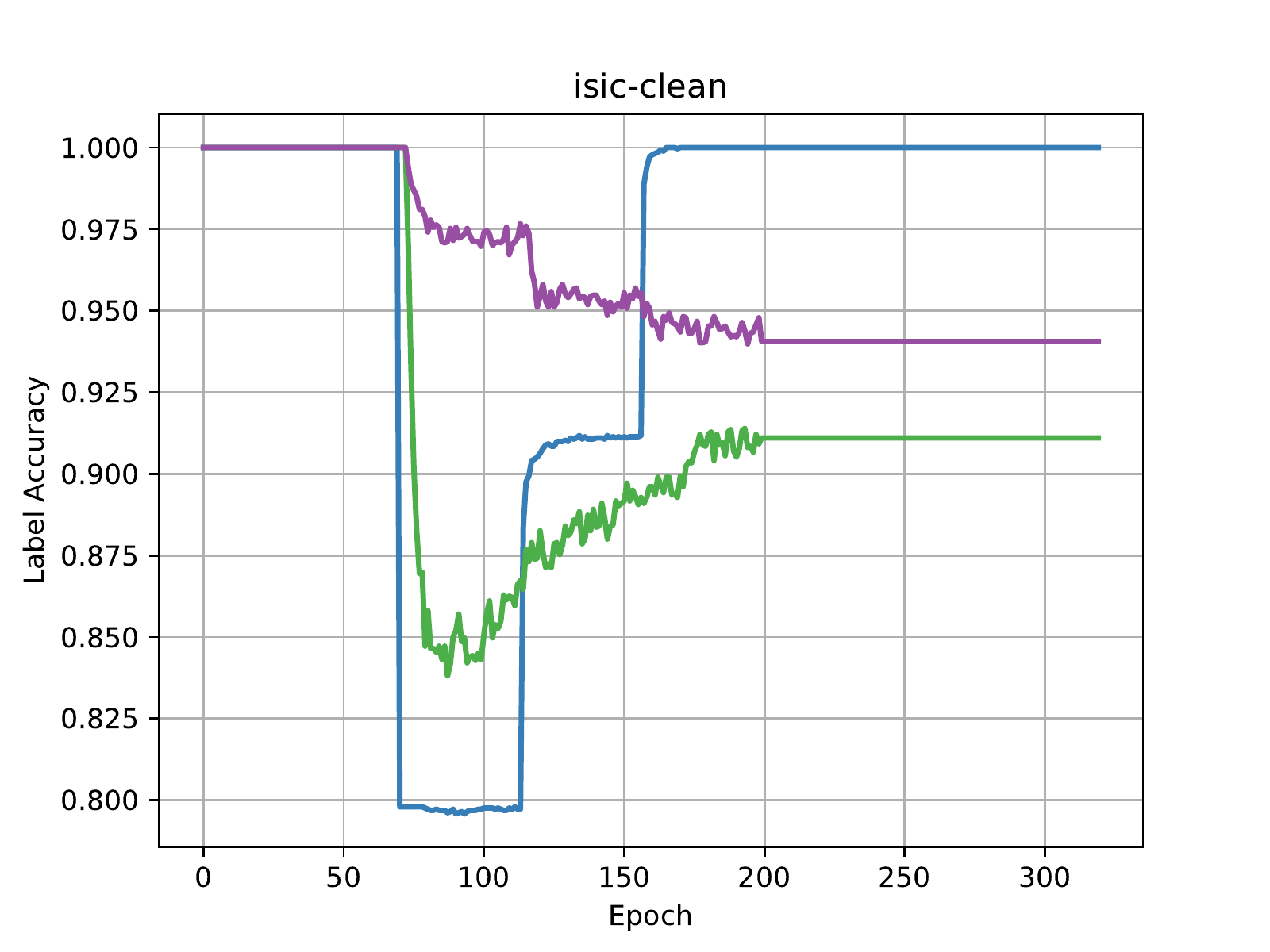} \label{fig:isic_ablation_label_1}
                } 
                \hspace{-6mm}
                \subfigure[0.05]{
                \includegraphics[width=2.9cm]{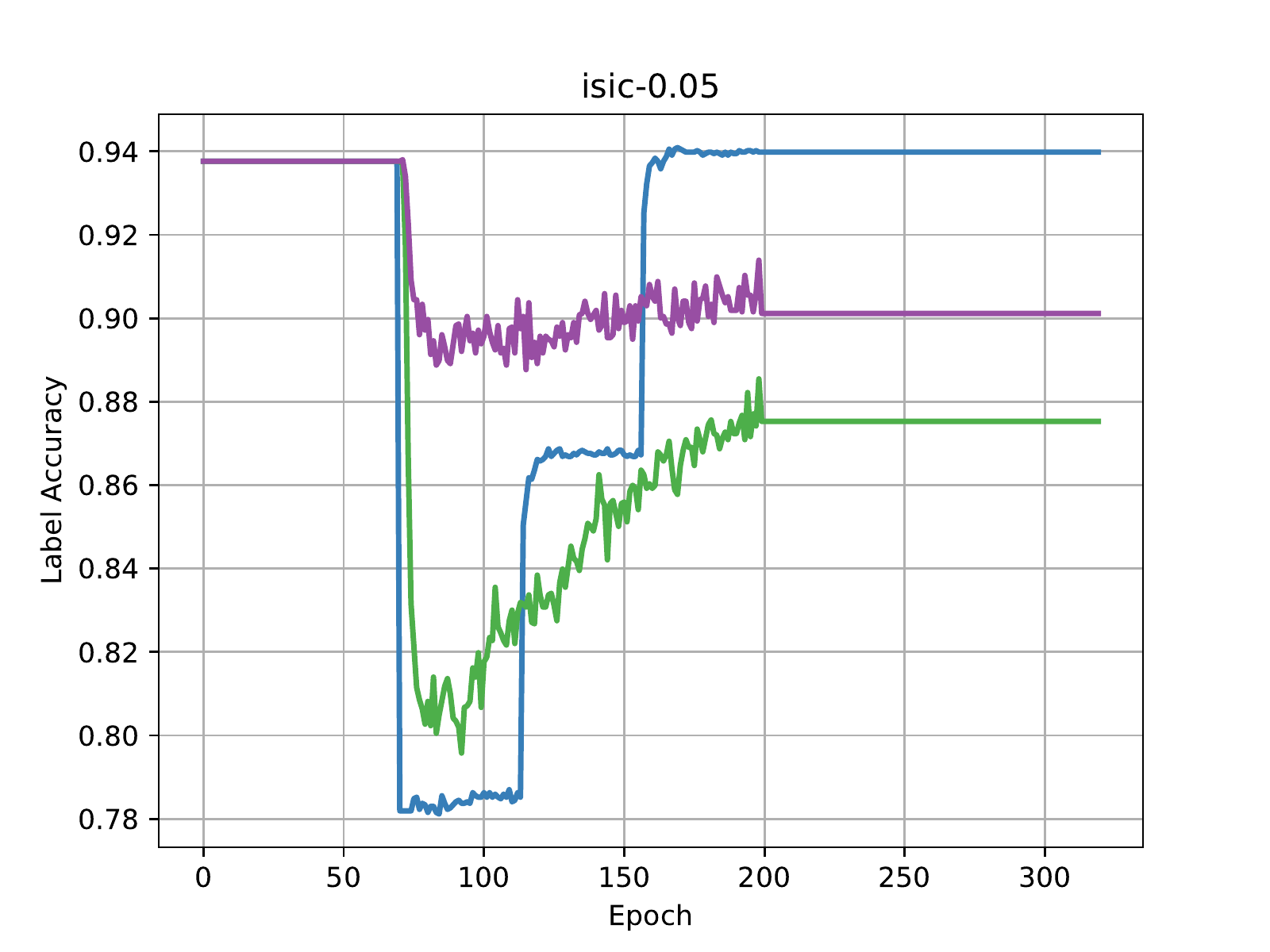} \label{fig:isic_ablation_label_2} 
                }
                \hspace{-6mm}
                \subfigure[0.1]{
                \includegraphics[width=2.9cm]{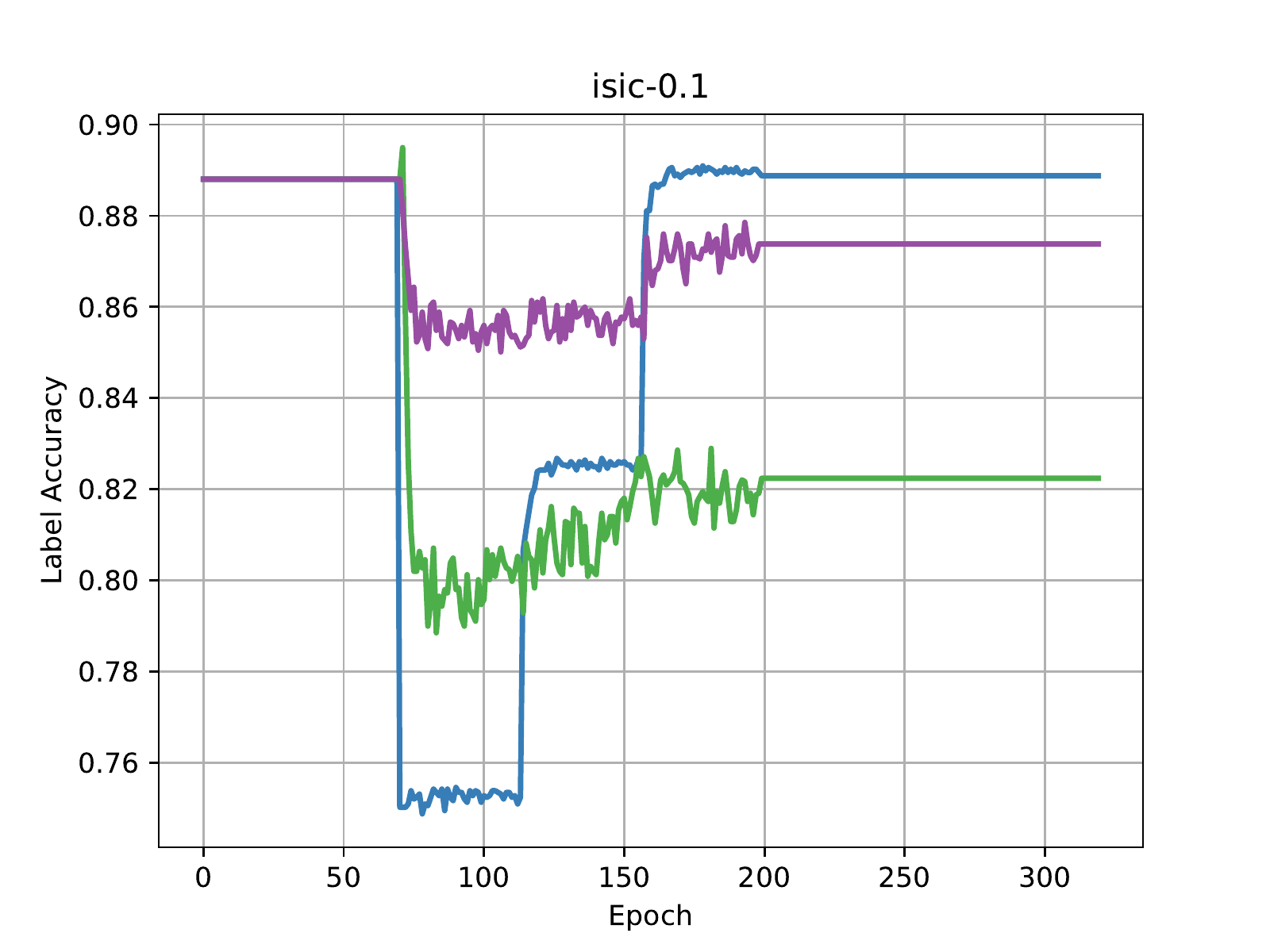}\label{fig:isic_ablation_label_3}
                }
                \hspace{-6mm}
                \subfigure[0.2]{
                \includegraphics[width=2.9cm]{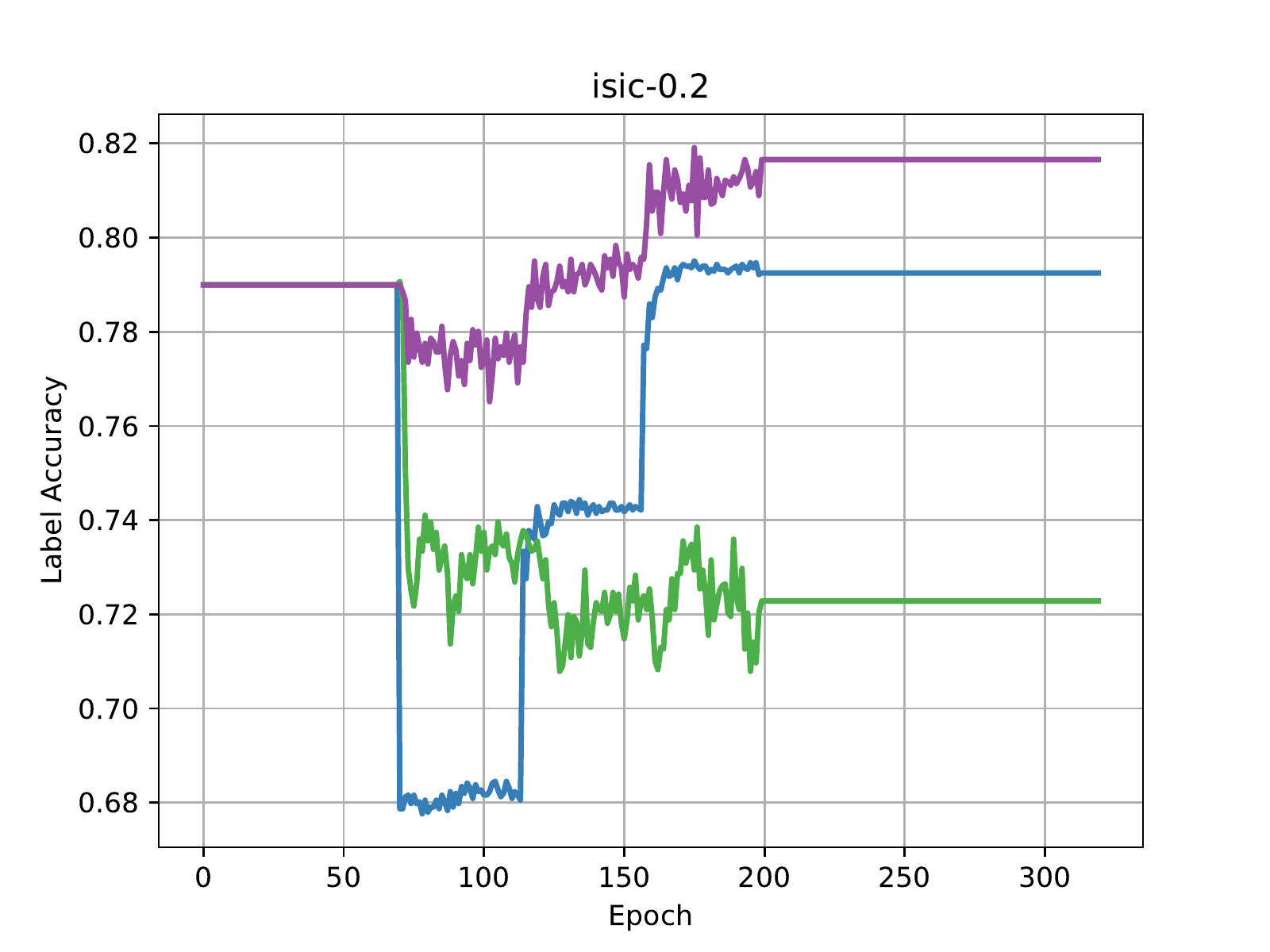}\label{fig:isic_ablation_label_4}
                }
                \hspace{-6mm}
                \subfigure[0.3]{
                \includegraphics[width=2.9cm]{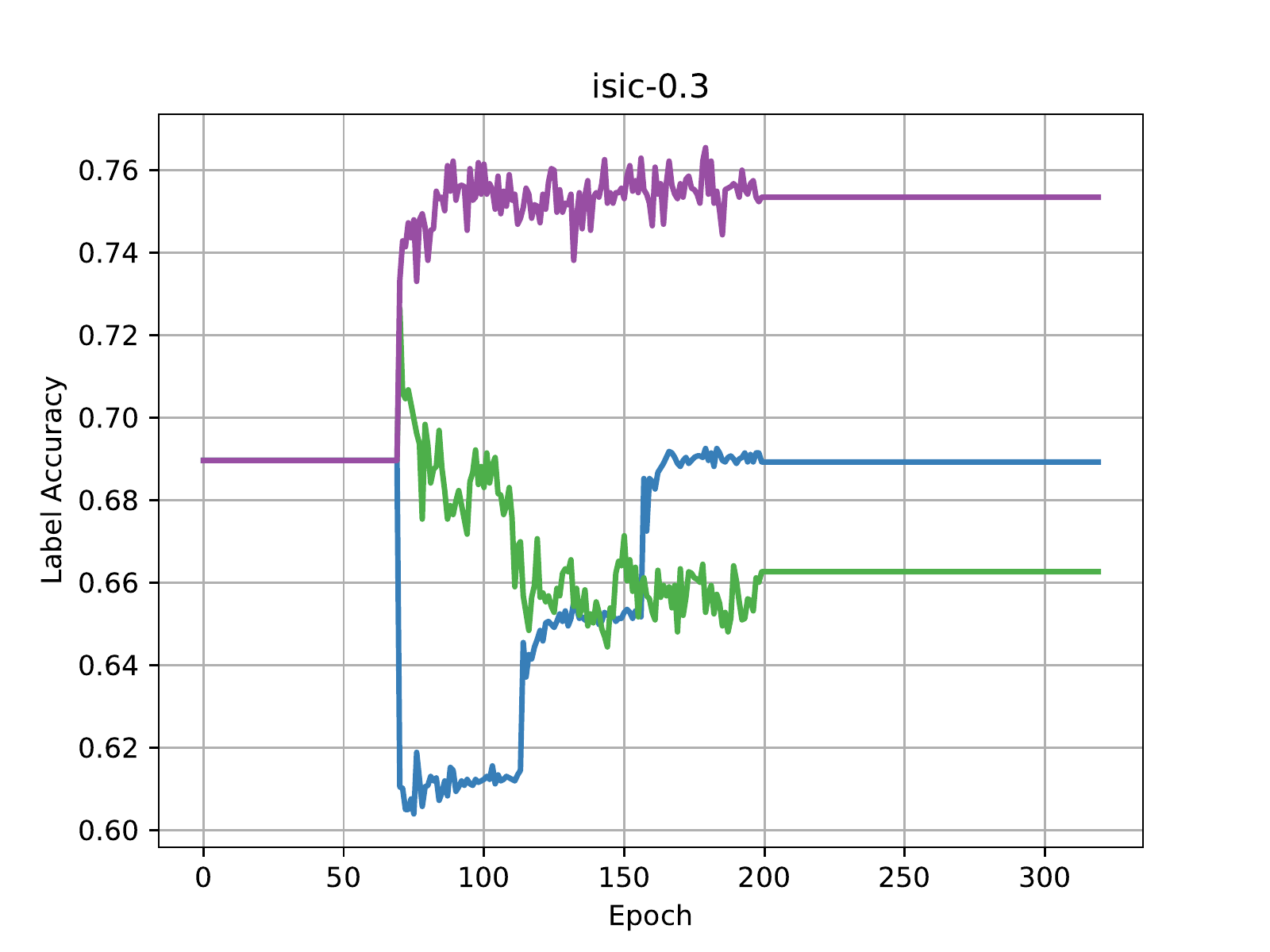}\label{fig:isic_ablation_label_5}
                }
                \hspace{-6mm}
                \subfigure[0.4]{
                \includegraphics[width=2.9cm]{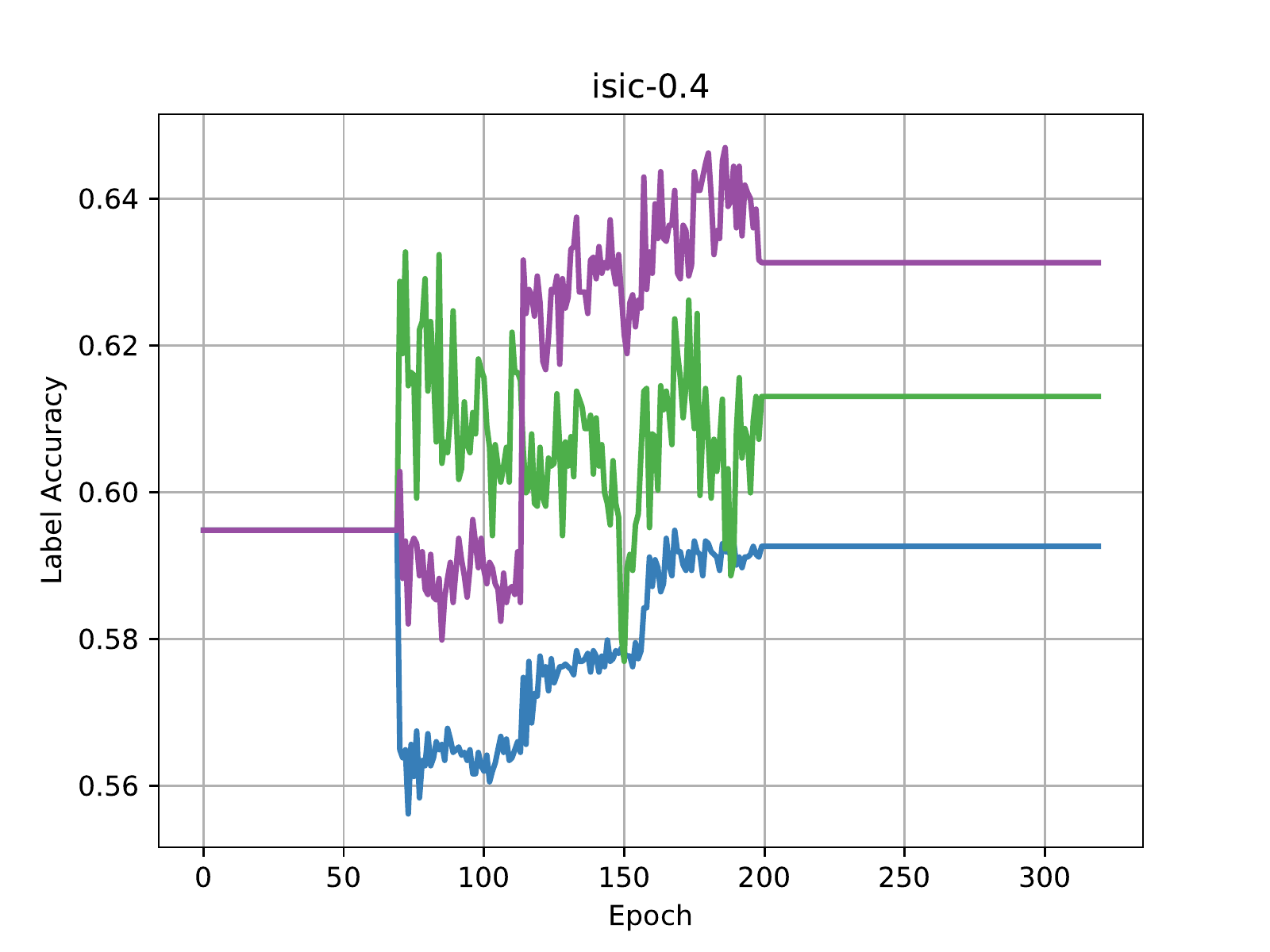}\label{fig:isic_ablation_label_6}
                }
                \hspace{-6mm}
                \subfigure{
                \includegraphics[width=6cm]{{legend_ablation}.pdf}\label{fig:isic_ablation_label_legend}
                }
                \caption{\ADDM{$ACC_{label}$ on ISIC-Archive.}}
                \label{fig:isic_ablation_label_accu}
            \end{figure}

            \begin{figure}[!htp]
                \centering
                \subfigure[$clean$]{
                \includegraphics[width=2.9cm]{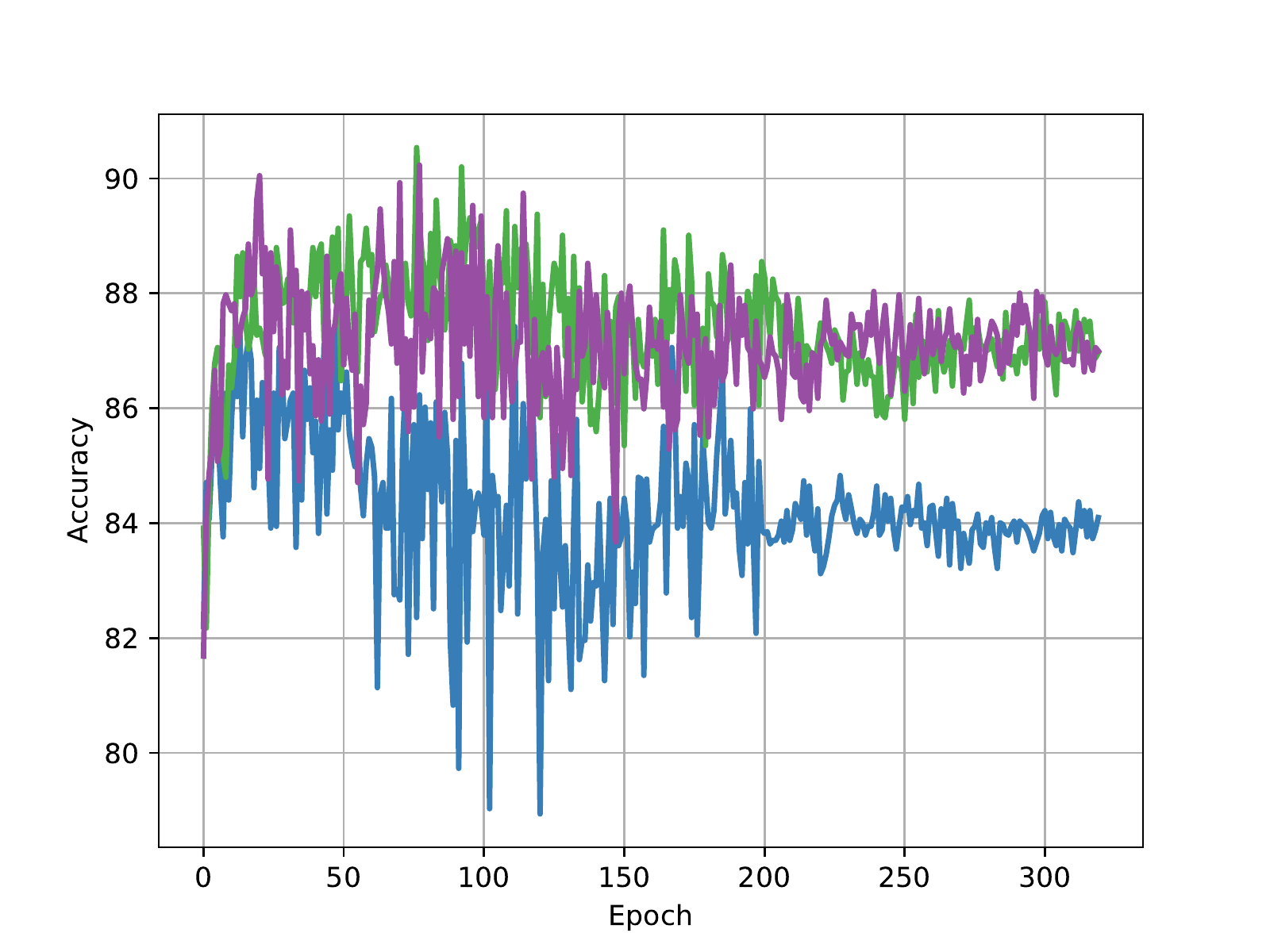} \label{fig:pcam_ablation_acc_1}
                }  
                \hspace{-6mm}
                \subfigure[$noise=0.05$]{
                \includegraphics[width=2.9cm]{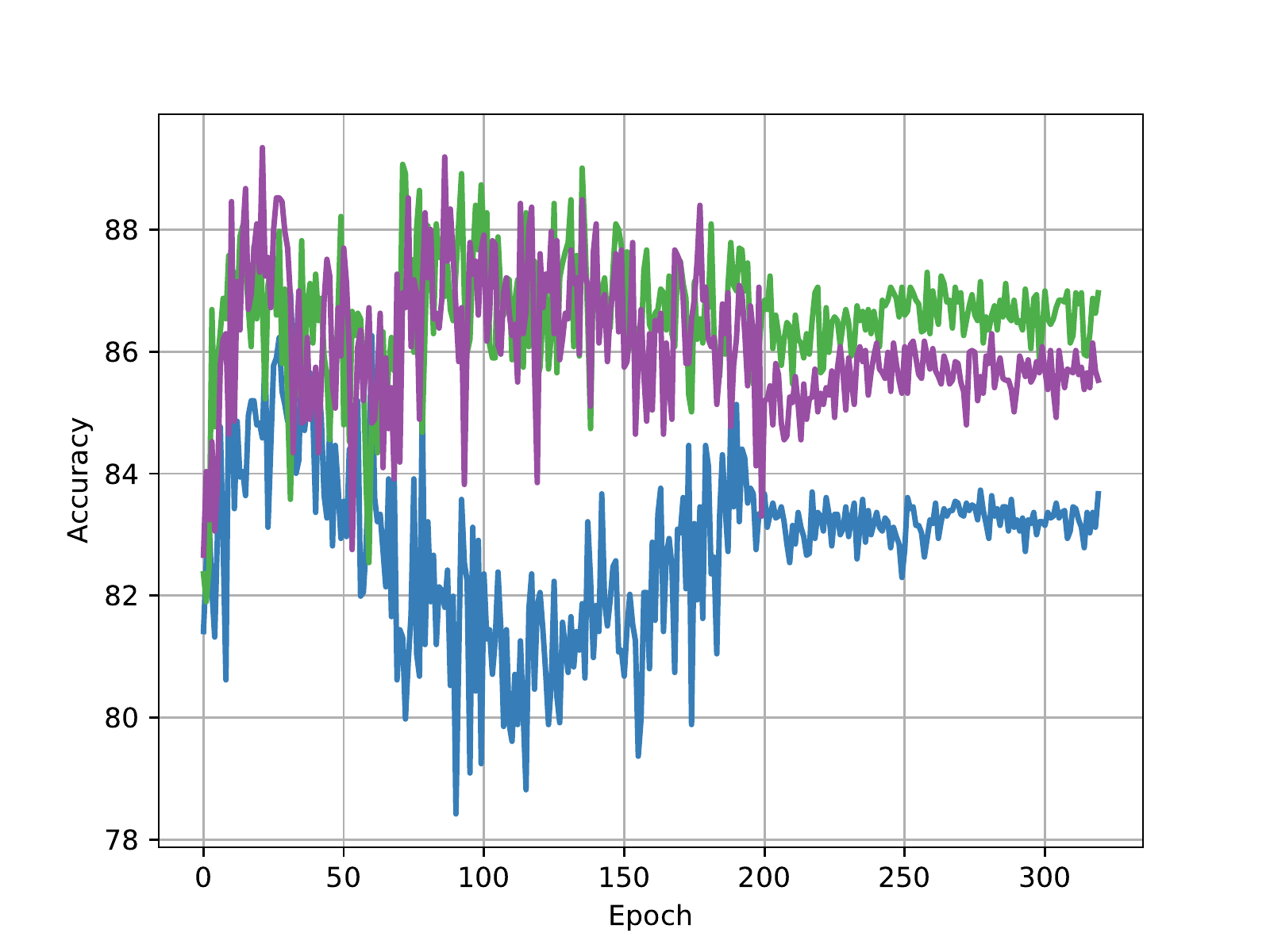} \label{fig:pcam_ablation_acc_2} 
                }
                \hspace{-6mm}
                \subfigure[$noise=0.1$]{
                \includegraphics[width=2.9cm]{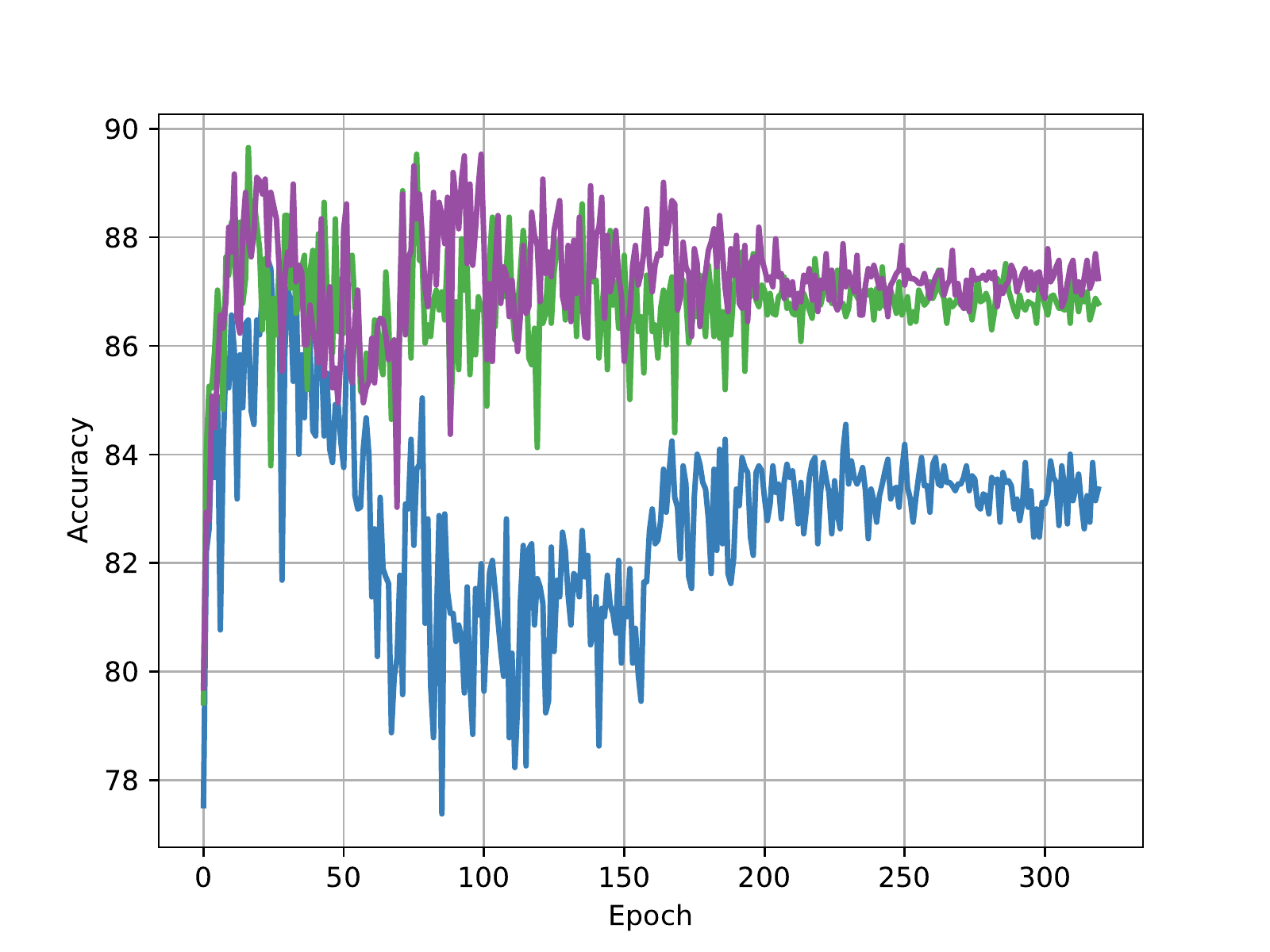}\label{fig:pcam_ablation_acc_3}
                }
                \hspace{-6mm}
                \vspace{-3mm}
                \subfigure[$noise=0.2$]{
                \includegraphics[width=2.9cm]{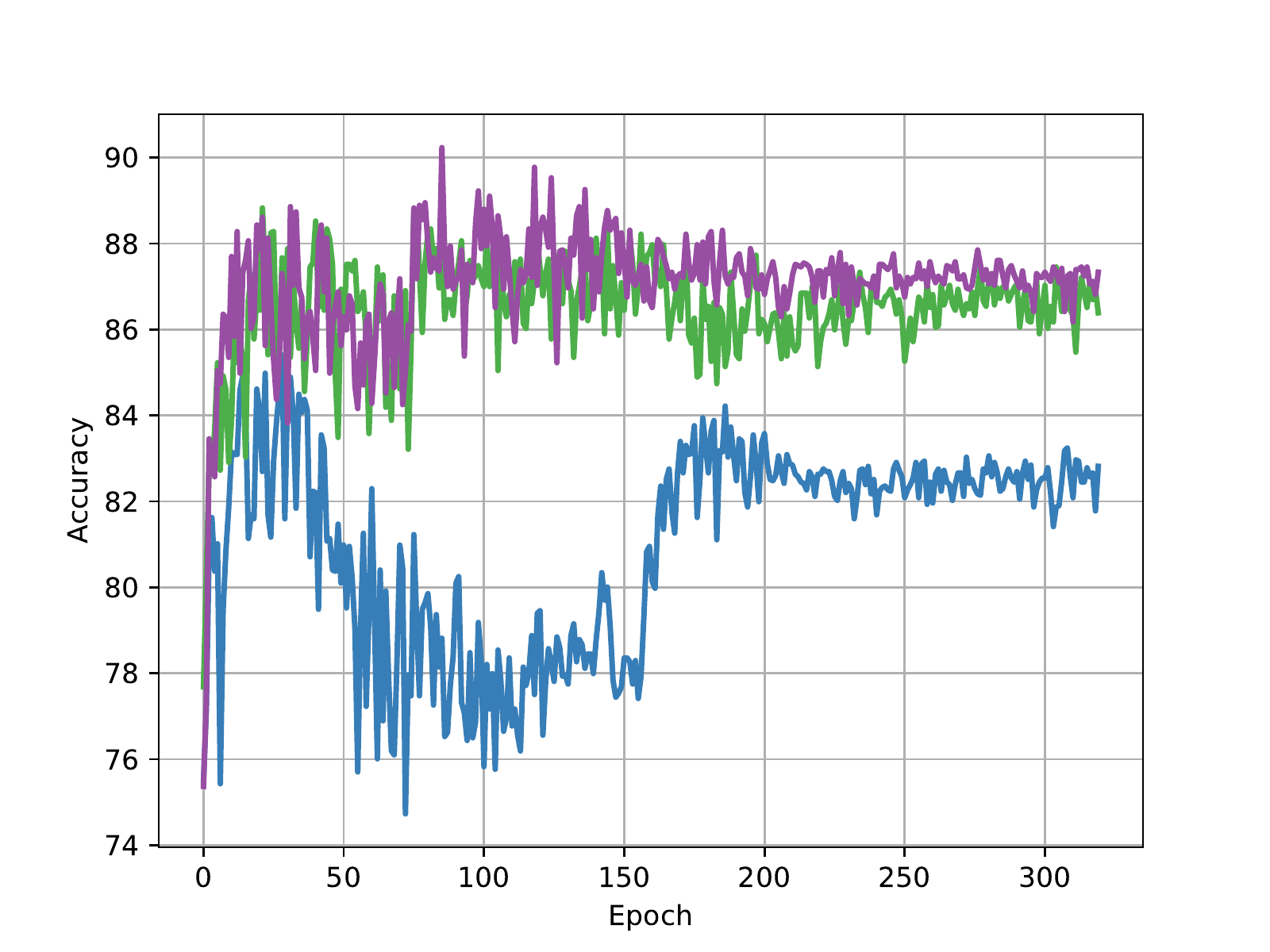}\label{fig:pcam_ablation_acc_4}
                }
                \hspace{-6mm}
                \subfigure[$noise=0.3$]{
                \includegraphics[width=2.9cm]{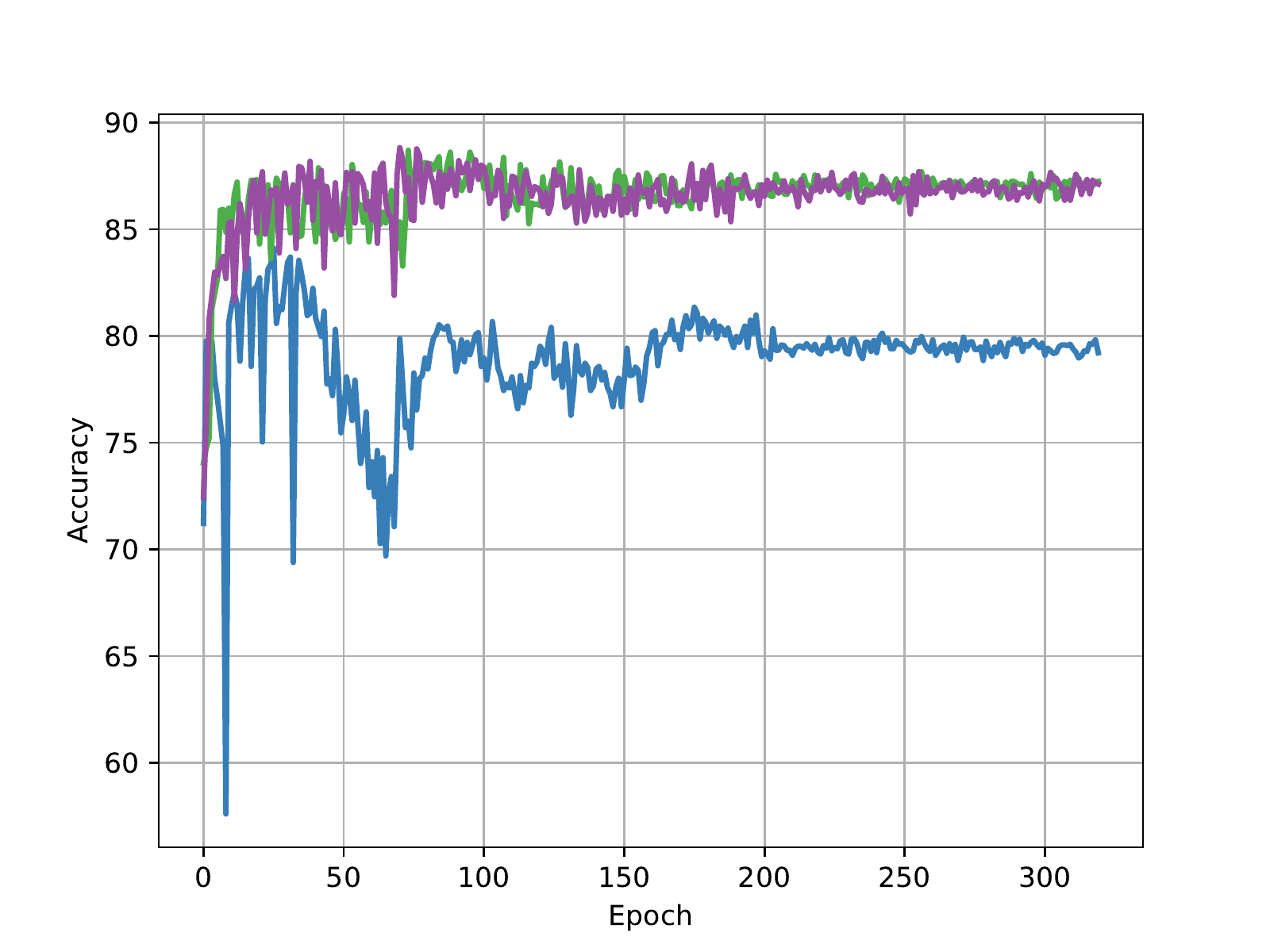}\label{fig:pcam_ablation_acc_5}
                }
                \hspace{-6mm}
                \subfigure[$noise=0.4$]{
                \includegraphics[width=2.9cm]{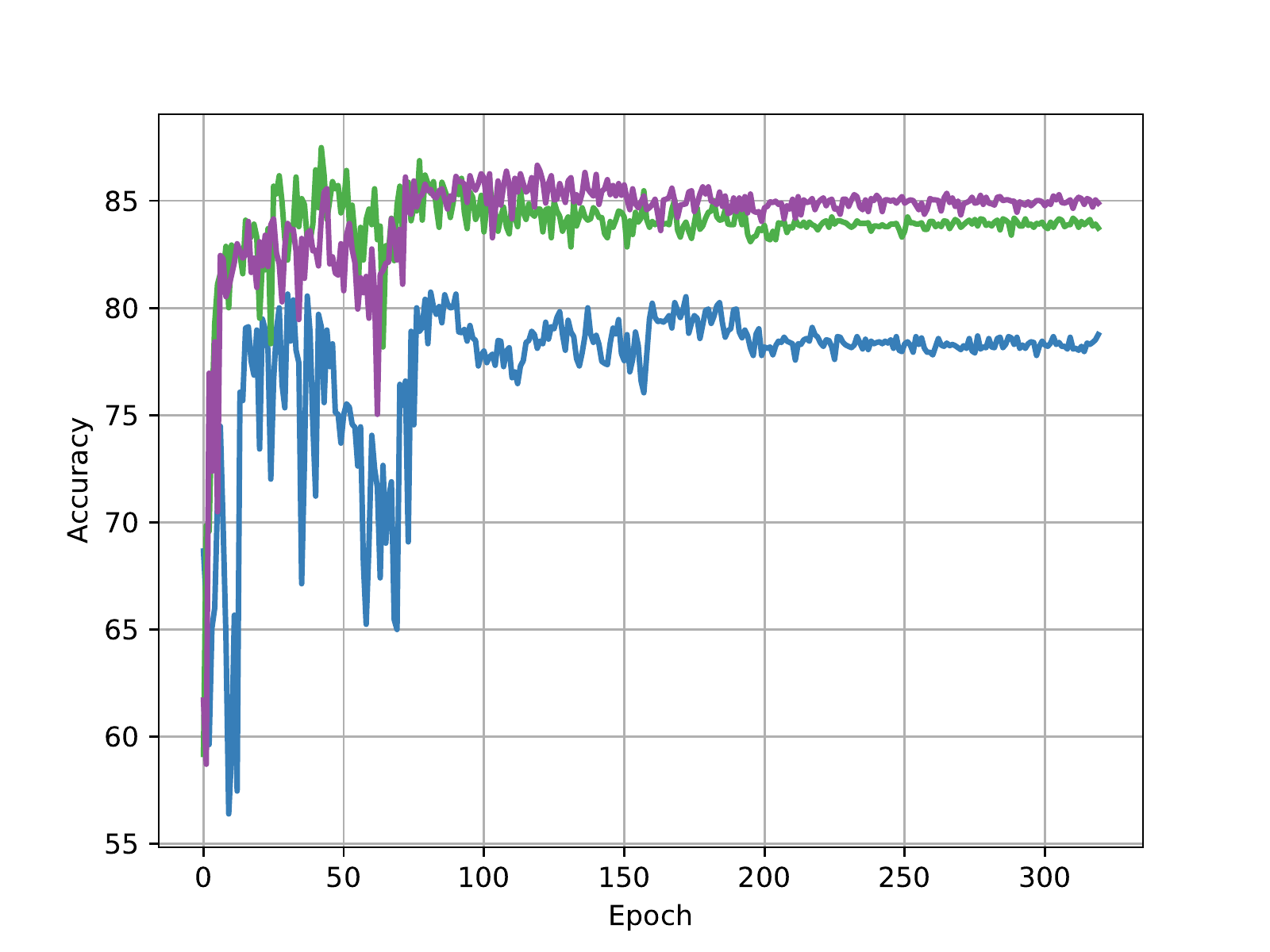}\label{fig:pcam_ablation_acc_6}
                }               
                \hspace{-6mm}
                \subfigure{
                \includegraphics[width=6cm]{{legend_ablation}.pdf}\label{fig:pcam_ablation_acc_legend}
                }
                \caption{$ACC_{class}$ on PatchCamelyon.}
                \label{fig:pcam_ablation_acc}
            \end{figure}
            \begin{figure}[!h]
                \centering
                \subfigure[clean]{
                \includegraphics[width=2.9cm]{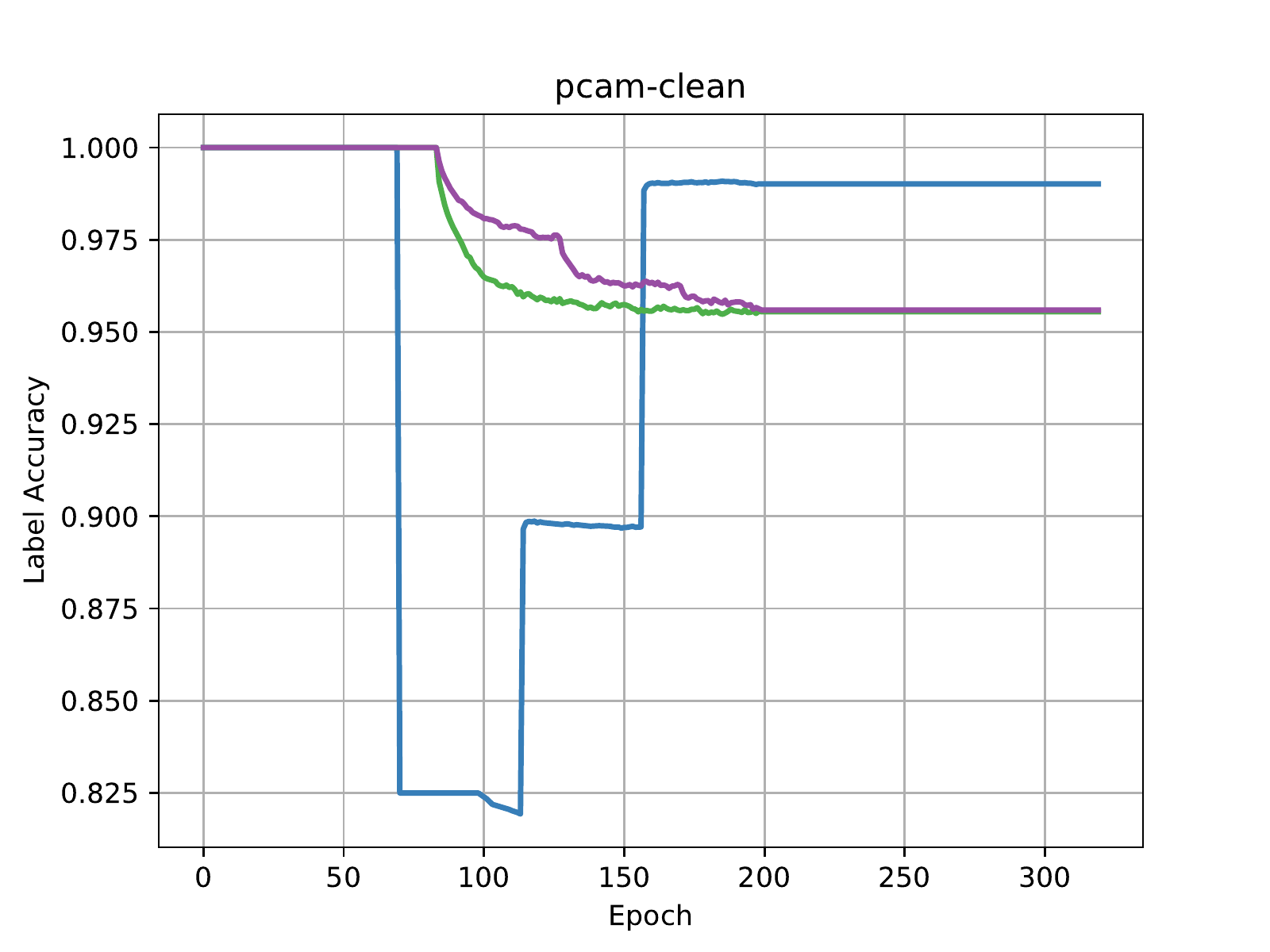} \label{fig:pcam_ablation_label_1}
                }  
                \hspace{-6mm}
                \subfigure[0.05]{
                \includegraphics[width=2.9cm]{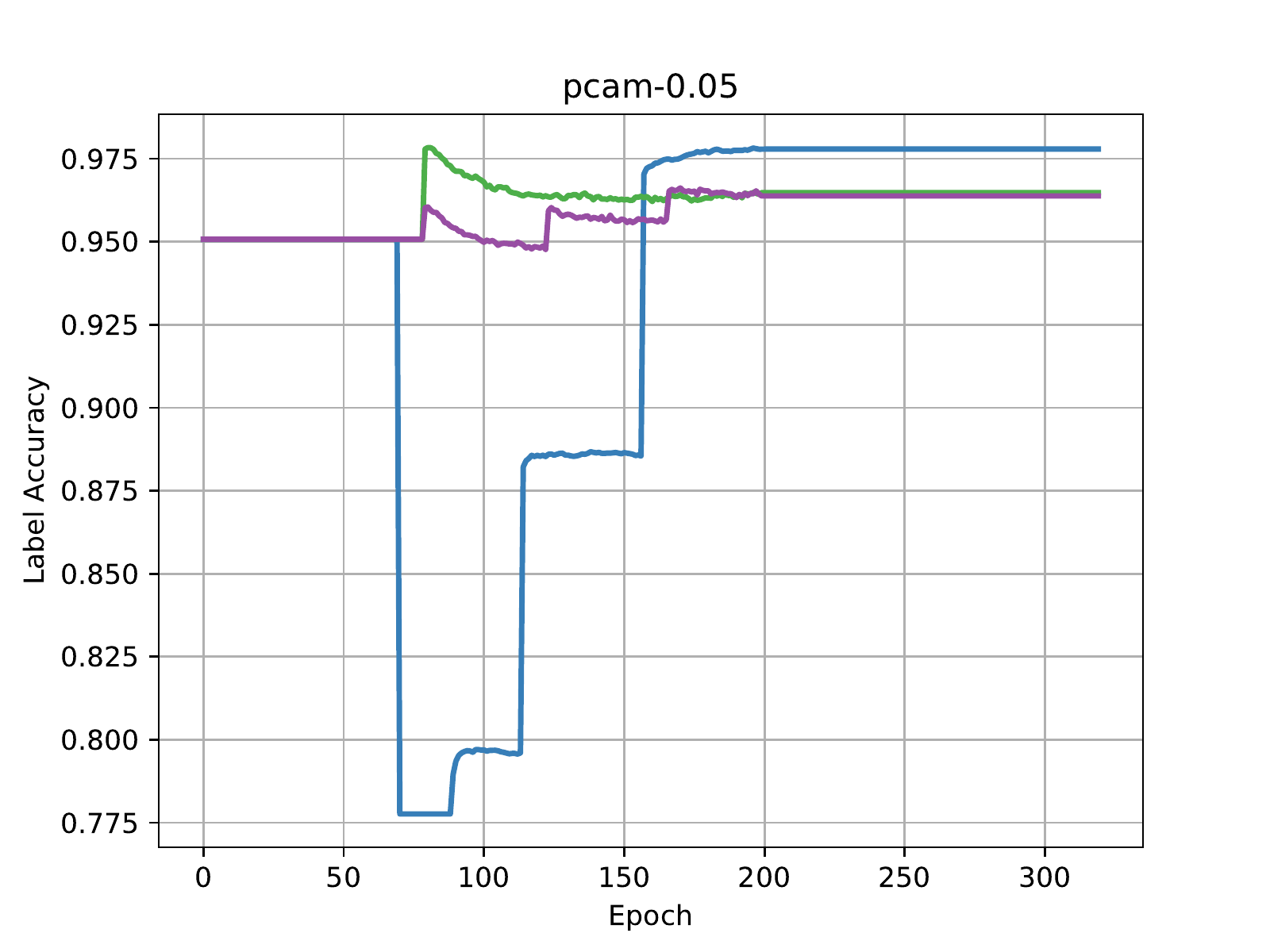} \label{fig:pcam_ablation_label_2} 
                }
                \hspace{-6mm}
                \subfigure[0.1]{
                \includegraphics[width=2.9cm]{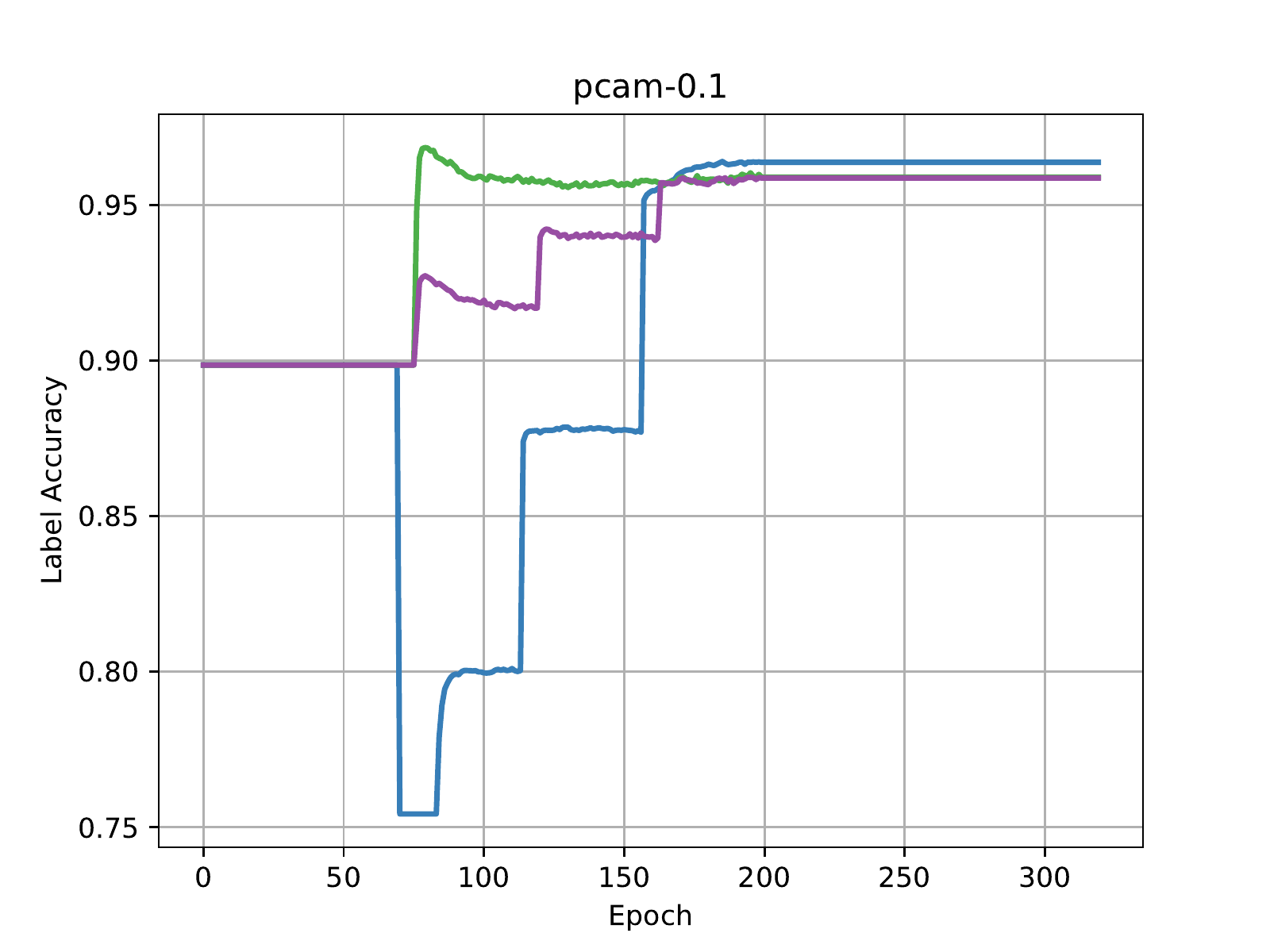}\label{fig:pcam_ablation_label_3}
                }
                \hspace{-6mm}
                \vspace{-3mm}
                \subfigure[0.2]{
                \includegraphics[width=2.9cm]{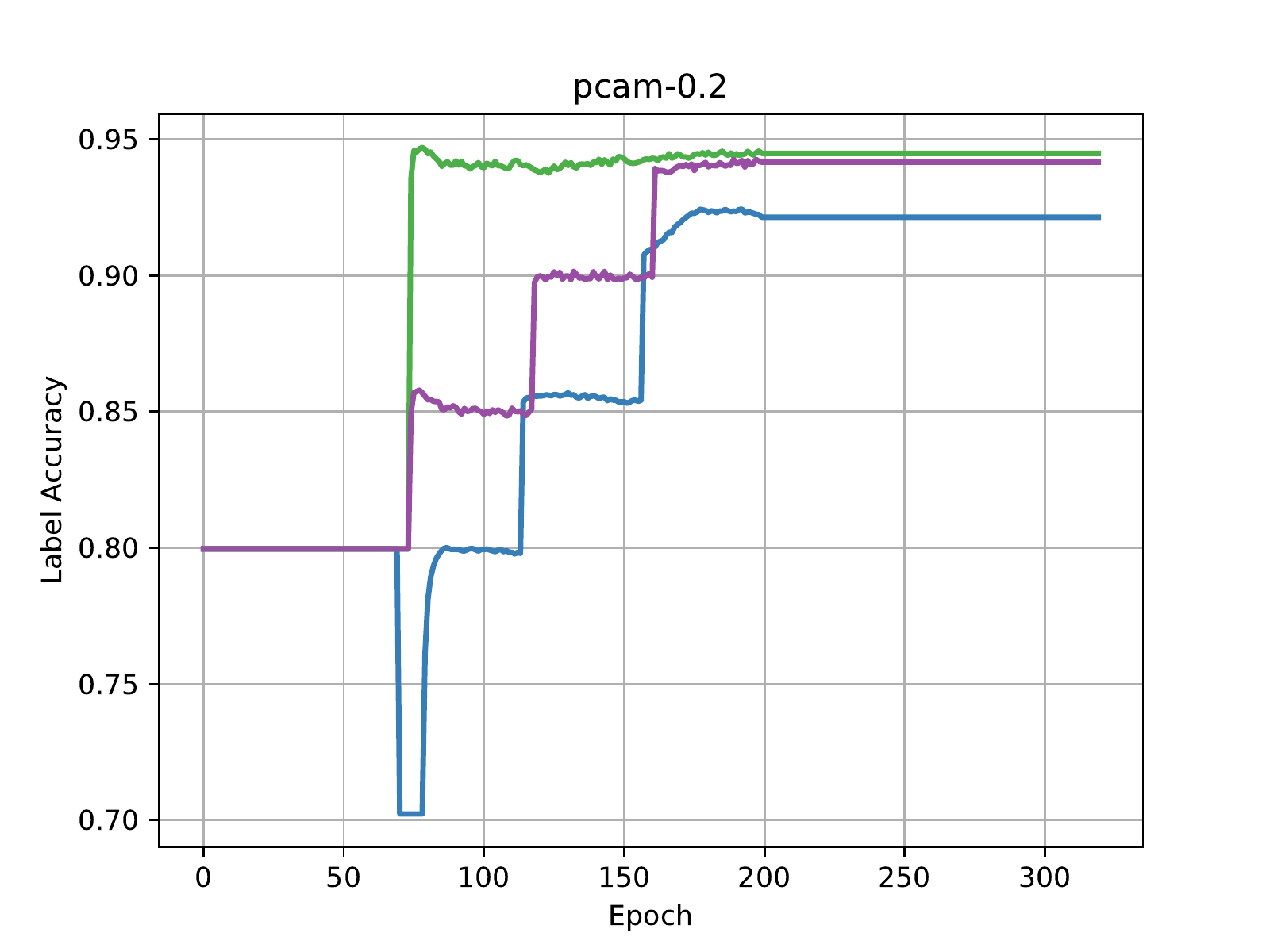}\label{fig:pcam_ablation_label_4}
                }
                \hspace{-6mm}
                \subfigure[0.3]{
                \includegraphics[width=2.9cm]{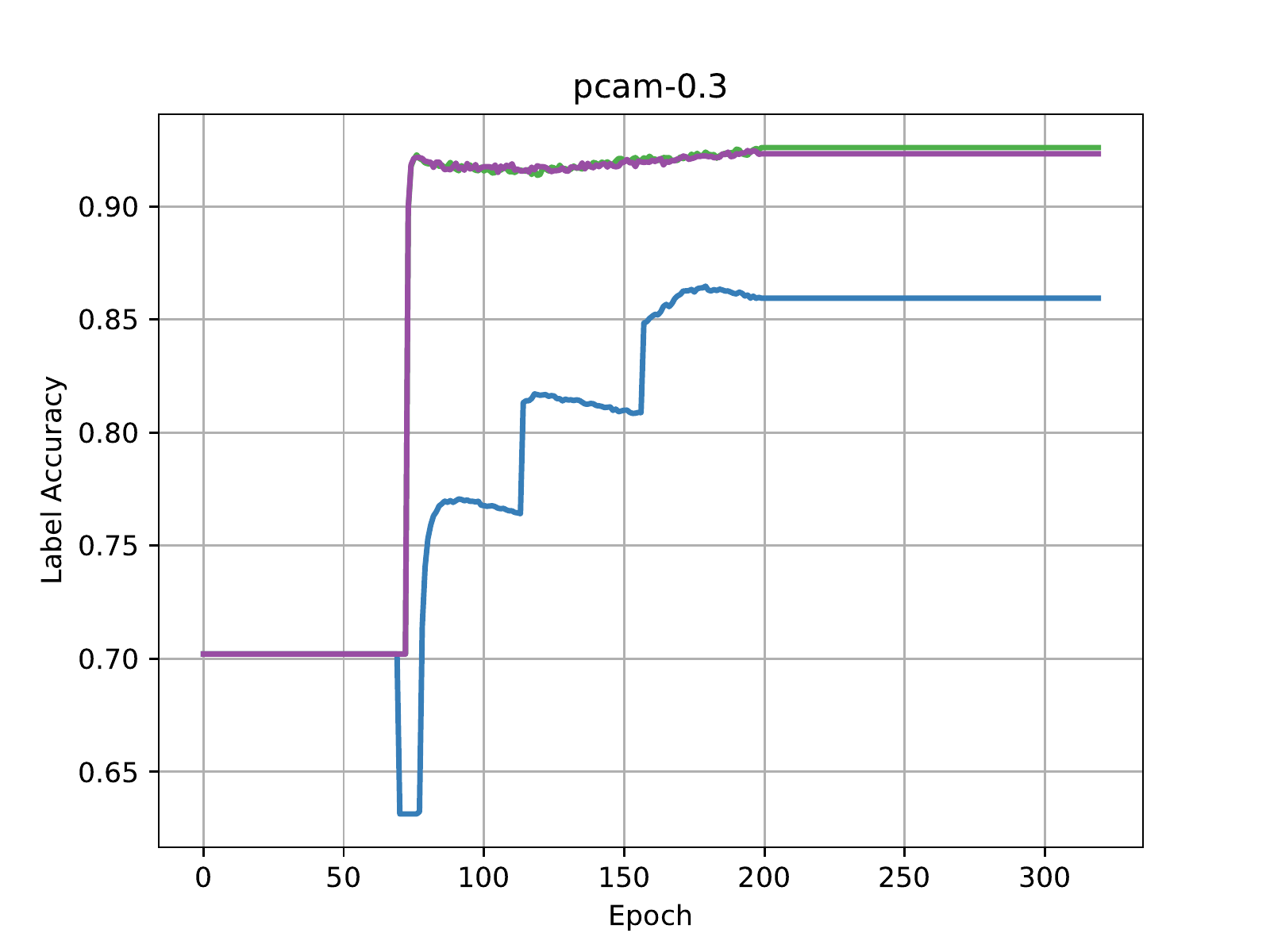}\label{fig:pcam_ablation_label_5}
                }
                \hspace{-6mm}
                \subfigure[0.4]{
                \includegraphics[width=2.9cm]{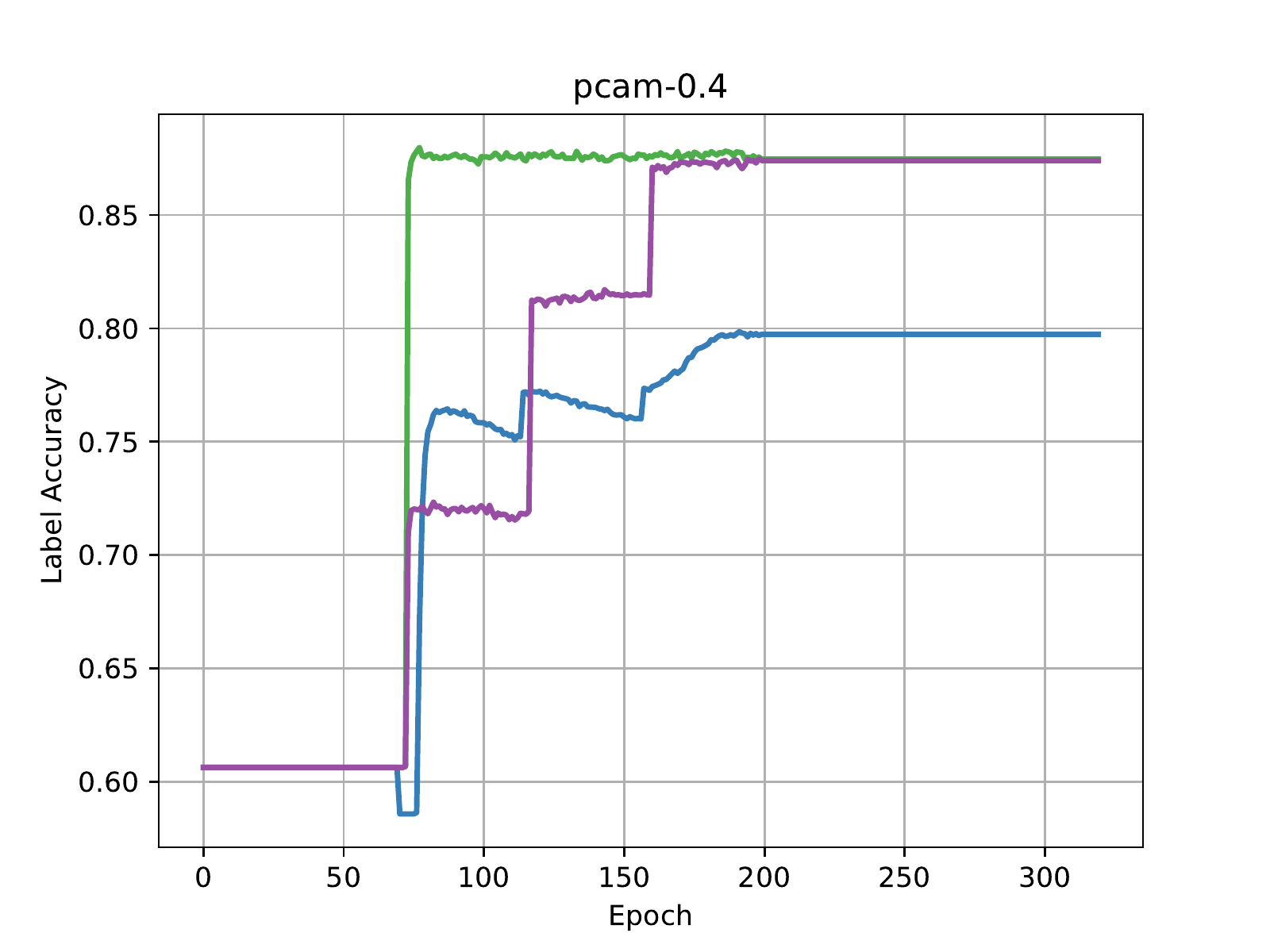}\label{fig:pcam_ablation_label_6}
                }
                \hspace{-6mm}
                \subfigure{
                \includegraphics[width=6cm]{{legend_ablation}.pdf}\label{fig:pcam_ablation_label_legend}
                }
                \caption{\ADDM{$ACC_{label}$ on PatchCamelyon. }}
                \label{fig:pcam_ablation_label_accu}
            \end{figure}
            
        \subsubsection{Analysis of the Dual network architecture}
            Dual-network architecture is effective. Basically, the classification accuracy of Co-Correcting is better than that of Co-Cor-A which is based on the single network architecture. It can be seen from the curve in Figure \ref{fig:isic_ablation_label_accu} and Figure \ref{fig:pcam_ablation_label_accu} that when the noise ratio is small, Co-Cor-A is better than Co-Correcting in labeling accuracy. But when the noise ratio is greater than 0.1, Co-Correcting has obvious advantages. In Figure \ref{fig:isic_ablation_acc} and Figure \ref{fig:pcam_ablation_acc}, the purple line is much higher than the blue line at the end. 
            
        \subsubsection{Analysis of the label correction curriculum}     
            In the ISIC-Archive experiments, the training data is small. Compared with not using the curriculum, using a curriculum in label correction has obvious improvements \ADDM{on both classification accuracy and labeling accuracy}.  In Figure \ref{fig:isic_ablation_acc} \ADDM{and Figure \ref{fig:isic_ablation_label_accu}},  the purple line is significantly higher than the green one in all the graphs. \ADDM{The curriculum can protect labels under the small noise ratio and facilitate label correction under the large noise ratio.}
            
            In the PatchCamelyon experiments where the training dataset is larger, the usage of the curriculum slightly improves the performance. We can see in Figure \ref{fig:pcam_ablation_acc} that the purple line representing Co-Correcting is very close to the green line representing Co-Cor-B. From these phenomenons it can be found that the curriculum mechanism is more helpful when the training samples are insufficient and the noise ratio is high. In such cases, Co-Cor-B is vulnerable to noises, See Figure \ref{fig:isic_ablation_acc_5}-\ref{fig:isic_ablation_acc_6}. Moreover, due to the usage of  curriculum, in Figure \ref{fig:isic_ablation_label_accu} and \ref{fig:pcam_ablation_label_accu}, Co-Correcting has a stepped curve, showing the procedure of label updating starting from easy labels and extending to hard ones. 
			
		\subsubsection{Analysis of time consumption}
			\ADD{We have conducted further comparative experiments on time consumption on the ISIC-Archive dataset. The results are shown in Table \ref{table:time_consumption}.  It can be found that Co-Correcting does not increase the training time and testing time. The training time and testing time are even shorter than those of Co-teaching, Co-teaching+, and DivideMix.}
            \begin{table}[htbp]
                \centering
                \caption{\ADD{Time consumption.}}
                \label{table:time_consumption}
                \begin{tabular}{cccc}
                    \toprule
                    \multirow{2}*{Method} & Total Time  & Training Time & Testing Time\\
                                  ~ & (ALL)             & (Batch Level) & (Batch Level) \\
                    \midrule
                    Standard        &075 min 01 sec     &0.128 sec  &0.045 sec\\
                    Joint Optim     &107 min 45 sec     &0.127 sec  & 0.067 sec\\
                    PENCIL          &078 min 02 sec     &0.132 sec  & 0.047 sec\\
                    Co-teaching     &145 min 08 sec     &0.251 sec  &0.092 sec\\
                    Co-teaching+    &146 min 06 sec     &0.253 sec  &0.092 sec\\
                    DivideMix       &411 min 09 sec     &0.859 sec  & 0.089 sec\\
                    Reweight        &874 min 08 sec     &1.895 sec  & 0.057 sec\\
                    Co-Correcting   &143 min 22 sec     &0.250 sec  & 0.078 sec\\
                    \bottomrule
                \end{tabular}            
            \end{table}

\ADD{\section{Experiments on MNIST}}
    \ADD{\subsection{MNIST Dataset}}
        \ADD{In order to prove the effectiveness of the proposed method and meanwhile better explain the reason why the proposed method is effective, we conducted experiments on MNIST. MNIST is a classic handwritten digit dataset. It contains a training set of 60,000 samples and a testing set of 10,000 samples. There are 10 classes in the dataset representing numbers from 0 to 9. Every image size is $28px*28 px$. For multiple image classifications, there are two approaches to generate noises: (1) Symmetrical flip: flip annotations to other labels with the same probability (2) Pairwise flip: simulate fine-grained classification with noisy labels, where annotator may only make mistakes within very similar classes.}
        
        \ADD{Experimenting on MNIST can better understand the characteristics of the comparison methods. Differed from the previous two medical image datasets, its size is larger. Moreover, the handwriting recognition problem is relatively simple, and there are very few cases where the sample is difficult to judge the category by human, which helps us to further verify the accuracy of the label correction. In this group of experiments, we use the backbone network defined in Co-teaching with 6 convolutional neural layers. We choose the SGD as the optimizer and set the momentum to 0.9, the weight decay to 0.005. All images are augmented by randomly perspective and color jittering. For the pairflip noise experiment, we set $\lambda=3000$ in $pairflip 0.2$ noise and set $\lambda=4000$ in $parflip 0.45$ noise.}
        
    \subsection{Results and analysis}
        \ADD{The overall results of the experiments are shown in Table \ref{table:mnist_acc_sn} and Table \ref{table:mnist_acc_pair}. Co-Correcting also obtains the best results in the multi-classification task on the general image classification dataset. On the whole, the methods of updating the labels have higher classification accuracy under different noise ratios than those of not updating the labels. }
        \begin{table}[htbp]
            \centering
            \caption{\ADD{$ACC_{class}$ on MNIST with symmetric noise.}}
            \label{table:mnist_acc_sn}
            \begin{tabular}{ccccccc}
                \toprule
                Methods             &   $clean$                                     &   0.05                                            &   0.1                                         &   0.2                                         &   0.3                                         &   0.4\\
                \midrule
                Standard            &   99.42                                       &   98.64                                       &   98.00                                       &   95.13                                       &   89.24                                       &   80.19\\
                Joint Optim   	  &   99.48                                       &   \textcolor{blue}{99.45}     &   99.53                                       &   \textcolor{blue}{99.46}     &   \textcolor{blue}{99.44}     &   99.33\\
                PENCIL            	 &   \textcolor{blue}{99.51}         &   99.45                                       &   \textcolor{blue}{99.53}     &   99.45                                       &   99.40                                       &   \textcolor{blue}{99.37}\\
                Co-teaching   	  &   94.53                                       &   94.28                                       &   94.17                                       &   93.70                                       &   92.51                                       &   90.03 \\
                Co-teaching+    &   98.76                                       &   99.26                                       &   99.17                                       &   98.48                                       &   97.1                                            &   94.73 \\
                DivideMix           &   95.66                                       &   99.04                                       &   99.08                                       &   99.09                                       &   99.02                                       &   99.05 \\
                Reweight			& 99.34												& 99.37											& 99.38											& 99.27												& 99.10												& 99.16 \\
                Co-Correcting   &   \textcolor{red}{\textbf{99.56}}&    \textcolor{red}{\textbf{99.50}}&    \textcolor{red}{\textbf{99.57}}&    \textcolor{red}{\textbf{99.53}}&    \textcolor{red}{\textbf{99.53}}&    \textcolor{red}{\textbf{99.52}} \\
                \bottomrule
            \end{tabular}            
        \end{table}  
        \begin{figure}[ht!]
            \centering
            \subfigure[$clean$]{
            \includegraphics[width=2.9cm]{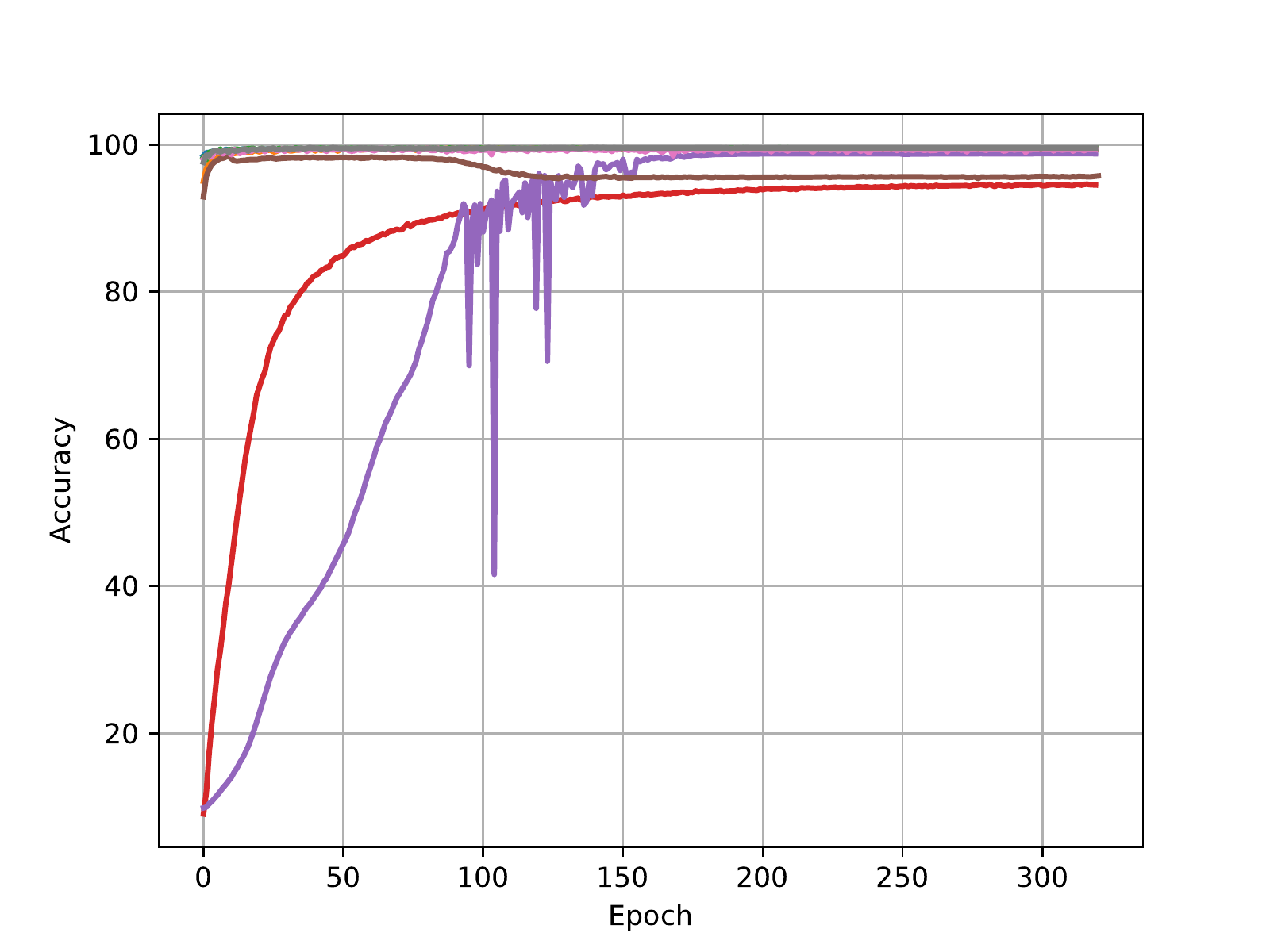} \label{fig:mnist_acc_1}
            }  
            \hspace{-6mm}
            \subfigure[$noise=0.05$]{
            \includegraphics[width=2.9cm]{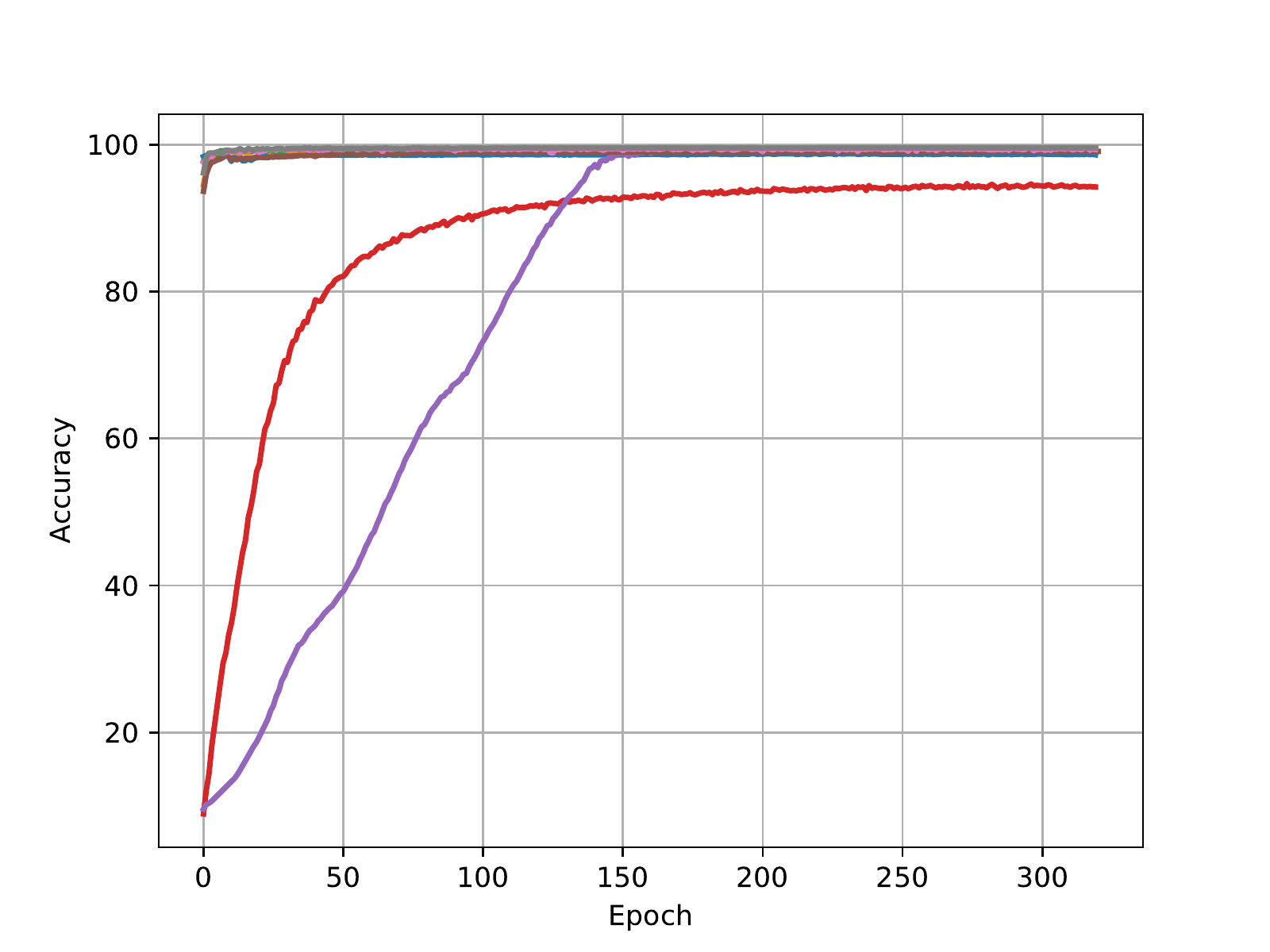} \label{fig:mnist_acc_2} 
            }
            \hspace{-6mm}
            \subfigure[$noise=0.1$]{
            \includegraphics[width=2.9cm]{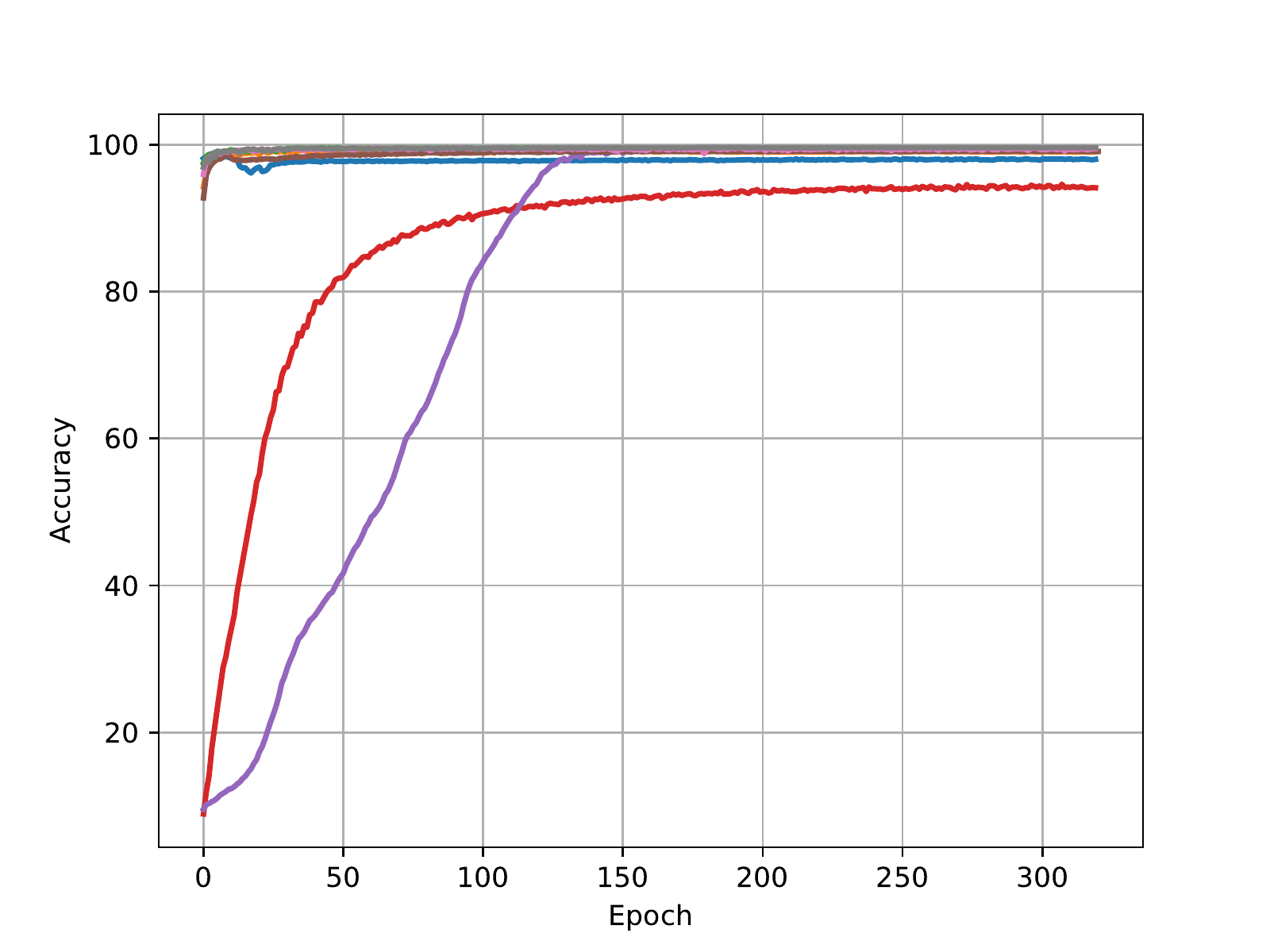}\label{fig:mnist_acc_3}
            }
            \hspace{-6mm}
            \subfigure[$noise=0.2$]{
            \includegraphics[width=2.9cm]{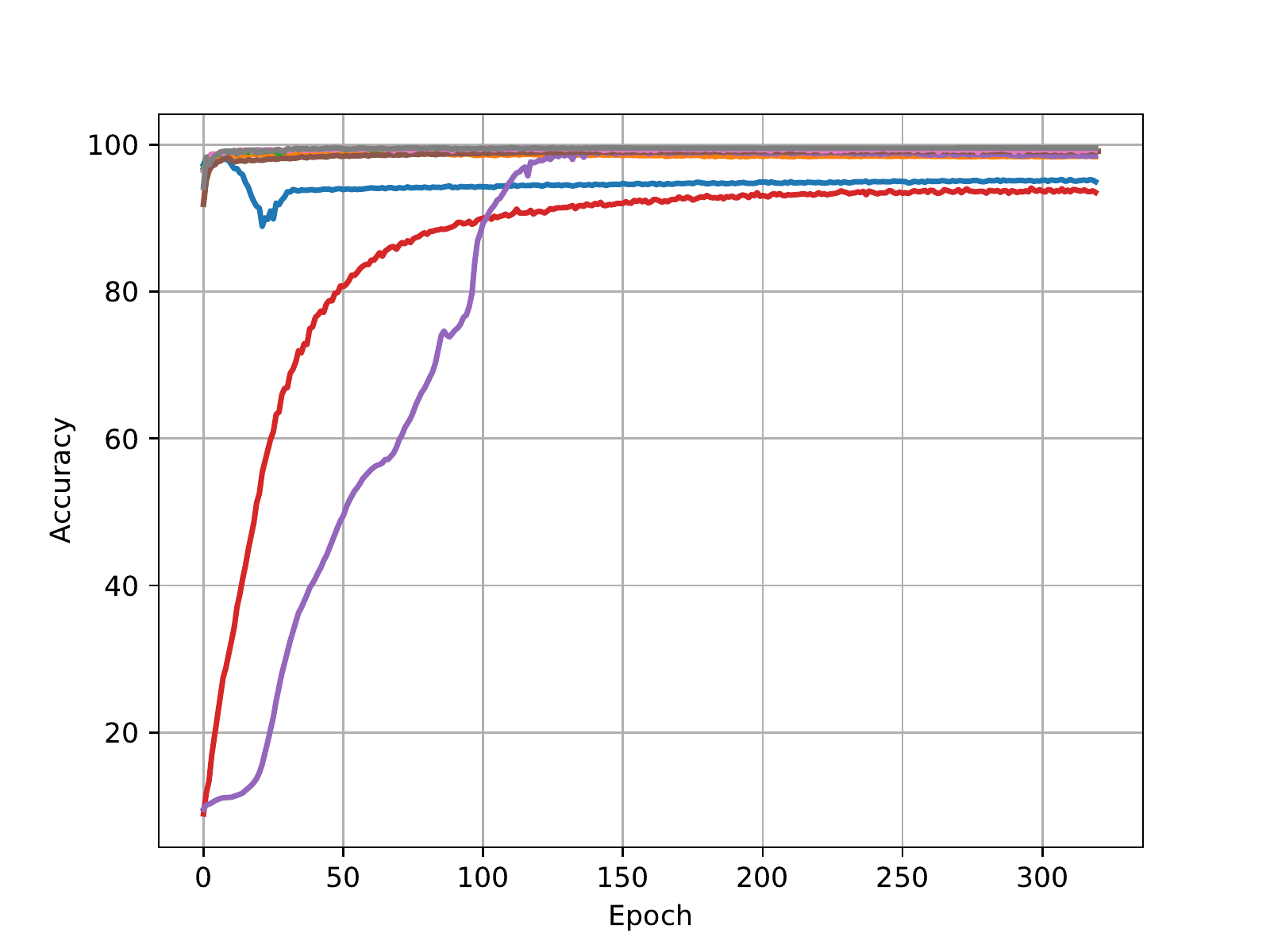}\label{fig:mnist_acc_4}
            }
            \hspace{-6mm}
            \subfigure[$noise=0.3$]{
            \includegraphics[width=2.9cm]{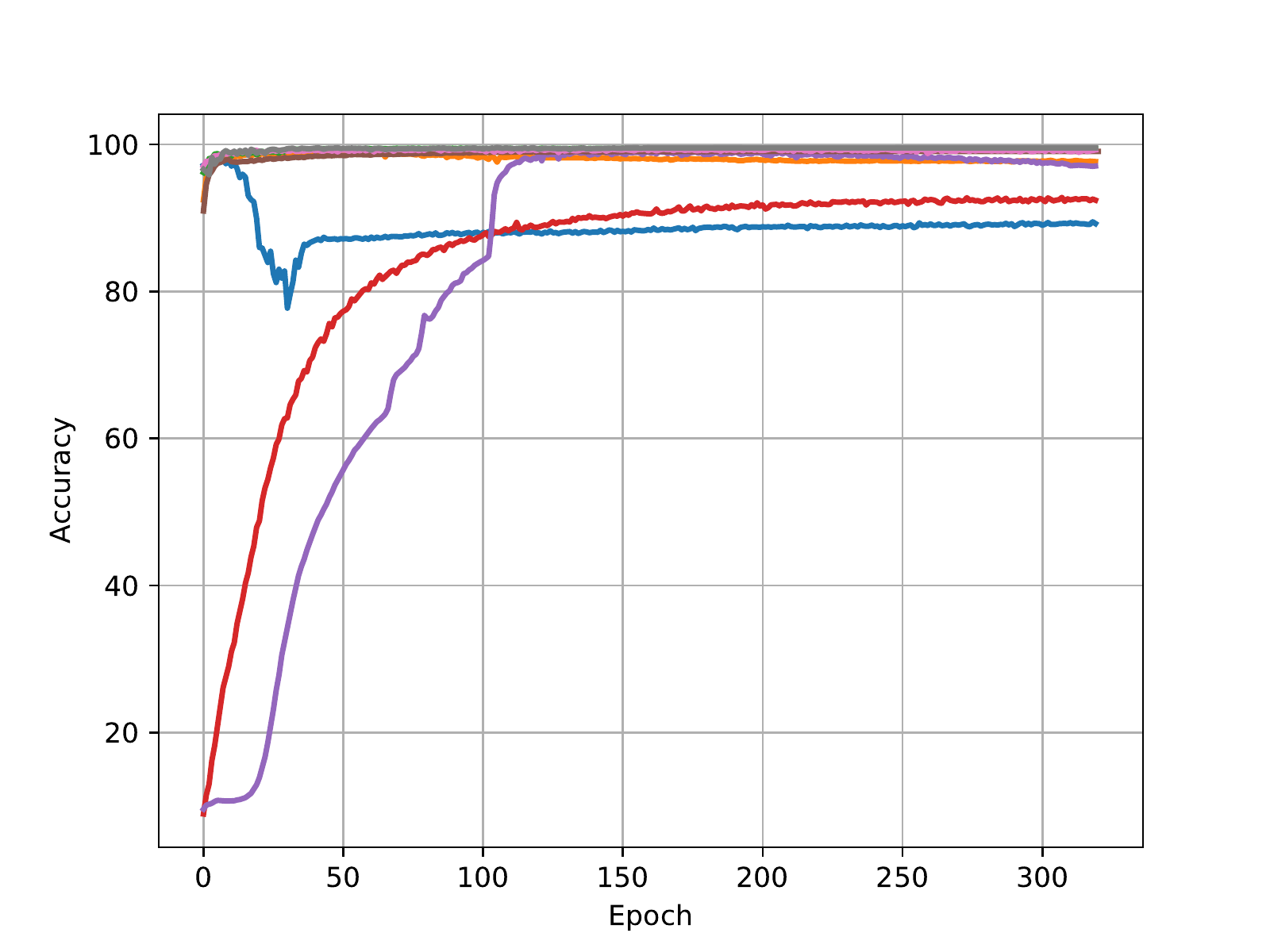}\label{fig:mnist_acc_5}
            }
            \hspace{-6mm}
            \subfigure[$noise=0.4$]{
            \includegraphics[width=2.9cm]{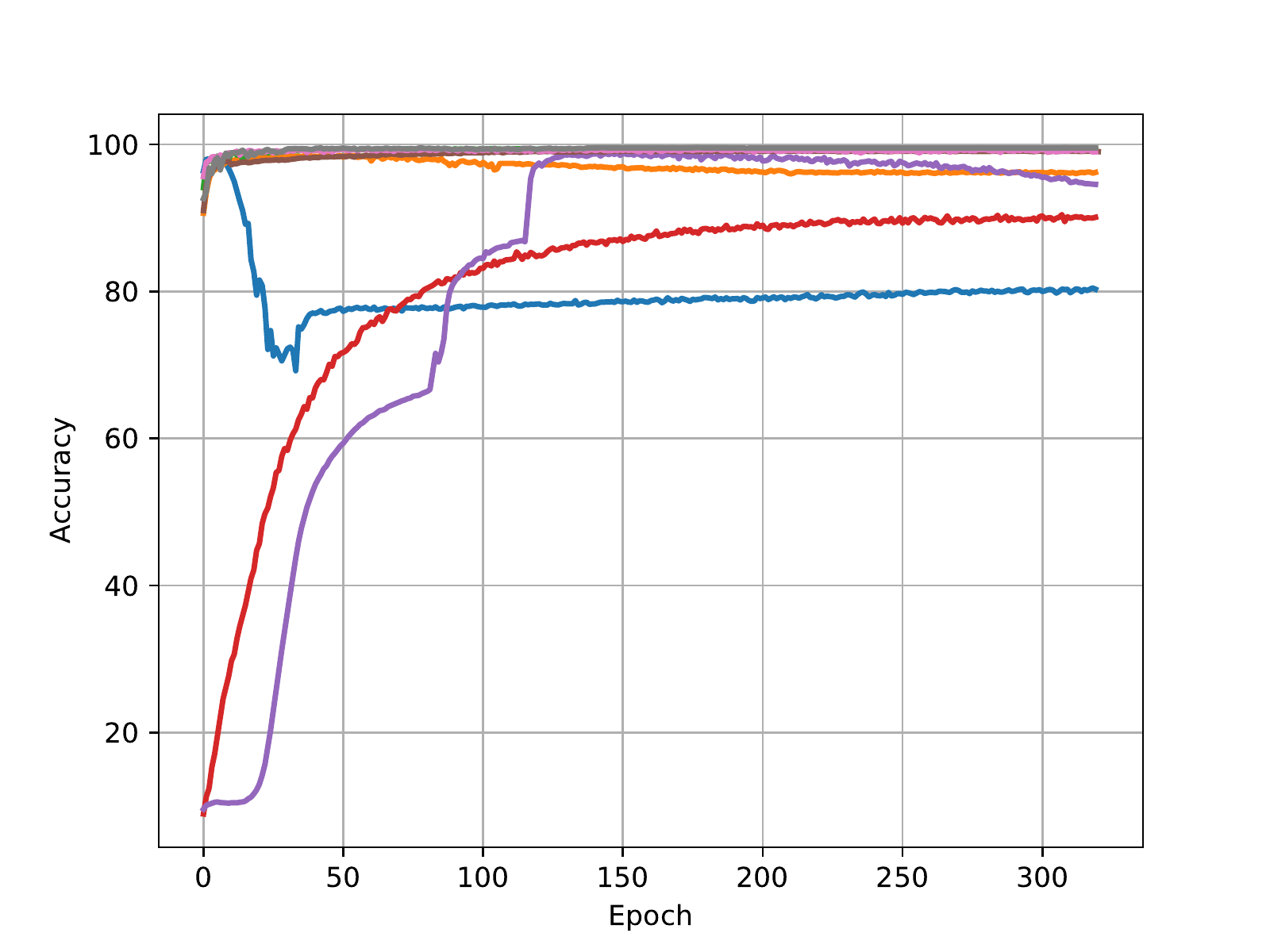}\label{fig:mnist_acc_6}
            }
            \subfigure{
            \includegraphics[width=8cm]{{legend_acc}.pdf}\label{fig:mnist_sy_acc_legend}
            }
            \caption{\ADD{$ACC_{class}$ on MNIST with symmetric noise.}}
            \label{fig:mnist_acc_sn}
        \end{figure}
        
        \ADD{On MNIST, the LNL methods have good performances in the cases of symmetrically adding the noises. This may be due to the more sufficient training data that makes the network easier to get more accurate predictions. }
        
\begin{figure}[htb!]
            \centering
            \subfigure[$noise=0.2$]{
            \includegraphics[width=2.9cm]{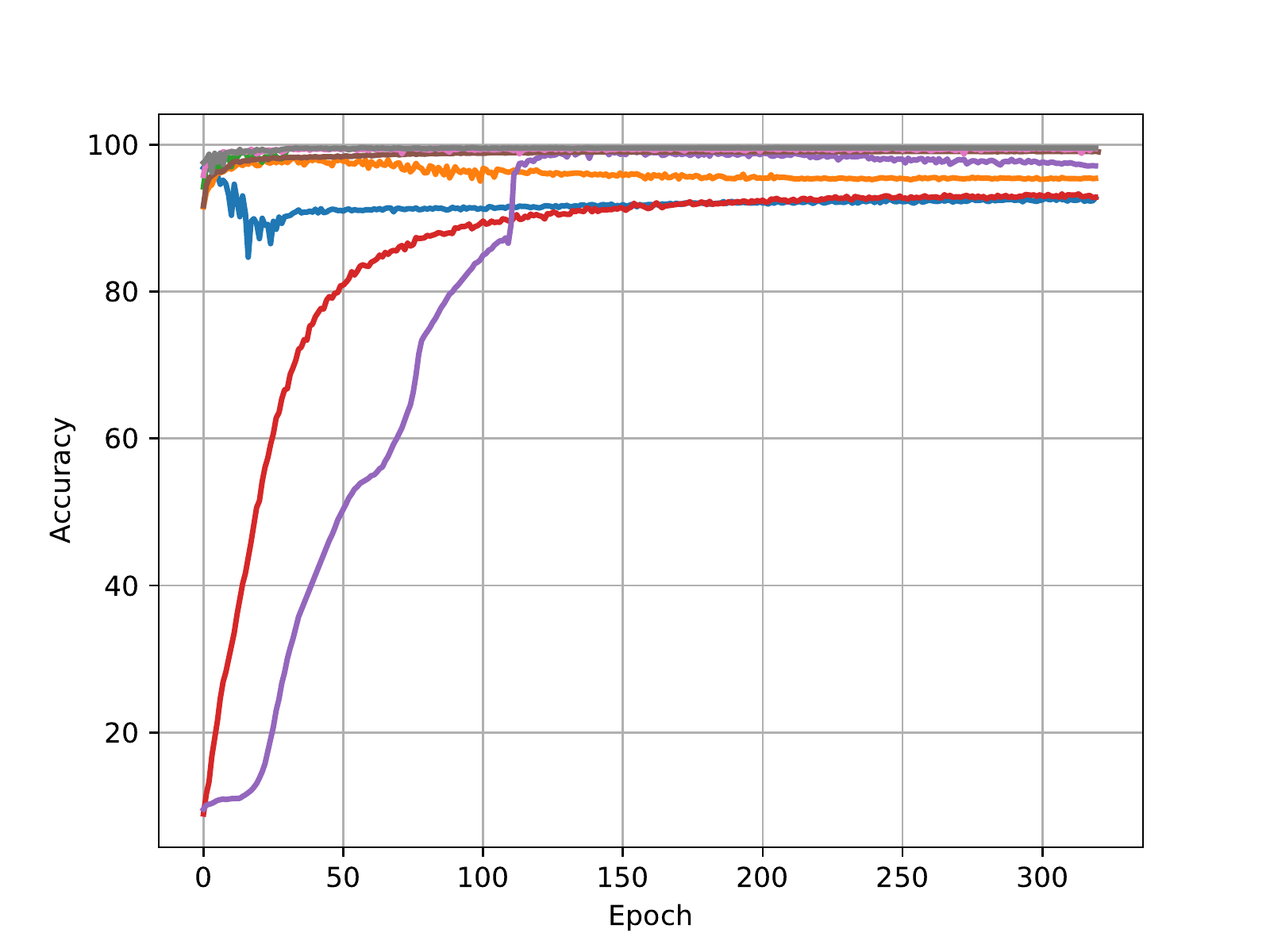}\label{fig:mnist_acc_7}
            }
            \hspace{-6mm}
            \subfigure[$noise=0.45$]{
            \includegraphics[width=2.9cm]{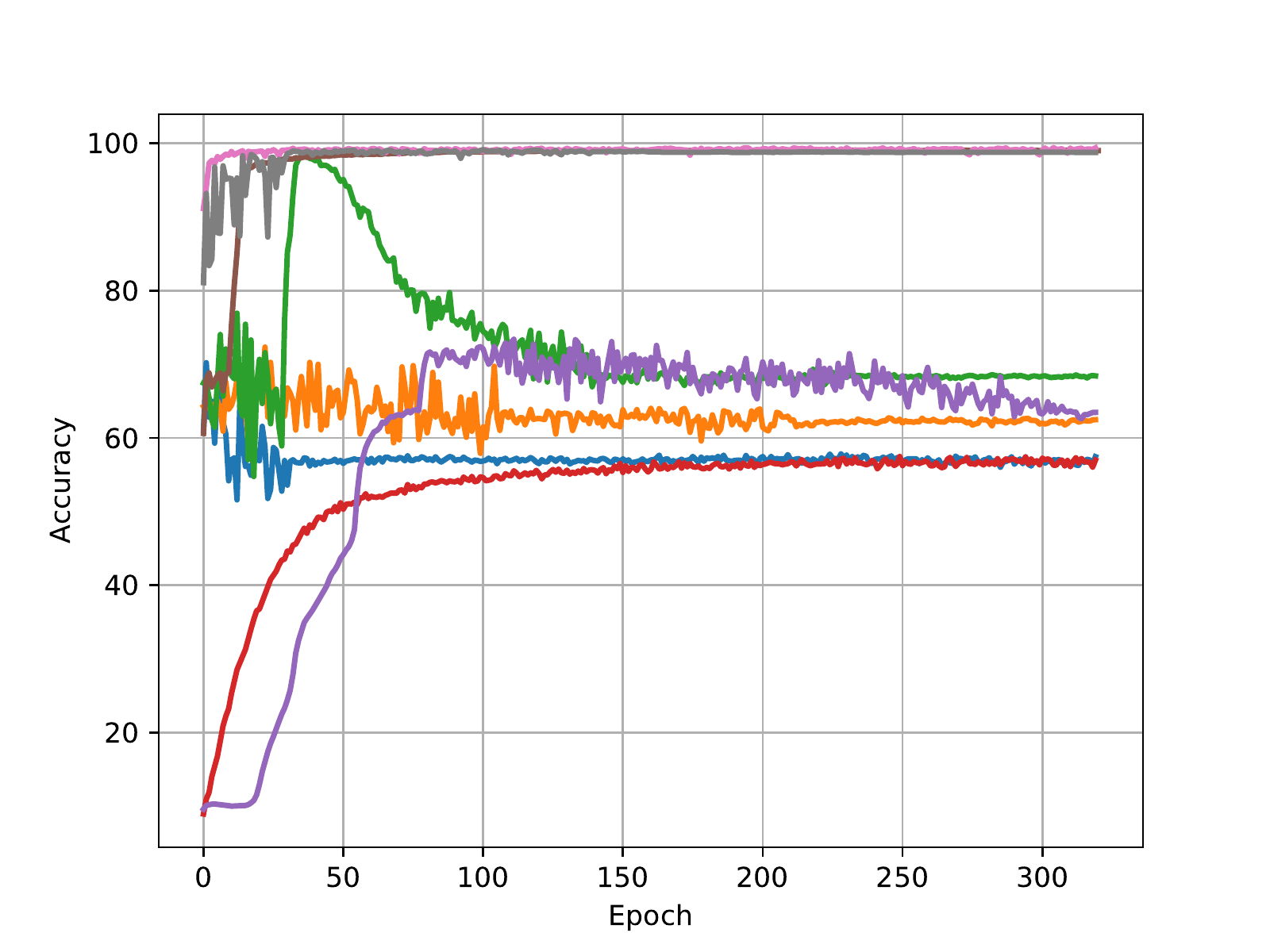}\label{fig:mnist_acc_8}
            }
            \hspace{-6mm}
            \subfigure{
            \includegraphics[width=8cm]{{legend_acc}.pdf}\label{fig:mnist_pair_acc_legend}
            }
            \caption{\ADD{$ACC_{class}$ on MNIST with pairflip noise.}}
            \label{fig:mnist_acc_pair}
        \end{figure}

        \ADD{In order to imitate the actual noise data, we used the pair-flip method to contaminate the samples, and carried out two sets of experiments, $pair0.2$ and $pair0.45$. As Figure \ref{fig:mnist_acc_pair}, We found that in these two sets of results, the gap in the performance of each method increases. Thus we pay more attention to analyzing these two sets of experiments.}
        \begin{table}[htb!]
            \centering
            \caption{\ADD{$ACC_{class}$ on MNIST with pairfilp noise.}}
            \label{table:mnist_acc_pair}
            \begin{tabular}{ccc}
                \toprule
                Methods             & 0.2                                & 0.45\\
                \midrule
                Standard            &          92.46                                       &   56.71\\
                Joint Optim            &   99.40                                       &   \textcolor{blue}{99.34}\\
                PENCIL               &   \textcolor{blue}{99.47}     &   68.37\\
                Co-teaching     &    93.01                                       &   56.71\\
                Co-teaching+    &   97.38                                       &   63.7\\
                DivideMix           & 99.04                                       &   99.05\\
                Reweight			&	99.31												&	99.23 \\
                Co-Correcting   &   \textcolor{red}{\textbf{99.53}}&    \textcolor{red}{\textbf{99.36}}\\
                \bottomrule
            \end{tabular}            
        \end{table}      

\begin{figure}[htb!]
            \centering
            \includegraphics[width=5cm]{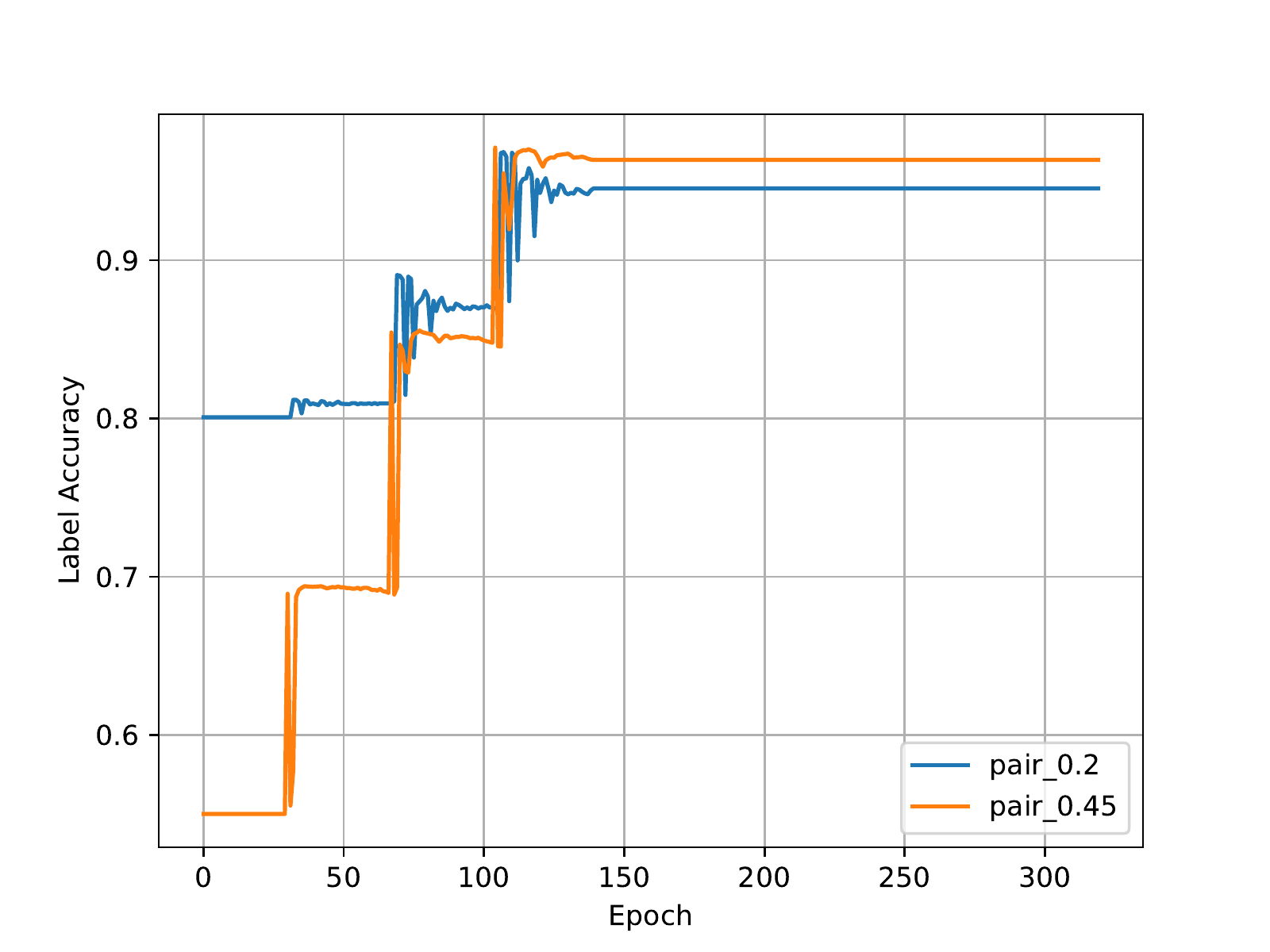}
            \caption{\ADD{$ACC_{label}$ in pairflip noise on MNIST.}}
            \label{fig:label_accu_mnist}
        \end{figure}

        \ADD{It shows that Co-Correcting still performs best, followed by Joint Optim or Dividemix. In the case of $pair 0.45$, PENCIL, Co-teaching and Co-teaching+ completely failed in keeping robust. One reason of our method performs best is that it does not accumulate errors as well as memorize noise. Another reason is that Co-Correcting can improve the label distribution, which can be verified from the label correction results on MNIST. As shown in Figure \ref{fig:label_accu_mnist}, curriculum-based label correction is a progressive learning process, in which the model in the next stage can stably update the noisy labels based on the knowledge verified in the previous stage. }
        
\begin{figure}[!h]
            \centering
            \includegraphics[width=8cm]{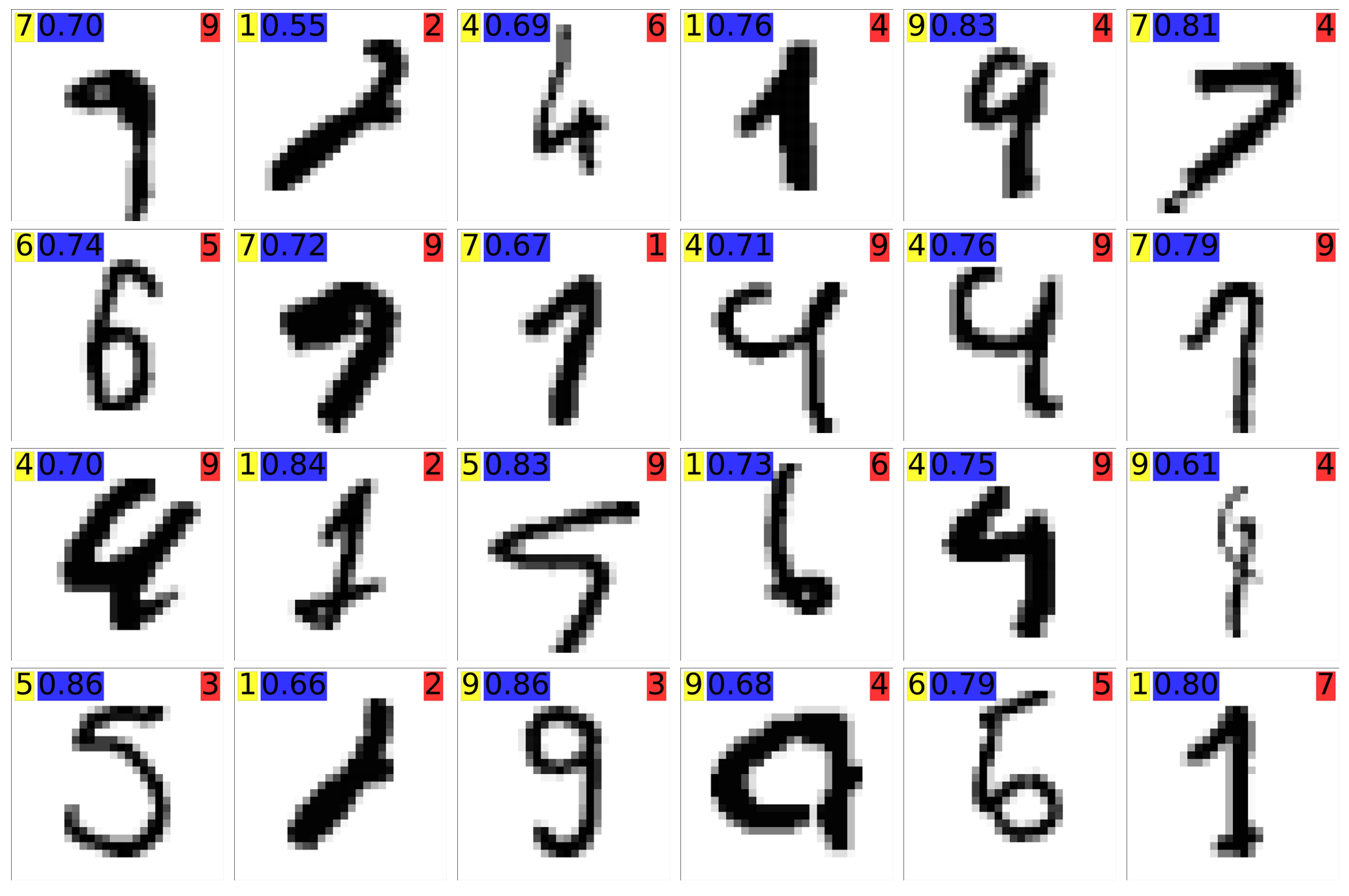}
            \caption{\ADD{Corrected samples in MNIST. The yellow value represents the corrected category and the blue is the corresponding probability. The red value represents the original category.}}
            \label{fig:mnist_sample}
        \end{figure}

        \ADD{The method we propose can often get more accurate annotations than ground truth labels. It can be found that there are a small number of labeling errors in the original dataset. Figure \ref{fig:mnist_sample} shows some instances of errors. In the graphs, the red texts are the original labels, and the yellow ones are revised labels. In these examples, it can be seen that our corrected labels are more accurate. This allows our method not only used to correct the annotations and improve the accuracy of the model, but also to evaluate the quality of the labels in the dataset.}
        
\section{Conclusion}
    In this paper, we present a framework for learning with noisy labels on the medical image dataset named Co-Correcting. The framework consists of three modules: 1) dual-network architecture, focus on preventing the network from being affected by noisy labels; 2) label probabilistic module, to correct the mislabeled data; 3) label correction curriculum, for improving the stability on label correction. We tested and verified the proposed method on two typical medical image datasets: ISIC-Archive and PatchCamelyon. To further verify its effectiveness, we also test it on the general dataset MNIST. Co-Correcting achieves the state-of-the-art result under different noise ratio/types of all datasets. As future work, the proposed method can be adjusted for adapting to the needs of imbalanced data.

\bibliography{bibfile}

\begin{thebibliography}{10}

\bibitem{cheplygina_crowd_2018}
V.~Cheplygina and J.~P.~W. Pluim, ``Crowd disagreement about medical images is
  informative,'' in {\em Intravascular Imaging and Computer Assisted Stenting
  and Large-Scale Annotation of Biomedical Data and Expert Label Synthesis}
  (D.~Stoyanov, Z.~Taylor, S.~Balocco, R.~Sznitman, A.~Martel, L.~Maier-Hein,
  L.~Duong, G.~Zahnd, S.~Demirci, S.~Albarqouni, S.-L. Lee, S.~Moriconi,
  V.~Cheplygina, D.~Mateus, E.~Trucco, E.~Granger, and P.~Jannin, eds.),
  (Cham), pp.~105--111, Springer International Publishing, 2018.

\bibitem{wang_tienet_2018}
X.~Wang, Y.~Peng, L.~Lu, Z.~Lu, and R.~M. Summers, ``{TieNet}: {Text}-{Image}
  {Embedding} {Network} for {Common} {Thorax} {Disease} {Classification} and
  {Reporting} in {Chest} {X}-{Rays},'' in {\em 2018 {IEEE}/{CVF} {Conference}
  on {Computer} {Vision} and {Pattern} {Recognition}}, (Salt Lake City, UT),
  pp.~9049--9058, IEEE, June 2018.

\bibitem{berthelot_mixmatch_2019}
D.~Berthelot, N.~Carlini, I.~Goodfellow, N.~Papernot, A.~Oliver, and C.~A.
  Raffel, ``Mixmatch: A holistic approach to semi-supervised learning,'' in
  {\em Advances in Neural Information Processing Systems} (H.~Wallach,
  H.~Larochelle, A.~Beygelzimer, F.~d\textquotesingle Alch\'{e}-Buc, E.~Fox,
  and R.~Garnett, eds.), vol.~32, pp.~5049--5059, Curran Associates, Inc.,
  2019.

\bibitem{arazo_pseudo-labeling_2020}
E.~{Arazo}, D.~{Ortego}, P.~{Albert}, N.~E. {O’Connor}, and K.~{McGuinness},
  ``Pseudo-labeling and confirmation bias in deep semi-supervised learning,''
  in {\em 2020 International Joint Conference on Neural Networks (IJCNN)},
  pp.~1--8, July 2020.

\bibitem{kuznetsova_open_2020}
A.~Kuznetsova and et~al., ``The {Open} {Images} {Dataset} {V4}: {Unified}
  {Image} {Classification}, {Object} {Detection}, and {Visual} {Relationship}
  {Detection} at {Scale},'' {\em International Journal of Computer Vision},
  vol.~128, pp.~1956--1981, July 2020.

\bibitem{dgani_training_2018}
Y.~Dgani, H.~Greenspan, and J.~Goldberger, ``Training a neural network based on
  unreliable human annotation of medical images,'' in {\em 2018 {IEEE} 15th
  {International} {Symposium} on {Biomedical} {Imaging} ({ISBI} 2018)},
  (Washington, DC), pp.~39--42, IEEE, Apr. 2018.

\bibitem{han_co-teaching_2018}
B.~Han, Q.~Yao, X.~Yu, G.~Niu, M.~Xu, W.~Hu, I.~W. Tsang, and M.~Sugiyama,
  ``Co-teaching: Robust training of deep neural networks with extremely noisy
  labels,'' in {\em Proceedings of the 32nd International Conference on Neural
  Information Processing Systems}, NIPS'18, (Red Hook, NY, USA),
  p.~8536–8546, Curran Associates Inc., 2018.

\bibitem{yu_how_2019}
X.~Yu, B.~Han, J.~Yao, G.~Niu, I.~Tsang, and M.~Sugiyama, ``How does
  disagreement help generalization against label corruption?,'' in {\em
  Proceedings of the 36th International Conference on Machine Learning}
  (K.~Chaudhuri and R.~Salakhutdinov, eds.), vol.~97 of {\em Proceedings of
  Machine Learning Research}, pp.~7164--7173, PMLR, 09--15 Jun 2019.

\bibitem{karimi_deep_2020}
D.~Karimi, H.~Dou, S.~K. Warfield, and A.~Gholipour, ``Deep learning with noisy
  labels: Exploring techniques and remedies in medical image analysis,'' {\em
  Medical Image Analysis}, vol.~65, p.~101759, 2020.

\bibitem{noauthor_isic_nodate}
ISIC-Archive, ``The international skin imaging collaboration: Melanoma
  project.'' Website, July 2020.

\bibitem{ehteshami_bejnordi_diagnostic_2017}
E.~B. et~al., ``Diagnostic {Assessment} of {Deep} {Learning} {Algorithms} for
  {Detection} of {Lymph} {Node} {Metastases} in {Women} {With} {Breast}
  {Cancer},'' {\em JAMA}, vol.~318, p.~2199, Dec. 2017.

\bibitem{veeling_rotation_2018}
B.~S. Veeling, J.~Linmans, J.~Winkens, T.~Cohen, and M.~Welling, ``Rotation
  equivariant cnns for digital pathology,'' in {\em Medical Image Computing and
  Computer Assisted Intervention -- MICCAI 2018} (A.~F. Frangi, J.~A. Schnabel,
  C.~Davatzikos, C.~Alberola-L{\'o}pez, and G.~Fichtinger, eds.), (Cham),
  pp.~210--218, Springer International Publishing, 2018.

\bibitem{esteva_dermatologist-level_2017}
A.~Esteva and et~al., ``Dermatologist-level classification of skin cancer with
  deep neural networks,'' {\em Nature}, vol.~542, pp.~115--118, Feb. 2017.

\bibitem{pranav_chexnet_2017}
P.~Rajpurkar and et~al., ``Chexnet: Radiologist-level pneumonia detection on
  chest x-rays with deep learning,'' {\em CoRR}, vol.~abs/1711.05225, 2017.

\bibitem{wang_effectiveness_2017}
L.~Perez and J.~Wang, ``The effectiveness of data augmentation in image
  classification using deep learning,'' {\em CoRR}, vol.~abs/1712.04621, 2017.

\bibitem{kermany_identifying_2018}
D.~S. Kermany and et~al., ``Identifying {Medical} {Diagnoses} and {Treatable}
  {Diseases} by {Image}-{Based} {Deep} {Learning},'' {\em Cell}, vol.~172,
  pp.~1122--1131.e9, Feb. 2018.

\bibitem{miyato_virtual_2019}
T.~Miyato, S.-I. Maeda, M.~Koyama, and S.~Ishii, ``Virtual {Adversarial}
  {Training}: {A} {Regularization} {Method} for {Supervised} and
  {Semi}-{Supervised} {Learning},'' {\em IEEE Transactions on Pattern Analysis
  and Machine Intelligence}, vol.~41, pp.~1979--1993, Aug. 2019.

\bibitem{patrini_making_2017}
G.~Patrini, A.~Rozza, A.~Krishna~Menon, R.~Nock, and L.~Qu, ``Making deep
  neural networks robust to label noise: A loss correction approach,'' in {\em
  Proceedings of the IEEE Conference on Computer Vision and Pattern Recognition
  (CVPR)}, July 2017.

\bibitem{jiang_mentornet_2018}
L.~Jiang, Z.~Zhou, T.~Leung, L.~Li, and L.~Fei-Fei, ``Mentornet: Learning
  data-driven curriculum for very deep neural networks on corrupted labels,''
  in {\em ICML}, 2018.

\bibitem{tanaka_joint_2018}
D.~{Tanaka}, D.~{Ikami}, T.~{Yamasaki}, and K.~{Aizawa}, ``Joint optimization
  framework for learning with noisy labels,'' in {\em 2018 IEEE/CVF Conference
  on Computer Vision and Pattern Recognition}, pp.~5552--5560, June 2018.

\bibitem{yi_probabilistic_2019}
K.~{Yi} and J.~{Wu}, ``Probabilistic end-to-end noise correction for learning
  with noisy labels,'' in {\em 2019 IEEE/CVF Conference on Computer Vision and
  Pattern Recognition (CVPR)}, pp.~7010--7018, June 2019.

\bibitem{li_dividemix_2020}
J.~Li, R.~Socher, and S.~C. Hoi, ``Dividemix: Learning with noisy labels as
  semi-supervised learning,'' in {\em International Conference on Learning
  Representations}, 2020.

\bibitem{bekker_training_2016}
A.~J. {Bekker} and J.~{Goldberger}, ``Training deep neural-networks based on
  unreliable labels,'' in {\em 2016 IEEE International Conference on Acoustics,
  Speech and Signal Processing (ICASSP)}, pp.~2682--2686, March 2016.

\bibitem{xue_robust_2019}
C.~{Xue}, Q.~{Dou}, X.~{Shi}, H.~{Chen}, and P.~{Heng}, ``Robust learning at
  noisy labeled medical images: Applied to skin lesion classification,'' in
  {\em 2019 IEEE 16th International Symposium on Biomedical Imaging (ISBI
  2019)}, pp.~1280--1283, April 2019.

\bibitem{le_pancreatic_2019}
H.~Le and et~al., ``Pancreatic cancer detection in whole slide images using
  noisy label annotations,'' in {\em Medical Image Computing and Computer
  Assisted Intervention -- MICCAI 2019}, (Cham), pp.~541--549, Springer
  International Publishing, 2019.

\bibitem{ren_learning_2019}
M.~Ren, W.~Zeng, B.~Yang, and R.~Urtasun, ``Learning to reweight examples for
  robust deep learning,'' in {\em Proceedings of the 35th International
  Conference on Machine Learning} (J.~Dy and A.~Krause, eds.), vol.~80 of {\em
  Proceedings of Machine Learning Research}, (Stockholmsmässan, Stockholm
  Sweden), pp.~4334--4343, PMLR, 10--15 Jul 2018.

\bibitem{liu_classification_2016}
T.~{Liu} and D.~{Tao}, ``Classification with noisy labels by importance
  reweighting,'' {\em IEEE Transactions on Pattern Analysis and Machine
  Intelligence}, vol.~38, pp.~447--461, March 2016.

\bibitem{yu_efficient_2018}
X.~Yu, T.~Liu, M.~Gong, K.~Batmanghelich, and D.~Tao, ``An {Efficient} and
  {Provable} {Approach} for {Mixture} {Proportion} {Estimation} {Using}
  {Linear} {Independence} {Assumption},'' in {\em 2018 {IEEE}/{CVF}
  {Conference} on {Computer} {Vision} and {Pattern} {Recognition}}, (Salt Lake
  City, UT), pp.~4480--4489, IEEE, June 2018.

\bibitem{gao_deep_2017}
B.-B. Gao, C.~Xing, C.-W. Xie, J.~Wu, and X.~Geng, ``Deep label distribution
  learning with label ambiguity,'' {\em Trans. Img. Proc.}, vol.~26,
  p.~2825–2838, June 2017.

\bibitem{bengio_curriculum_2009}
Y.~Bengio, J.~Louradour, R.~Collobert, and J.~Weston, ``Curriculum learning,''
  in {\em Proceedings of the 26th {Annual} {International} {Conference} on
  {Machine} {Learning} - {ICML} '09}, (Montreal, Quebec, Canada), pp.~1--8, ACM
  Press, 2009.

\bibitem{guo_curriculumnet_2018}
S.~Guo, W.~Huang, H.~Zhang, C.~Zhuang, D.~Dong, M.~R. Scott, and D.~Huang,
  ``Curriculumnet: Weakly supervised learning from large-scale web images,'' in
  {\em Computer Vision -- ECCV 2018} (V.~Ferrari, M.~Hebert, C.~Sminchisescu,
  and Y.~Weiss, eds.), (Cham), pp.~139--154, Springer International Publishing,
  2018.

\end{thebibliography}

\clearpage
\section{Appendix}
	\subsection{Variables}
		\ADD{We list all the variables as well as their descriptions and suggested values in Table \ref{table:all_param}.}

\begin{table*}[htbp!]
            \centering
            \caption{\ADD{Description of variables.}}
            \label{table:all_param}
            \begin{tabularx}{\textwidth}{cp{6cm}ccp{6cm}}
                \toprule
                Parameter & Description & Module & Position & Suggestion \\
                \midrule
                $b$ & Mini-batch size.  & -- & Section \ref{section:dual-networks} & Setting as Standard.\\
                $p_i$ &  Prediction of $i_{th}$ sample.  & -- & Section \ref{section:dual-networks} & \makecell[c]{--}\\
                $P$ & A set of predictions for a mini-batch of samples.  & -- & Section \ref{section:dual-networks} & \makecell[c]{--}\\
                $e$ & Current epoch number.  & -- & Section \ref{section:dual-networks} & \makecell[c]{--}\\
                $c$ & Number of classes.  & -- & Section \ref{section:loss} & \makecell[c]{--}\\
                $\bar{D}$ & Samples with consistent prediction by dual-networks.  & Dual-networks & Section \ref{section:dual-networks} & Selected by Formula \ref{equ:div}.\\
                $E_k$ & Number of epochs to decrease $R(t)$.  & Dual-networks & Section \ref{section:dual-networks} & $E_k$ should set to a small value corresponding to the convergence speed.  In our Experiments, $E_k=10$.\\
                $\tau$ & Decrease range of $R(t)$. & Dual-networks & Section \ref{section:dual-networks} &$\tau$ should equal to the noise rate.\\
                $R(t)$ & Memory rate.  & Dual-networks & Section \ref{section:dual-networks} & Calculated by Formula \ref{equ:rt}.\\
                $m$ & Network id.  & Dual-networks & Section \ref{section:curriculum} & \makecell[c]{--}\\
                $y^d$ & Label distribution. & Label update & Section \ref{section:label_update} & Updated by Formula \ref{equ:init_yd}.\\
                $\hat{y}$ & Original noisy labels. & Label update & Section \ref{section:label_update} & \makecell[c]{--}\\
                $\tilde{y}$ & An auxiliary variabel to assit the label update. & Label update & Section \ref{section:label_update} & Initialized by Formula \ref{equ:init_tilde_y} and Updated by Formula \ref{equ:udpate_y}.\\
                $K$ & A constant value for initializing label distribution. & Label update & Section \ref{section:label_update} & $K=10$\\
                $\lambda$ & Label correction rate. & Label update & Section \ref{section:label_update} & $\lambda$ should increasing with class number and noise rate.\\
                $\delta$ & Gradient. & Label update & Section \ref{section:label_update} & \makecell[c]{--}\\
                $l_c$ & Classification loss.  & Label update & Section \ref{section:loss} & Calculated by Formula \ref{equ:l_stage1} or \ref{equ:l_stage4}.\\
                $l_o$ & Compatibility loss. & Label update & Section \ref{section:loss} & Calculated by Formula \ref{equ:l_o}.\\
                $l_e$ & Entropy loss. & Label update & Section \ref{section:loss} & Calculated by Formula \ref{equ:l_e}.\\
                $\alpha$ & Weight to control compatibility loss $l_o$.  & Label update & Section \ref{section:loss} & Reduce $\alpha$ in extreme noise.\\
                $\beta$ & Weight to control compatibility loss $l_e$.  & Label update & Section \ref{section:loss} & $\beta=0.1$\\
                $D$ & Confusion matrix in feature space for all samples measured by Euclidean distance. & Curriculum & Section \ref{section:curriculum} & Calculated by Formula \ref{equ:d}.\\
                $k$ & The threshold for $d_c$ selection. & Curriculum & Section \ref{section:curriculum} & $k\in(50,70)$\\
                $d_c$ & The $k_th$ smallest value of distance matrix D.  & Curriculum & Section \ref{section:curriculum} & \makecell[c]{--}\\
                $s_i$ & A density of sample in curriculum setting. & Curriculum & Section \ref{section:curriculum} & Calculated by Formula \ref{equ:s}.\\
                $\varepsilon$ & A distance for set curriculums.  & Curriculum & Section \ref{section:curriculum} & Calculated by Formula \ref{equ:distance}.\\
                \bottomrule
            \end{tabularx}
        \end{table*} 

\end{document}